%% file: Manuscript-r02-Arxiv.tex
\documentclass[12pt, draftclsnofoot, onecolumn]{Components/MetaFiles/IEEEtran}

\input{Components/MetaFiles/packages}
\input{Components/MetaFiles/acronyms}

\begin{document}

\title{On the Energy Efficiency of MIMO Hybrid Beamforming for Millimeter Wave Systems with
Nonlinear Power Amplifiers}
\author{Nima~N.~Moghadam,~\IEEEmembership{Member,~IEEE,}
        G\'{a}bor~Fodor,~\IEEEmembership{Senior~Member,~IEEE,}
        Mats~Bengtsson,~\IEEEmembership{Senior~Member,~IEEE,}
    and~David~J.~Love,~\IEEEmembership{Fellow, IEEE}
\thanks{The work of N.~N.~Moghadam and G.~Fodor was partially financed by Ericsson Research through the HARALD project.
The work of D.~J.~Love was supported in part by the National Science Foundation under grant NSF CCF1403458.}
\thanks{N.~N.~Moghadam and M. Bengtsson are with
the School of Electrical Engineering, KTH Royal Institute of Technology, 100 44 Stockholm, Sweden (e-mail: nimanm@kth.se; mats.bengtsson@ee.kth.se).}
\thanks{G. Fodor is with the School of Electrical Engineering, KTH Royal Institute of Technology, 100 44 Stockholm, Sweden, and also with Ericsson Research, 164 83 Kista, Sweden (e-mail: gaborf@kth.se).}
\thanks{D. J. Love is with the School of Electrical and Computer
Engineering, Purdue University, West Lafayette, IN 47906 USA (e-mail: djlove@purdue.edu).}
}
\input{Components/MetaFiles/commands}

\maketitle

\begin{abstract}
Multiple-input multiple-output (MIMO) millimeter wave (mmWave) systems are vulnerable to hardware impairments due to operating at high frequencies and employing a large number of radio-frequency (RF) hardware components. In particular, nonlinear power amplifiers (PAs) employed at the transmitter distort the signal when operated close to saturation due to energy efficiency considerations. In this paper, we study the performance of a MIMO mmWave hybrid beamforming scheme in the presence of nonlinear PAs.
First, we develop a statistical model for the transmitted signal in such systems and show that the spatial direction of the inband distortion is shaped by the beamforming filter. This suggests that even in the large antenna regime, where narrow beams can be steered toward the receiver, the impact of nonlinear PAs should not be ignored. Then, by employing a realistic power consumption model for the PAs, we investigate the trade-off between spectral and energy efficiency in such systems. Our results show that increasing the transmit power level when the number of transmit antennas grows large can be counter-effective in terms of energy efficiency. Furthermore, using numerical simulation, we show that when the transmit power is large, analog beamforming leads to higher spectral and energy efficiency compared to digital and hybrid beamforming schemes.
\end{abstract}


\section{Introduction}
\ac{LS-MIMO} systems involving
an order of magnitude greater number of antenna elements
than in the early releases of wireless standards {are
key enablers} of next generation mobile broadband services~\cite{Larsson2014}.
Theoretically, a fully digital \ac{LS-MIMO} beamforming architecture employing a large number of digital
transmit and receiver chains can yield optimal performance in terms of energy and spectral
efficiency~\cite{Han2015}.

However, deploying \ac{LS-MIMO} systems in traditional cellular frequency bands is
problematic due to the large physical size of the antenna arrays and related environmental
concerns of the general public.
Therefore, higher frequency bands, including the
\ac{mmWave} bands have recently emerged as an appealing alternative for
the commercial deployment of \ac{LS-MIMO} systems \cite{Rangan2014}.
Indeed, in \ac{mmWave} bands, the physical array size can be greatly reduced, and, as an
additional advantage, vast amount of unused spectrum can be utilized for
attractive and bandwidth-demanding services \cite{Pi:11}, \cite{Hur2013}.

Deploying a large number of antennas with the associated fully digital beamforming architecture
incurs high cost and increased power consumption, due to the excessive demand for a large
number of transceiver chains.
Therefore, \ac{LS-MIMO} systems with hybrid analog and digital
beamforming for \ac{mmWave} deployment have attracted much attention from the research and engineering
communities, and a great number of promising hybrid architectures
and associated technologies such
as training sequence and codebook designs
have been proposed and tested
in practice \cite{Ayach2014,Alkhateeb2014, Kutty2016, Noh2016, Song2017,He2017}.
The results of the marriage of \ac{LS-MIMO} and hybrid
beamforming include significant gains in terms of spectral and energy efficiency, and a cost-efficient
technology for accessing large amount of unused spectrum \cite{Han2015, Shokri-Ghadikolaei2016, Noh2016}.


In practice, the performance and scalability of \ac{LS-MIMO} systems are confined
by a variety of hardware limitations and impairments that distort the transmitted and
received signals \cite{Studer2010, Qi2010, Studer2011, Bjoernson2012}.
The recognition of the importance of analysing and overcoming the impact of non-ideal hardware
and, in particular, nonlinear \acp{PA} on \ac{LS-MIMO} performance has triggered intensive research
resulting in valuable insights.

First, the distortion introduced in the transmit signal by an \ac{LS-MIMO} transmitter
is mainly caused by \ac{RF}
impairments, such as in-phase/quadrature-phase imbalance, crosstalk, and, predominantly,
by high power amplifier (HPA) nonlinearity, especially when HPAs operate close to saturation
\cite{Ghannouchi2009,Qi2010, Qi2012}.
Conventionally, applying a large back-off from the saturation power of a \ac{PA}
has been considered as a solution for decreasing the nonlinear distortion
since reducing the transmit power allows the \acp{PA} to operate in their
linear operating region \cite{Zavjalov:16}.
A serious disadvantage of this solution is that backing off from the saturation level causes
\acp{PA} to work less energy efficiently, because the \ac{PA}'s 
ability to generate \ac{RF} energy decreases when operating away from the saturation point
\cite{Schenk2008}.
Secondly, the negative effect of nonlinear distortion can be mitigated by employing waveforms
with low \ac{PAPR}, because signals with a low \ac{PAPR} are less sensitive to distortion
than signals with higher \ac{PAPR}.
Unfortunately, \ac{PAPR} reduction typically reduces the spectral efficiency,
that can only partially be compensated by increased
complexity and cost at the receivers \cite{Lucciardi:16}.

These two observations imply that there is an inherent trade-off between the targeted
energy and spectral efficiency and the distortion generated at the transmitter, as has
been investigated in \cite{Ochiai2013}.
To find near optimum operating points for \ac{LS-MIMO} systems built on a hybrid beamforming
architecture within the constraints of this trade-off is challenging, and
requires an accurate model of the
distortions caused by hardware impairments including the non-linearities of \acp{PA}.

To this end, a common approach
is to
represent the spatial properties of the
distortion as additive white
Gaussian noise (AWGN)
signals at different antenna elements
\cite{Studer2010,Studer2011,Bjoernson2012,Bjornson2013,Bjoernson2014,Bjoernson2014a,Brandt2014}.
This model assumes that the distortion signals are independent across the different antenna elements
and that the distortion power at each antenna element is
a monotonically increasing function of the signal power fed to the corresponding antenna branch.
These assumptions hold only after sufficient calibrations and compensations where the combined residual of a wide range of independent hardware impairments give rise to an additive distortion signal.
Unfortunately, the AWGN-based distortion signal model may not be appropriate when
the distortion is predominantly generated by the transmitter's \acp{PA} working close to saturation
aiming at high spectral and energy efficiency targets.
In particular, as pointed out in~\cite{Moghadam2012}, the spatial direction of
the transmitted distortion is dependent on the spatial direction of the transmitted signal,
while the AWGN model fails to capture this dependency.

Therefore, in this paper our main objective is to formulate a model that
provides a more precise characterization of the statistical properties of the distortion, than the AWGN-based distortion signal model.
We use this model to determine the achievable rate and energy efficiency of \ac{LS-MIMO} systems
built on a hybrid analog-digital architecture and operating in \ac{mmWave} frequency bands in the presence of nonlinear distortion.
The analysis is based on the assumption that the PAs have the same transfer function, for all the transmitter branches. Moreover, in general we assume that the crosstalk between the antenna branches is negligible due to proper isolation. However, in Section~\ref{sec: Crosstalk}, we extend our model to describe the system impaired with crosstalk as well.
In particular, we formulate the problem of maximizing the energy efficiency of this system
as an optimization task in the digital and analog precoding matrices subject
to sum-power constraints.


The rest of the paper is structured as follows.
Section~\ref{sec: Related Works} presents a summary of the related work.
Section~\ref{sec: system model} describes the system model that we used in this paper.
In Section~\ref{sec: Nonlinear Power Amplification}, we derive a model for a nonlinearly
amplified signal at a multiantenna transmitter. In this section, we further extend our model to describe the system impaired with crosstalk.
Section~\ref{sec: Spectral Efficiency}
and Section~\ref{sec: Energy Efficiency} study the spectral and energy efficiency of the system, respectively.
We present simulation results in Section~\ref{sec: Numerical Results},
followed by concluding remarks
in Section~\ref{sec: Conclusions}.

\emph{Notations:}
Capital bold letters denote matrices and lower bold letters denote vectors.
The superscripts $\vec{X}^*$, $\vec{X}\tran$, $\vec{X}\herm$ stand for the conjugate, transpose, transpose conjugate of $\vec{X}$, respectively.
$[\mathbf{X}]_{ij}$ is the entry of $\mathbf{X}$ at row $i$ and column $j$.
$|x|$ is the absolute value of $x$.
$\mathbf{X}\odot \mathbf{Y}$ denotes the Hadamard (entry-wise) product of matrices $\mathbf{X}$ and $\mathbf{Y}$.
$\vec{I}_x$ is an $x\times x$ identity matrix and $\diag(\mathbf{x})$ is a diagonal matrix with entries of $\mathbf{x}$ on its principal diagonal.
The set of \ac{PSD} matrices of size $n$ is denoted by $\mathbb{S}^n$ and
$\mathbb{R}^{+}$ represents the set of nonnegative real numbers.

\subsection{Related Works and Contributions of the Present Paper}
\label{sec: Related Works}

\subsubsection{Papers Analyzing the Combined Effects of Hardware Impairments}
{A large body of research} {has investigated} the aggregate
impacts of  \ac{RF} hardware impairments on the performance of \ac{MIMO} systems,
see for example ~\cite{Schenk2008,Studer2010,Studer2011,Bjoernson2013,Bjoernson2014,Bjoernson2015,Bjoernson2012,Brandt2014,Sabbaghian2013}.
The effects of transmit-receive hardware
impairments on the capacity of the \ac{MIMO} channel and, in particular, \ac{MIMO} detection algorithms are studied in \cite{Studer2010}.
This analysis is based on an independent and
identically distributed (i.i.d.) Gaussian model for the distortion caused by the
hardware impairments.
The system-level implications of residual transmit-\ac{RF} impairments in MIMO systems
are studied in \cite{Studer2011} using a similar modeling approach as in \cite{Studer2010}.
In \cite{Bjoernson2013}, it is shown that the physical \ac{MIMO} channel has a finite
upper capacity limit for any channel distribution and \ac{SNR}, while the results in
\cite{Bjoernson2014} indicate that the hardware impairments create finite ceilings
on the channel estimation accuracy and on the downlink/uplink capacity of each served
\ac{UE} in cellular \ac{MIMO} systems. The aggregate effects of hardware imperfections
including phase-noise, non-linearities, quantization errors, noise amplification and
inter-carrier interference are formulated as practical hardware scaling laws
in \cite{Bjoernson2015}, which proposes circuit-aware design of \ac{LS-MIMO} systems.
In~\cite{Sabbaghian2013}, an information theoretic approach is used in order to bound
the capacity of a point-to-point single-antenna system,
with nonlinearities at both transmitting and receiving sides.

Multicell coordinated beamforming algorithms in the presence of the aggregate effects of
hardware impairments are studied in \cite{Bjoernson2012} and \cite{Brandt2014}.
These works suggest that impairments-aware beamforming algorithms and resource allocation
are feasible and yield superior performance as compared with algorithms that assume
ideal hardware.


\subsubsection{Papers Focusing on Dominant Impairment Effect}
The nonlinearity of high power \ac{RF} amplifiers is often the predominant hardware
impairment and has a crucial effect on the performance of \ac{MIMO} systems, as was
emphasized in \cite{Qi2010,Qi2012,Fozooni2015}, which characterize
the effect of memoryless nonlinear hardware on the performance of \ac{MIMO} systems.
In particular, \cite{Qi2010} investigated the performance of \ac{MIMO} orthogonal space-time
block coding systems in the presence of nonlinear \acp{HPA}, and proposed a sequential Monte Carlo-based
compensation method for the \ac{HPA} nonlinearity.
Subsequently, the optimal transmit beamforming scheme
in the presence of nonlinear \acp{HPA} is found in \cite{Qi2012}
using a general nonlinearity model for the transmitter \ac{RF}-chains.
However, the suggested strategy is not practical as the precoders depend
on the transmitted signal and hence need to be designed prior to each channel use.
Furthermore, an accurate knowledge about the nonlinearity model of the transmitters
is needed, which makes the design of the precoders complicated.

More recently, the inherent trade-off between nonlinearity distortions and power efficiency was studied
in \cite{Fozooni2015}.
That paper uses a polynomial model for the transmitter \acp{PA}, and -- following the approach in~\cite{Schenk2008}
for modeling the nonlinear distortion -- derived the ergodic rate for \ac{MIMO} systems.

\subsubsection{Papers Dealing with mmWave Systems}
Specifically,
in the framework of \ac{mmWave} communications, \cite{Wu2016,Khansefid2016,Yan2017}
have studied the effect of hardware impairments on the performance of \ac{MIMO} systems.
The results of \cite{Wu2016} show that single-carrier frequency domain equalization is
more robust against impairments from nonlinear power amplifiers than \ac{OFDM} in typical
mmWave system configurations. On the other hand, the results reported in \cite{Khansefid2016}
show a slight bit error rate performance advantage of \ac{OFDM} over single-carrier frequency domain equalization
under nonlinear \ac{RF} distortions, and suggest that subcarrier spacing is a crucial parameter in
mmWave massive \ac{MIMO} systems.

\subsubsection{Papers Related to Power Minimization and Energy Efficiency}
References \cite{Persson2013,Persson2014,Bjoernson2015} provide insights related to
the energy efficiency of \ac{MIMO} systems.
Reference \cite{Persson2013} proposes a \ac{PA}-aware power allocation scheme that takes into account
the power dissipation at the \acp{PA} in \ac{MIMO} systems, and results in substantial
gains in terms of data rate and consumed power compared with non-\ac{PA}-aware
power allocation schemes.
Subsequently, a low computational complexity algorithm that finds the minimum consumed power for any given
mutual information is developed in \cite{Persson2014}.
This algorithm gives significant rate and total
consumed power gains in comparison with non-\ac{PA}-aware algorithms.
Energy efficient optimal designs of multi-user \ac{MIMO} systems are developed in \cite{Bjoernson2015},
where the number of antennas, active (scheduled) users and transmit power levels are part of the design
and operation parameters. However, in this latter paper the impact of hardware impairments are not taken
into account.
Additionally, the impact of regulatory electromagnetic exposure constraints has also been taken into account when designing multiple transmit antenna signals in~\cite{Ying2015,Ying2017,Castellanos2016}.
Recently, the interplay between waveforms, amplifier efficiency, distortion and performance in the
massive \ac{MIMO} downlink was studied in \cite{Mollen2016}.
In that work, it was found that in terms of the consumed power by the \acp{PA}, \ac{OFDM} and
single-carrier transmission have similar performance over the hardened massive \ac{MIMO} channel,
and low-\ac{PAPR} precoding at massive \ac{MIMO} base stations can significantly increase the
power efficiency as compared with \ac{PAPR}-unaware precoders.

\subsubsection{Contributions of the Present Paper}
In this paper, we consider a multi-antenna transmit signal model that incorporates the distortion
generated by each \ac{PA}. 
Under the assumption that the \acp{PA} in the different
antenna branches have the same input-output relation and follow a memoryless polynomial model, we show that the nonlinear distortion vector is a zero mean complex random vector and
derive its covariance matrix in closed form.
Since the resulting statistics
of the nonlinear distortion vector is
a function of the covariance matrix of the beamformed signal, it is therefore affected by the transmit beamforming filters.
Next, for the special case of a single \ac{RF} chain, we derive a
closed form expression both for the maximum spectral efficiency {and for a lower bound on the achievable
rate}. We then consider the problem of optimizing the energy efficiency of the system as a
function of the consumed power per information bit using a realistic power consumption model for the transmit PAs.

\section{Signal and System Model}
\label{sec: system model}
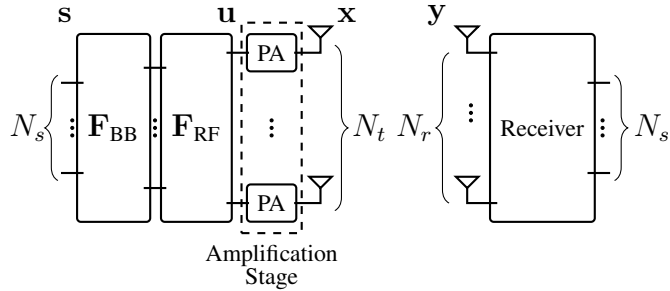
\begin{figure}
  \centering
  \input{Components/Figures/transmitter_structure}
  \caption{System model.}\label{fig: system model}
\end{figure}

\subsection{System Model}
Consider a single-carrier \ac{mmWave} system where a transmitter
with $N_t$ antennas and $N_{\RF}\ll N_t$ RF-chains communicates with a receiver equipped with $N_{r}$ antennas.
We assume that the receiver is equipped with $N_r$ RF-chains and has an all-digital structure.
The transmitter is intended to convey a complex symbol vector denoted by
${\mathbf{s}\sim\mathcal{CN}(\mathbf{0},\mathbf{I}_{N_s})}$ to the receiver, where $N_s\le N_{\RF}$ is the number of transmitted streams.
The symbol is beamformed in the baseband by a beamforming matrix ${\mathbf{F}_{\BB}\in\mathbb{C}^{N_{\RF}\times N_s}}$ and in the analog domain using a network of phase-shifters with transfer matrix $\mathbf{F}_{\RF} \in
\mathbb{C}^{N_t\times N_{\RF}}$.
Therefore, the beamformed signal is
$\mathbf{u} = [u_1,\dots,u_{N_t}]\tran \define \mathbf{F}_{\RF}\mathbf{F}_{\BB}\mathbf{s} \in \mathbb{C}^{N_t}$
and is distributed as $\mathcal{CN}(\mathbf{0},\mathbf{C}_{\mathbf{u}})$, where
\begin{equation}
  \mathbf{C}_{\mathbf{u}} \define \E\left\{\mathbf{u}\mathbf{u}\herm\right\} = \mathbf{F}_\RF\mathbf{F}_\BB\mathbf{F}_\BB\herm\mathbf{F}_\RF\herm\in \mathbb{C}^{N_t\times N_t}\;.
\end{equation}
The beamformed signal then goes through the amplification stage, where at each antenna branch a \ac{PA}, with transfer function $f(.)$, amplifies the signal before transmission.
{We will elaborate further on the function $f(.)$ in Section~\ref{sec: PA Model}.}
We represent the transmitted signal collectively by $\mathbf{x} \define [f(u_1),\dots,f(u_{N_t})]\tran$,
where we have assumed that all the PAs have the same transfer function and there is no coupling between the different antenna branches.
Therefore,
the received signal is
\begin{equation}\label{eq: received signal}
  \mathbf{y} = \mathbf{H}\mathbf{x}
  + \mathbf{n} \in \mathbb{C}^{N_r},
\end{equation}
where $\mathbf{H}\in\mathbb{C}^{N_r\times N_t}$ represents the channel and $\mathbf{n} \sim \mathcal{CN}(\boldsymbol{0},\sigma_n^2 \mathbf{I}_{N_r})$ is the receiver thermal noise.
Fig.~\ref{fig: system model} illustrates the system model\footnote{
The transmitter structure used in this paper is also suggested in several other works including~\cite{Kutty2016,Mendez-Rial2016,Han2015}.}.

\subsection{PA Model}
\label{sec: PA Model}
Behavioural modeling of PAs using polynomials
is a low-complexity, mathematically tractable and yet accurate method which has long been used
in the RF PA design literature (see, e.g., \cite{Kenington2000,Cripps2002,Schenk2008}).
Accordingly, in this paper we adapt a memoryless polynomial model of order $2M+1$ to describe the
nonlinear behavior of the transmitter PAs.
Note that by adjusting the model parameters, this model
can provide an arbitrarily exact approximation of any
other well-known (memoryless) models that has been introduced for PAs in the literature (e.g., see~\cite[Chapter 6]{Schenk2008}).
Clearly, the dynamic behavior of a PA due to its memory effect is not captured in the memoryless polynomial model,
and the investigation of this effect on the performance of the system is out of the scope of this work\footnote{The dynamic behaviour of PAs has been considered in some of the previous works such as~\cite{Yan2017}.}.

Furthermore, we assume that
{the \ac{PA}s in the different antenna branches} follow the same input-output relation.
This assumption is widely used in the literature~\cite{Schenk2008,Mollen2016,Mollen2017}.
In this case, the equivalent baseband output signal of the $n^{th}$
\ac{PA} is
\begin{equation}\label{eq: polynomial model}
  x_{n} = f(u_n) \define \sum\limits_{m=0}^{M} \beta_{2m+1} \left|u_{n}\right|^{2m}u_{n}\;,
\end{equation}
where $\beta_{2m+1}$'s are the model parameters and take complex values in general.
Usually, only a limited number of terms in this model suffices for modeling the \emph{smooth} nonlinear PAs at the RF front-ends.
Observe that in this model the even order terms are omitted as they only contribute to the out-of-band distortion and lead to spectrum regrowth~\cite{Schenk2008}.

Using~\eqref{eq: polynomial model}, we define the instantaneous (amplitude) gain of the $n^{th}$ \ac{PA} as
\begin{equation} \label{eq: inst. gain}
g_{n} \define \frac{x_n}{u_n} = \sum\limits_{m=0}^{M} \beta_{2m+1}\; \left|u_{n}\right|^{2m}\;.
\end{equation}
This equation implies that both the absolute value and phase of the \ac{PA}'s instantaneous gain depends on the input signal's amplitude $|u_n|$. In the literature, the effect of the signal's amplitude on the absolute value and phase of the PA's gain are referred to as \ac{AM-AM} and \ac{AM-PM} characteristics of the PA, respectively.
In practical PAs, the \ac{AM-AM} is a monotonically decreasing function\footnote{Note that although the \ac{AM-AM} gain of a PA is a monotonically decreasing function of the input amplitude, the output amplitude increases with the input signal's amplitude.}
of the input's amplitude
while the \ac{AM-PM} is only slightly changing at high amplitudes.

\subsection{Channel Model}
We consider a cluster channel model~\cite{Ayach2014} with $L$ paths between the transmitter and the receiver.
{Let} {$\psi_{\ell}$ denote the complex gain of path $\ell$ between the transmitter and the receiver,
which includes both} {the} {path-loss and small-scale fading.}
In particular, for the given large-scale fading, $\{\psi_{\ell}\}$ for all $\ell\in\{1,\ldots,L\}$ are i.i.d. random variables
drawn from distribution $\mathcal{CN}(0,10^{-0.1 \rm{PL}})$ where $\rm{PL}$ is the path-loss in dB~\cite{Akdeniz2014}.
The path-loss consists of a constant attenuation, a distance dependent attenuation, and a large scale log-normal fading.
The channel matrix between the transmitter and the receiver is
\be \label{eq:geometry-based channel model}
\mathbf{H} = \sqrt{\frac{N_t N_r}{L}} \, \sum\limits_{\ell=1}^{\Np}{\psi_{\ell}\,{\mathbf{a}_{r}\left(\theta_{\ell}\right)} \mathbf{a}_{t}\herm\left(\phi_{\ell}\right) } = \mathbf{A}_{r} \mathbf{\Psi} \mathbf{A}_{t}\herm \in
\mathbb{C}^{N_r\times N_t} \:,
\ee
where $\theta_\ell$ and $\phi_\ell$ are the \ac{AoA} and \ac{AoD} corresponding to path $\ell$ of the channel, respectively.
Vectors $\mathbf{a}_t \in \mathbb{C}^{N_t}$ and $\mathbf{a}_{r} \in \mathbb{C}^{N_r}$ represent the unit-norm array response vectors of the transmitter and the receiver antenna arrays, respectively, $\mathbf{A}_{t} =
[\mathbf{a}_{t}(\phi_1),\dots,\mathbf{a}_{t}(\phi_{L})]$,
$\mathbf{A}_{r} = [\mathbf{a}_{r}(\theta_1),\dots,\mathbf{a}_{r}(\theta_{L})]$, and $\mathbf{\Psi}  \in \mathbb{C}^{L \times L}$ is a diagonal matrix whose $\ell$-th diagonal entry is $\psi_{\ell}=\sqrt{N_t N_r/L}$.
We assume that both of the transmitter and the receiver are equipped with \acp{ULA} with array responses
\begin{align}\label{}
  \mathbf{a}_t(\phi) &\!=\! \frac{1}{\sqrt{N_t}}\!
  \left[1\!,\!e^{-j2\pi D_t\!\sin(\phi)}\!,\dots,\!e^{-j2\pi(N_t-1)D_t\!\sin(\phi)}\right]\tran\!\!, \\
  \mathbf{a}_r(\theta) &\!=\! \frac{1}{\sqrt{N_r}}\!
  \left[1\!,\!e^{-j2\pi D_r\!\sin(\theta)}\!,\dots,\!e^{-j2\pi(N_r-1)D_r\!\sin(\theta)}\right]\tran\!\!,
\end{align}
where $D_t$ and $D_r$ represent the antenna spacing of the transmitter and receiver, respectively, normalized to the carrier wavelength.

\section{Nonlinear Power Amplification}
\label{sec: Nonlinear Power Amplification}

\subsection{Nonlinear Distortion}
Due to the nonlinear behaviour of the \ac{PA}s in the amplification stage, the transmitted signal is an amplified and distorted version of the input signal{, $\mathbf{u}$.
On {the} one hand, using the PA model {of} Section~\ref{sec: PA Model},
the transmitted signal is a function of $\mathbf{u}$ as represented in $\mathbf{x}=[f(u_1),\dots,f(u_{N_t})]\tran$, where $f(.)$ is defined in \eqref{eq: polynomial model}.
On the other hand, following the approach in~\cite{Dardari2000} and extending it to the multiantenna case, the same signal can be represented as a linearly amplified version of the input signal $\mathbf{u}$ contaminated with the nonlinear distortion. That is }
\begin{equation}\label{eq: transmitted signal}
  \mathbf{x} = \widebar{\mathbf{G}}\;\mathbf{u} + \mathbf{d} \in \mathbb{C}^{N_t},
\end{equation}
where $\widebar{\mathbf{G}}$ denotes the average linear gain of the amplification stage and $\mathbf{d} = [d_1,\dots,d_{N_t}]\tran$ in which $d_n$ is the distortion generated by the $n^{th}$ \ac{PA}.
According to the definition in~\cite{Dardari2000}, the distortion generated at the output of each \ac{PA} is uncorrelated with the input signal to that \ac{PA}, i.e., $\E\{u_n^*d_n^{}\} = 0$ for $n=1,\dots,N_t$.
Subsequently, we can conclude that $\E\{u_n^*d_k^{}\} = 0$ for any $k,n\in \{1,\dots,N_t\}$.
Furthermore, we assume that the antenna branches are perfectly isolated from each other and therefore the coupling between them is negligible. Hence, $\widebar{\mathbf{G}}$ is assumed to be a diagonal matrix.


By collectively representing the instantaneous gain of the power amplification stage by $\mathbf{G} = \diag(g_1,\dots,g_{N_t})$, the transmitted signal can be alternatively represented as $\mathbf{x} =
\mathbf{G}\mathbf{u}$.
Correspondingly, by substituting $\mathbf{x}=\mathbf{G}\mathbf{u}$ into \eqref{eq: transmitted signal}, the nonlinear distortion can be expressed as
\begin{equation}\label{eq: distortion}
\mathbf{d} = (\mathbf{G}-\Gav)\mathbf{u}.
\end{equation}

Let us denote the average power of the input signal to the $n^{th}$ \ac{PA} by $P_n \define \E\left\{|u_n|^2\right\} = [\Cu]_{nn}$,
the following two propositions characterize the average linear gain and the nonlinear distortion signal.

\begin{prop}\label{prop: average linear gain}
The average linear gain $\widebar{\mathbf{G}}$ of the power amplification stage in \eqref{eq: transmitted signal} is
\begin{equation}\label{eq: ave. gain}
\Gav = \diag\left(\widebar{g}(P_1),\dots,\widebar{g}(P_{N_t}) \right)\;,
\end{equation}
where $\widebar{g}(P_n) = \sum\limits_{m=0}^{M} \beta_{2m+1}\; P_n^{m} (m+1)!$\;.
\end{prop}
A sketch of proof for Proposition~\ref{prop: average linear gain} is given in the Appendix.

\begin{prop}\label{prop: nonlinear distorion covariance matrix}
The nonlinear distortion vector $\mathbf{d}$ in~\eqref{eq: transmitted signal} is a zero-mean complex random vector with covariance matrix
\begin{equation}\label{eq: Cd}
\Cd \!=\! \sum_{m=1}^{M}
\mathbf{\Gamma}_m
\overbrace{\left(\Cu \odot \dots \odot \Cu \right)} ^
{(m+1)\ \text{times}} \odot
\overbrace{\left(\Cu^T \odot \dots \odot \Cu^T \right)}^
{m\ \text{times}}
\mathbf{\Gamma}_m\herm,
\end{equation}
where $\mathbf{\Gamma}_m = \diag\left(\gamma_{m}(P_1),\dots,\gamma_{m}(P_{N_t})\right)$ and
\begin{equation}\label{eq: gamma}
\gamma_{m}(P_n) = \sqrt{\frac{1}{m+1}} \sum_{q=m}^{M} \beta_{2m+1}{{q}\choose{m}}(q+1)!\; P_{n}^{(q-m)}\;.
\end{equation}
\end{prop}
\begin{IEEEproof}
A proof is given in the Appendix.
\end{IEEEproof}
As Proposition~\ref{prop: nonlinear distorion covariance matrix} implies,
the spatial direction of the nonlinear distortion is dependent on the direction of the beamformed signal.
Therefore, {an important} intuition from this proposition is that by beamforming the desired signal,
the distortion is also beamformed toward the receiver.
{As we will see in the next sections, this phenomenon affects the spectral and energy efficiency of the system,
especially} {when the PAs} {are pushed to work in their energy efficient, but nonlinear, regions.}
The following example elaborates further on this intuition.

\begin{figure}[t]
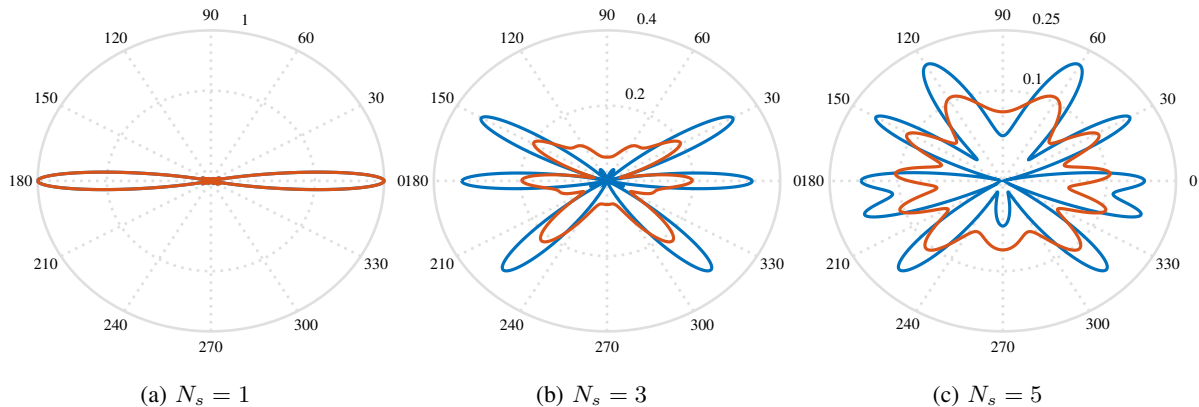

\begin{subfigure}[b]{0.31\columnwidth}
  \centering
  \input{Components/Figures/BeamPattern_NRF=1}
  \caption{$N_s = 1$}\label{fig: beampattern NRF=1}
\end{subfigure}
\begin{subfigure}[b]{0.31\columnwidth}
  \centering
  \input{Components/Figures/BeamPattern_NRF=3}
  \caption{$N_s = 3$}\label{fig: beampattern NRF=3}
\end{subfigure}
\begin{subfigure}[b]{0.31\columnwidth}
  \centering
  \input{Components/Figures/BeamPattern_NRF=5}
  \caption{$N_s = 5$}\label{fig: beampattern NRF=5}
\end{subfigure}
\caption{Normalized beampattern of the radiated desired signal (blue) and the radiated distortion signal (red).
Note that when only one stream is transmitted, i.e. $N_s=1$, distortion and desired signal have the same beampattern.}\label{fig: beampattern}
\end{figure}

\begin{example}
\label{ex: distortion direction}
Consider a \ac{mmWave} system as described in Section~\ref{sec: system model} with $N_t = 8$ and the \ac{PA} model parameters stated in Table~\ref{table: Simulation parameters}.
Assume that $N_s = N_\RF$ and no baseband beamforming is applied, i.e., $\mathbf{F}_{\BB} = \frac{1}{\sqrt{N_s}}\mathbf{I}_{N_s}$.
Figure~\ref{fig: beampattern} illustrates the simulated beampattern of the transmitted signal when the analog beamformer
$\mathbf{F}_{\RF} = [\mathbf{a}_t(\phi_1),\dots,\mathbf{a}_t(\phi_{N_s})]$ is used for $N_s=1,3,5$. In this figure, the AoDs, i.e. $\phi_i,\ i=1,\dots,5$, are $0,-\pi/4,\pi/6,\pi/3,-\pi/12$, respectively.
As the figure implies, the peak power of the distortion signal is steered in the same direction as the desired beamformed signal. However, as the number of transmitted streams increases, the distortion signal behaves
more like an omnidirectional noise. Mathematically, we can also see that by noting that as the number of transmitted streams from antenna branches increase, the off-diagonal elements of $\Cd$ get smaller compared to the diagonal elements.
\end{example}

In the case where $N_s < N_{\text{RF}}$
and the signal is digitally beamformed in the baseband, the effect of $N_s$ on the directionality of radiated distortion signal is not easily tractable.
In general, the directionality of the distortion signal depends
on the hybrid beamformer $\mathbf{F}_{\text{RF}}\mathbf{F}_{\text{BB}}$, and subsequently on $\mathbf{C}_{\mathbf{u}}$,
as Proposition~\ref{prop: nonlinear distorion covariance matrix} implies.
This proposition shows that as the beamformed signals transmitted from different antenna branches
become more uncorrelated (i.e., the off-diagonal elements of $\mathbf{C}_{\mathbf{u}}$ become smaller),
the distortion behaves more like an omni-directional signal with almost equal power transmitted in different directions. Below, we show the effect of $N_s$ on the radiated distortion in a simple example.

\begin{figure}[t!]
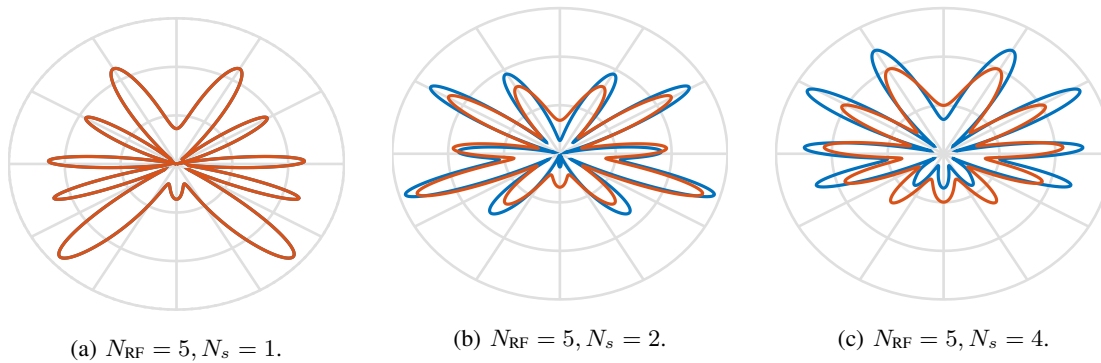

    \begin{subfigure}{0.3\columnwidth}
        \centering
        \input{Components/Figures/Ns=1-Nrf=5}
        \caption{$N_{\text{RF}} = 5,N_s = 1$.}\label{fig: Beampattern Nrf=5,Ns=1}
    \end{subfigure}
    \begin{subfigure}{0.3\columnwidth}
        \centering
\input{Components/Figures/Ns=2-Nrf=5}
\caption{$N_{\text{RF}} = 5,N_s = 2$.}\label{fig: Beampattern Nrf=5,Ns=2}
    \end{subfigure}
    \begin{subfigure}{0.3\columnwidth}
        \centering
\input{Components/Figures/Ns=4-Nrf=5}
\caption{$N_{\text{RF}} = 5,N_s = 4$.}\label{fig: Beampattern Nrf=5,Ns=4}
    \end{subfigure}
    \caption{Normalized beampattern of the radiated desired signal (blue) and the
radiated distortion signal (red). Note that when only one stream is transmitted,
i.e. $N_s = 1$, distortion and desired signal have the same beampattern.}\label{fig: beampattern Ns<Nrf}
\end{figure}

\begin{example}
\label{ex: directionallity ve. Ns}
Consider the system of Example~\ref{ex: distortion direction} where $N_s<N_{\text{RF}}$ streams are beamformed using a hybrid beamformer $\mathbf{F}_{\text{RF}}\mathbf{F}_{\text{BB}}$.
Fig.~\ref{fig: beampattern Ns<Nrf} illustrates
the simulated beampattern of the transmitted desired signal as well as the radiated distortion for different numbers of streams,
$N_s$, when $N_{\text{RF}} = 5$.
In this figure, the entries of $\mathbf{F}_{\text{BB}}$ are i.i.d. Gaussian distributed.
As the figure shows, when $N_s = 1$ (and consequently all the signals transmitted from different antennas are fully correlated)
then
the distortion signal is transmitted in the direction of desired signal, similarly to Example~\ref{ex: distortion direction}.
However unlike Example~\ref{ex: distortion direction}, increasing $N_s$ from 2 to 4 while keeping $N_{\text{RF}}$ constant
does not necessarily lead to lower directionality in the distortion signal.
\end{example}

\subsection{Nonlinear Crosstalk}
\label{sec: Crosstalk}
Another impairment that is observed in multi-antenna systems is \emph{crosstalk}, which is due to coupling of the signal from one antenna branch to another.
If we make the assumption that the antenna branches are sufficiently isolated from each other,
the coupling can be modeled as a linear crosstalk between different antenna branches~\cite{Raeesi2017}.
The linear coupling of the signals after the amplification stage can in principle be seen as part of the channel and therefore is not studied separately in this paper.
However, the coupling before the amplification stage results in a nonlinear crosstalk impairment.
In this case, the input signal to the amplification stage will be
\begin{equation}\label{}
  \widetilde{\mathbf{u}} = \mathbf{B}_{\text{TX}}\mathbf{u},
\end{equation}
where $\mathbf{B}_{\text{TX}}\in \mathbb{C}^{N_t\times N_t}$ represents the transmit coupling matrix.
Moreover, when coupling exists, both the average linear gain $\widebar{\mathbf{G}}$ and the
distortion vector $\mathbf{d}$ will be affected by the coupling matrix through the covariance matrix of $\widetilde{\mathbf{u}}$ which is
\begin{equation}
\mathbf{C}_{\widetilde{\mathbf{u}}} =
\E\left\{\mathbf{B}_{\text{TX}}\mathbf{u}\mathbf{u}\herm\mathbf{B}_{\text{TX}}\herm\right\} = \mathbf{B}_{\text{TX}}\mathbf{C}_{\mathbf{u}}\mathbf{B}_{\text{TX}}\herm.
\end{equation}
Replacing $P_n = \left[\mathbf{C}_{\mathbf{u}}\right]_{nn}$ by $\widetilde{P}_n = [{\mathbf{C}}_{\widetilde{\mathbf{u}}}]_{nn},\ n=1,\dots,N_t$
in \eqref{eq: ave. gain} and \eqref{eq: gamma}
and replacing $\mathbf{C}_{\mathbf{u}}$ by $\mathbf{C}_{\widetilde{\mathbf{u}}}$ in \eqref{eq: Cd} gives the average linear gain, $\widetilde{\mathbf{G}}$, and
the covariance of the distortion signal, $\mathbf{C}_{\widetilde{\mathbf{d}}}$, in systems with coupling.

In the sequel, we ignore the crosstalk impairment and focus on the effects of the distortion on the system performance unless otherwise stated. In the next section, we investigate the performance of the system in terms of achievable rate and the consumed power per information bit.

\section{Spectral Efficiency}\label{sec: Spectral Efficiency}
The distortion signal is a self-interference which is generated by the desired signal itself. Therefore, it carries information about the desired signal. Nonetheless, extracting information from it relies on two impractical conditions. First, a precise knowledge about the nonlinear behavior of the system should be available. Second, a complicated nonlinear receiver should be employed.
In practice, it is easier to treat the received distortion as noise and discard the information buried in it.
Furthermore, the received distortion is not necessarily Gaussian distributed.
However, by noticing that among different distributions of the additive noise, the Gaussian distribution leads to the smallest possible spectral efficiency~\cite{Goldsmith1997},
we define the (worst case) spectral efficiency of the system (in bits/sec/Hz) as
\begin{equation} \label{eq: SE}
SE \!\define\! \log_2\det\!\left(\mathbf{I}_{N_r} \!\!+\! \left(\mathbf{H}\Cd\mathbf{H}\herm \!+\! \sigma_n^2 \mathbf{I}_{N_r}\right)^{\!-1}\! \mathbf{H} \Ctu \mathbf{H}\herm\!\right),
\end{equation}
where $\Ctu \define \Gav\Cu\Gav\herm$ is the covariance matrix of the transmitted desired signal.
Note that the spectral efficiency in~\eqref{eq: SE} gives a lower-bound on the capacity of the nonlinear channel, since part of the transmitted information is regarded as undesired distortion at the receiver.
The following proposition gives the maximum spectral efficiency, optimized over the beamforming vector $\mathbf{F}_{\text{RF}}$, for the special case where the transmitter is equipped with one RF chain.

\begin{prop}\label{prop: maximum SE one RF chain}
In the case where $N_{\RF} = 1$, the maximum spectral efficiency of the system described in Section~\ref{sec: system model}, maximized over the beamforming vector $\mathbf{F}_{\text{RF}}$, is
\begin{equation}\label{eq: SE one RF chain}
\begin{split}
  \widebar{SE} \!\!=
  \! \log_2\!\det\!\left(\!\mathbf{I}_{N_r} \!\!\!+\!\!
  \left(\!\widetilde{\mathbf{H}}\widetilde{\mathbf{H}}\herm \widebar{g}_{d}\!\left(\!\frac{P}{N_t}\!\right) \!+\! \frac{\sigma_n^2}{P}\! \right)^{\!\!\!-1}\!\! \widetilde{\mathbf{H}}\widetilde{\mathbf{H}}\herm \widebar{g}_{s}\!\!\left(\!\frac{P}{N_t}\!\right)
  \!\right),
\end{split}
\end{equation}
 where $P \define \E\{\|\mathbf{u}\|^2\} = \sum_{n=1}^{N_t} P_n$ is the total input power into the amplification stage, $\widetilde{\mathbf{H}}\define 1/\sqrt{N_t}\;\mathbf{H}\mathbf{a}_{t}(\phi_{\max})$ is the effective channel between the transmitter and the receiver, $\phi_{\max}$ is the \ac{AoD} corresponding to the path with the largest small scale fading gain, and
\begin{align}
\widebar{g}_{s}(P/N_t) &\define \left|\widebar{g}\left({P}/{N_t}\right)\right|^2, \\
\widebar{g}_{d}(P/N_t) &\define \sum_{m=1}^{M}\left|\gamma_{m}\left({P}/{N_t}\right)\right|^2 \left({P}/{N_t}\right)^{2m}.
\end{align}
\end{prop}
\begin{IEEEproof}
A proof is given in the Appendix.
\end{IEEEproof}
\begin{corollary}\label{cor: lower-bound on SE one RF chain}
A lower-bound on the achievable rate of the system described in~\eqref{eq: received signal} when $N_\RF = 1$ can be found as
\begin{equation}\label{eq: R_LB}
  \widebar{SE} \ge \log_2\left( 1 + \frac{\widebar{g}_s \!\left(\frac{P}{N_t}\right) }
  {\widebar{g}_d \left(\frac{P}{N_t}\right) + \frac{\sigma_n^2}{\delta P}} \right).
\end{equation}
where
$\delta \!=\! \left|1/\sqrt{N_t N_r}\left(\mathbf{a}_{r}(\theta_{\max})\right)\herm\mathbf{H}\mathbf{a}_{t}(\phi_{\max})\right|^2$
is the effective channel gain and $\theta_{\max}$ is the \ac{AoA} corresponding to the path with the largest small scale fading gain. This bound is tight when $L = N_\RF$.
\end{corollary}
The proof of Corollary~\ref{cor: lower-bound on SE one RF chain} is a straightforward application of Lemma~\ref{lemma: h<f} in the Appendix.

Corollary~\ref{cor: lower-bound on SE one RF chain} clearly shows that the benefit of increasing the number of transmit antennas $N_t$ on the spectral efficiency of the system is two-fold. On one hand, by coherently transmitting the signal, an array gain proportional to $N_t$ can be obtained. This gain is reflected in the effective channel gain $\delta$. On the other hand, the input signal power to each \ac{PA} decreases inversely with $N_t$ which leads to a higher linear gain for the desired signal and lower distortion power.

\section{Energy Efficiency}\label{sec: Energy Efficiency}
Spectrally efficient modulation techniques, such as OFDM, lead to signals with a high \ac{PAPR}, which are more prone to the distortion, specially when the PAs in the amplification stage are working close to saturation.
One conventional technique to avoid distortion is to apply a large input back-off (IBO) at the input of the \ac{PA}s. By applying IBO, the input powers are decreased to ensure that the \ac{PA}s are operating in their linear region even when the signals are at their peaks.

Although the smaller input power leads to less distortion at the output of a \ac{PA}, reducing the input power at the same time decreases the power efficiency of the \ac{PA} leading to more power dissipation in the system. In fact, there is a trade-off between the spectral and energy efficiency of the system on one side and the generated distortion on the other side~\cite{Ochiai2013}. In order to investigate this trade-off in our system, we first need to find the total power consumption of the system.

Let us denote
the power efficiency of the $n^{th}$ \ac{PA} by
\begin{equation}\label{eq: PA efficiency}
  \eta(P_n) \define \frac{P_{\rm{rad},n}}{P_{\rm{cons},n}}\;,
\end{equation}
where $P_{\rm{rad},n} \define [\Ctu]_{nn}+ [\Cd]_{nn}$ is the radiated power from the $n^{th}$ antenna and
$P_{\rm{cons},n}$ is the consumed power by the \ac{PA} including both the radiated power and the dissipated power.
Note that not all the radiated power from the antenna is useful at the receiver as part of it belongs to the transmitted distortion signal.
Following the approach in~\cite{Persson2014}, the consumed power by the $n^{th}$ \ac{PA} can be expressed as
\begin{equation}\label{eq: cons. power}
P_{\rm{cons},n} = \frac{\sqrt{P_{\max}}}{\eta_{\max}}\sqrt{P_{\rm{rad},n}}\;,
\end{equation}
where
$P_{\max}$ is the maximum output power
and $\eta_{\max}$ is the maximum efficiency of the \ac{PA}.
Therefore the total power consumption\footnote{
Since the focus of this paper is on the impact of nonlinear \acp{PA} on the system performance, by considering that a large portion of the consumed power in communication systems is used by \acp{PA},
we do not include the power consumed by other components in our calculations.}
is $P_{\rm cons} \define \sum_{n=1}^{N}P_{{\rm cons},n}$.

\begin{remark}
Although the maximum efficiency that a PA can achieve is constant and depends on its physical structure,
the efficiency of a PA is changing with its input power.
In some works such as~\cite{Han2015,Bjoernson2015}, the efficiency of the transceiver PAs is assumed to be constant and independent from the input power.
This can potentially lead to an inaccurate calculation of the consumed power and consequently the
energy efficiency of the overall system.
\end{remark}

To characterize the actual energy that is used to transmit one information bit
from the transmitter to the receiver we define
the energy efficiency of the system (in bits/Joul) as
\begin{equation}\label{eq: EE definition}
  EE \define \frac{BW\times SE}{P_{{\rm cons}}},
\end{equation}
where $BW$ is the total bandwidth of the system used for data transmission.
Using~\eqref{eq: EE definition}, the optimal beamforming strategy for maximizing the energy efficiency of
system can be found by solving the following problem:
\begin{alignat*}{2}\tag{P1}\label{eq: optimization problem}
  &\underset{\mathbf{F}_{\BB},\mathbf{F}_{\RF}}{\text{maximize}} \quad&& EE \\
  &\text{subject to}\quad&& P_{\rm cons}\le P_0 \\
  &                 \quad&& \left|[\mathbf{F}_{\RF}]_{i,j}\right| = {\sqrt{1/N_t}},
  \quad \forall i,j.
\end{alignat*}
The Problem~\ref{eq: optimization problem} is not convex and is not likely to be solvable in polynomial time.
{However, in the special case when $N_\RF~=~1$,}
{the dimension of Problem~(P1)} {reduces to one.
Therefore, in this case the problem is tractable and can be solved using numerical approaches.
By studying this special case, we can gain some insight} {into}
{the impact of the input power on the spectral and energy efficiency of the system (see Fig.~\ref{fig: SE vs P},~\ref{fig: energy efficiency} and~\ref{fig: EE vs SE} for a quick insight).
The following proposition gives the equivalent problem of (P1) when $N_\RF~=~1$.}

\begin{prop}\label{prop: lower bound for EE}
In the case where $N_\RF = 1$, Problem~(P1) is equivalent to the following problem:
\begin{alignat*}{2}\tag{P2}\label{eq: optimization problem one RF chain}
  &\underset{P}{\text{maximize}} \quad&& \widebar{SE}/\widebar{P}_{\rm cons} \\
  &\text{subject to}\quad&& \widebar{P}_{\rm cons}\le P_0
\end{alignat*}
where $\widebar{SE}$ is given by Proposition~\ref{prop: maximum SE one RF chain} and
\begin{equation}\label{eq: cons. power one RF chain}
  \widebar{P}_{\rm cons} = \frac{\sqrt{P_{\max}}}{\eta_{\max}}\sqrt{\left(\widebar{g}_s\left(\!\frac{P}{N_t}\!\right)
  +\widebar{g}_d\left(\!\frac{P}{N_t}\!\right)\right)PN_t}
\end{equation}
\end{prop}
\begin{IEEEproof}
A proof is given in Appendix.
\end{IEEEproof}
{Note that \eqref{eq: optimization problem one RF chain} has only one dimension
and can efficiently be solved in practice by using, for example, the Newton-Raphson method.}

\section{Numerical Results}\label{sec: Numerical Results}
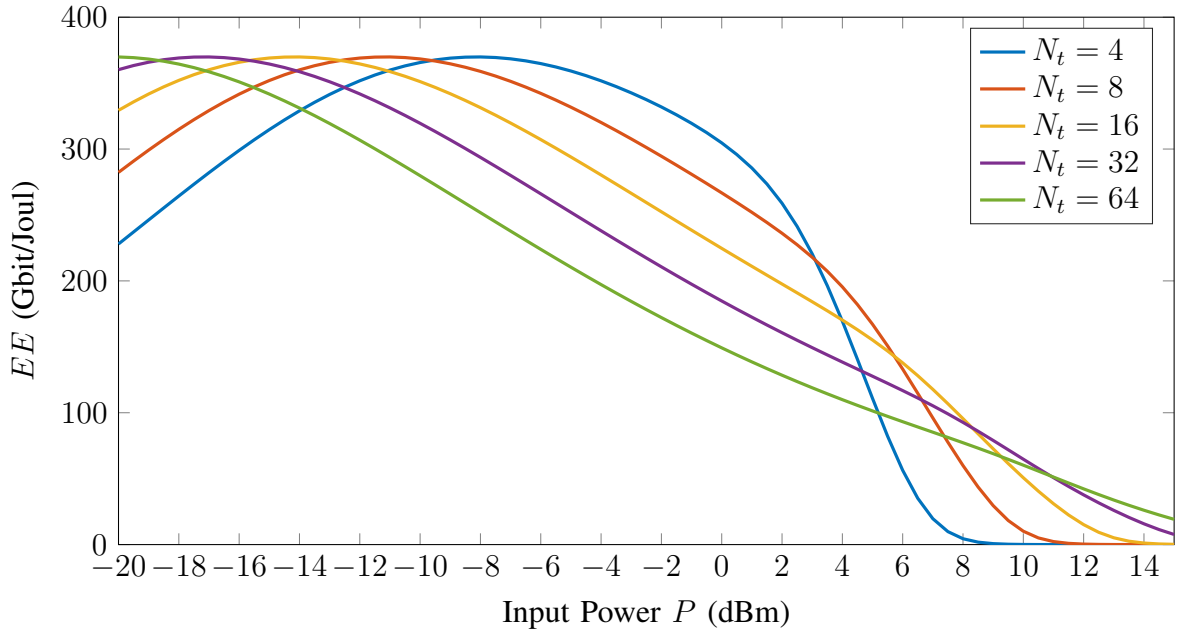
\begin{figure}[t!]
  \centering
  \input{Components/Figures/EEvsP_Nt.tex}
  \caption{Energy efficiency of the system in (Gbit/Joul) with varying input power and number of transmit antennas.}\label{fig: energy efficiency}
\end{figure}

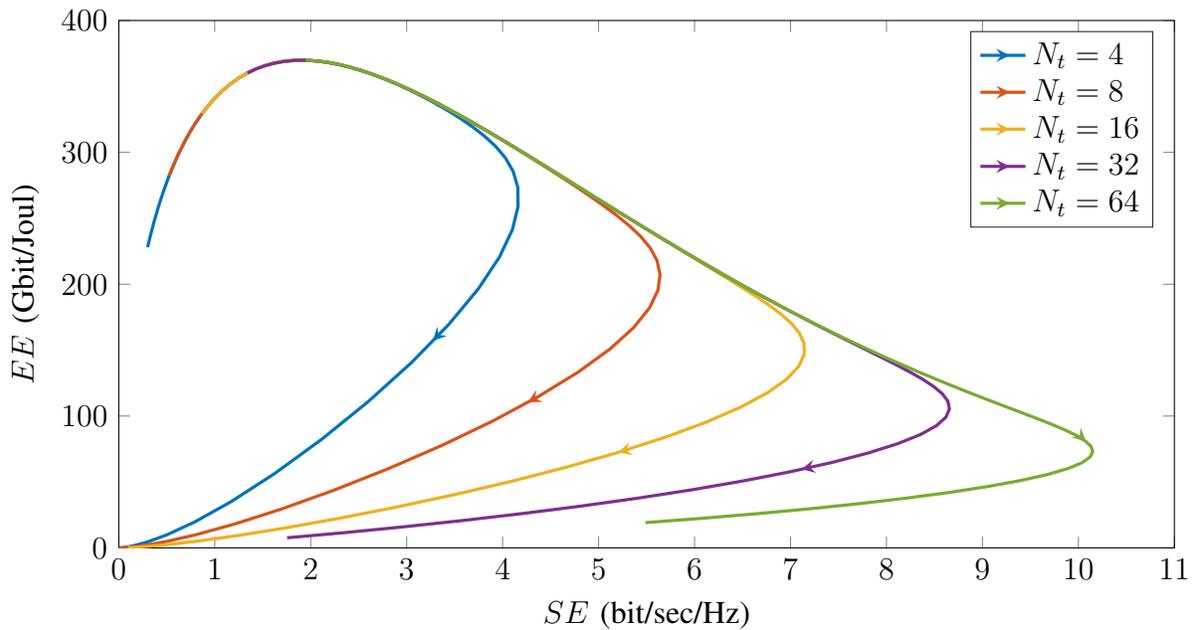
\begin{figure}[t!]
  \centering
  \input{Components/Figures/EEvsSE_2.tex}
  \caption{System energy efficiency vs. spectral efficiency. The input power, $P$, increases in the direction of arrows.}\label{fig: EE vs SE}
\end{figure}

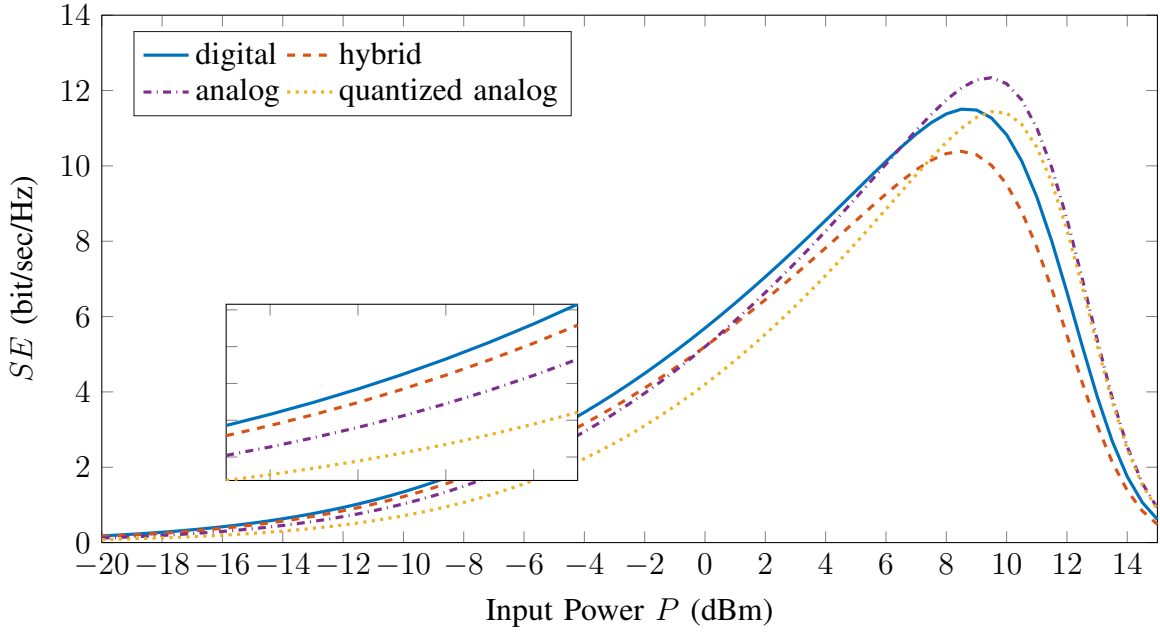
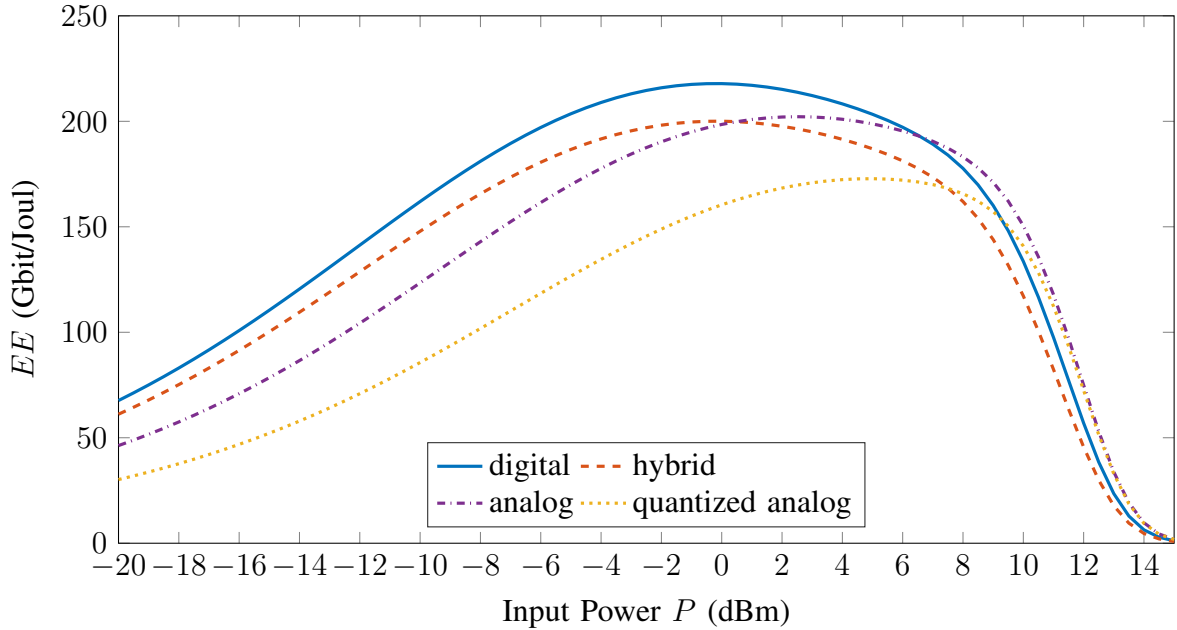
\begin{figure}[t!]
    \begin{subfigure}{1\columnwidth}
        \centering
        \input{Components/Figures/SEvsP_different_beamforming.tex}
        \caption{System spectral efficiency.}\label{fig: SE different beamforming}
    \end{subfigure}
    \begin{subfigure}{1\columnwidth}
        \centering
        \input{Components/Figures/EEvsP_different_beamforming.tex}
        \caption{System energy efficiency.}\label{fig: EE different beamforming}
    \end{subfigure}
    \caption{System performance for different transmit beamforming schemes.}\label{fig: different beamforming approaches}
\end{figure}

In this section, we present simulation results for a MIMO mmWave system with $N_r = 16$ receiving antennas,
and a variable number of transmit antennas.
The transmitter and receiver are $15$ meters apart.
We assume that the number of paths between the transmitter and receiver is $L=5$.
In Fig.~\ref{fig: SE vs Nt}-\ref{fig: EE vs SE}, both the transmitter and the receiver
are equipped with $N_{\RF} = 1$ RF chain, while in Fig.~\ref{fig: different beamforming approaches}
the number of RF chains is $N_{\RF}=5$.
The rest of the (fixed) simulation parameters are presented in Table~\ref{table: Simulation parameters}.

\begin{table}[t]
\centering
\caption{Fixed Parameters in the Numerical Evaluation.}
\label{table: Simulation parameters}
\begin{tabular}[t!]{|c|c|}
\hline
\textbf{Parameters} & \textbf{Values} \\
\hline
Carrier Frequency / Bandwidth  & 73 GHz / 1 GHz\\
\hline
Noise Power (8 dB noise figure) & -105 dBm  \\
\hline
Small Scale Fading Distribution & $\mathcal{CN}(0,1)$ \\
\hline
Large Scale Fading (dB) &  $\zeta \sim \mathcal{CN}(0,8)$ \\
\hline
Path Loss (dB) at Distance $d$ (m) & $86.6 + 24.5\log_{10}(d) + \zeta $ \\
NLOS~\cite{Akdeniz2014} &  \\
\hline
 & $\beta_1 = 2.96 $\\
PA Model Parameters~\cite{Faulkner1992} & $\beta_3 = 0.1418e^{-j2.816}$ \\
 & $\beta_5 = 0.003e^{j0.39}$
\\
\hline
Maximum PA Efficiency $\eta_{\max}$ & $0.3$ \\
\hline
Maximum PA Output Power $P_{\max}$ & 6 dBm \\
\hline
\end{tabular}
\end{table}

Fig.~\ref{fig: SE vs Nt} illustrates the system spectral efficiency as a function of the number of transmit antennas.
In this figure, the total input power to the amplification stage, $P$, is $10$~dBm and is fixed for different values of $N_t$. In this case, $\mathbf{F}_{BB} = 1$ and only an analog beamforming $\mathbf{F}_{\RF}$ designed using Proposition~\ref{prop: maximum SE one RF chain} is applied at the transmitter.

\begin{figure}
  \centering
  \input{Components/Figures/SEvsNt}
  \caption{Spectral efficiency as a function of the number of transmit antennas.
  In the Nonlinear System the PAs follow a memoryless polynomial model with the parameters
  stated in Table~\ref{table: Simulation parameters}. In the Linear System, PAs are linear.}
  \label{fig: SE vs Nt}
\end{figure}
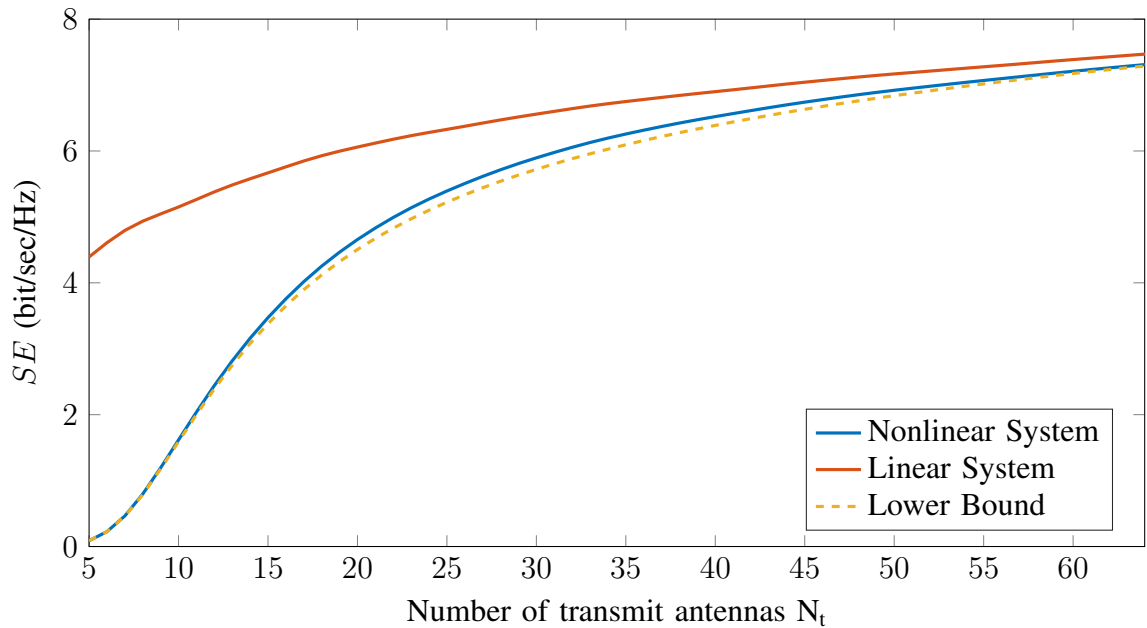
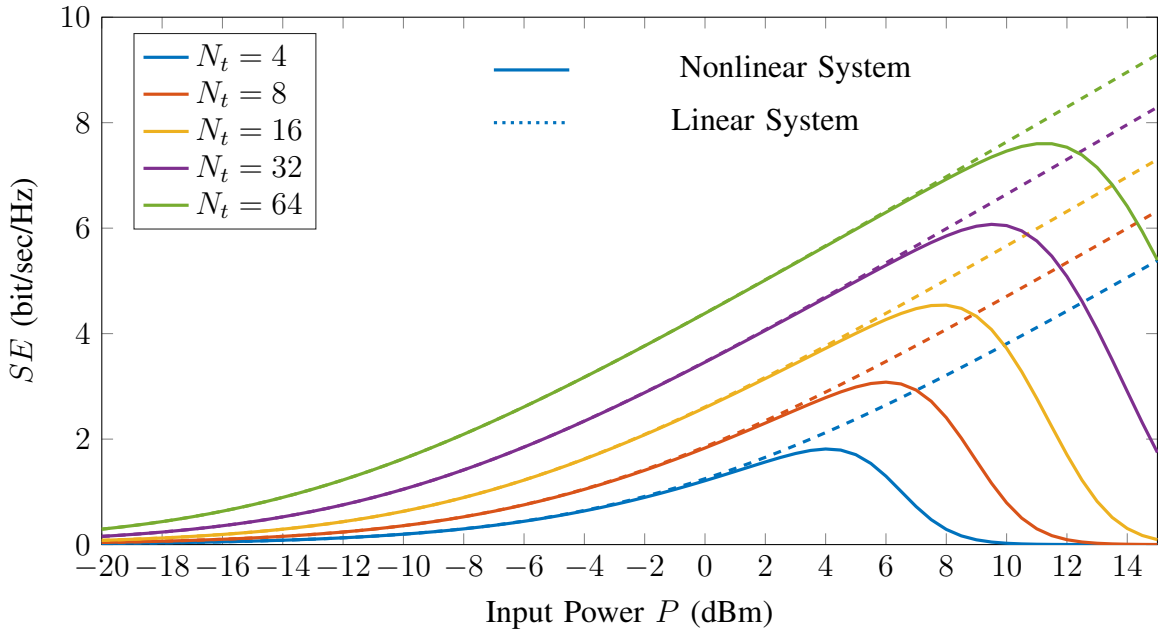
\begin{figure}[t!]
  \centering
  \input{Components/Figures/SEvsP_2.tex}
  \caption{Spectral efficiency as a function of the input power to the amplification stage.
  In the Nonlinear System the PAs follow a memoryless polynomial model with the parameters stated in Table~\ref{table: Simulation parameters}.
  In the Linear System, PAs are linear.}
  \label{fig: SE vs P}
\end{figure}

This figure also shows the maximum spectral efficiency of the system with linear PAs (i.e., when $\beta_{2m+1} = 0,~\forall m>0$).
In addition, the lower bound found in Corollary~\ref{cor: lower-bound on SE one RF chain} is also plotted in this figure.
As the figure indicates,
by increasing the number of transmit antennas, due to the increase in the array gain,
the spectral efficiency of both the linear and nonlinear systems improve.
However, this improvement is steeper for the nonlinear system, especially when the number of antennas is small.
This happens because assuming a fixed transmit power budget,
when $N_t$ is small, the input power to each PA is larger, and consequently the PAs are pushed harder toward saturation,
which leads to more distortion radiation from the transmitter. The amount of generated distortion decreases
as the input powers decrease and the PAs move toward the linear region.

Fig.~\ref{fig: SE vs P} shows the spectral efficiency of the system as a function of input power $P$
for different numbers of transmit antennas.
In this figure, the same simulation parameters as the ones in Fig.~\ref{fig: SE vs Nt}, are used.
As the figure suggests, the spectral efficiency in nonlinear systems
is not a strictly increasing  function of the transmit power.
In fact, after a certain threshold, any increase in the input power degrades
the performance of the system due to increasing distortion.
Another observation from this figure is that increasing the number of transmit antennas always leads to higher spectral efficiency for a fixed transmit power (even at high $P$).

Fig.~\ref{fig: energy efficiency} illustrates the energy efficiency of the system described in Section~\ref{sec: system model}
versus the input power to the amplification stage, $P$, for various values of $N_t$. The energy efficiency in this figure is computed using~\eqref{eq: EE definition}.
As the figure implies, at low and high input powers, increasing $N_t$ improves the energy efficiency
while at medium values of $P$, energy efficiency decreases as $N_t$ increases.
Another observation from this figure, which might look counter-intuitive, is that the energy efficiency of the system is small when $P$ is large. This is against the common rule of thumb that by increasing the input power to a PA, it will work more efficiently (see the definition of PA efficiency, $\mu(.)$, in~\eqref{eq: PA efficiency}). However, we should note that although the PAs are working more efficiently in their nonlinear region they also distort the signal more severely. Hence, part of the radiated power is in fact the distortion signal power which in turn negatively affects the $SE$ and leads to
a degradation of the energy efficiency at the system level,
i.e., to a degradation of~$EE$~(see the definition of $EE$ in~\eqref{eq: EE definition}).

It can clearly be observed
from Fig.~\ref{fig: SE vs P} and Fig.~\ref{fig: energy efficiency}
that although the spectral efficiency increases monotonically with power within the whole linear region of the PAs (which can be determined in Fig.~\ref{fig: SE vs P} by the range of $P$ where curves corresponding to Nonlinear System and Linear System match), $EE$ starts to decline before the PAs enter their nonlinear region.
The reason for that will be clear by noting that when $N_{{\rm RF}}=1$ and $P/N_t$ is small
using \eqref{eq: cons. power one RF chain} and Corollary~\ref{cor: lower-bound on SE one RF chain}, the relationship of the system spectral and energy efficiency with the input power can be determined as $\widebar{SE} \approxeq \log_2(1+\frac{\delta P}{\sigma_n^2})$ and
$\widebar{EE} \propto \widebar{SE}/\sqrt{N_t P}$.
This in fact is in line with the results of the previous works, such as the ones in \cite{Han2015} where by considering the consumed power in a linear system (and not only the transmitted power) it is shown that the $EE$-$SE$ relationship is not always monotonic. The $EE$-$SE$ relationship is investigated further in Fig.~\ref{fig: EE vs SE}.


Fig.~\ref{fig: EE vs SE} illustrates the trade-off between the spectral and energy efficiency in our system.
One observation that can be made based on this figure is that, in a system with $N_{\RF} = 1$ RF chain,  increasing the number of transmit antennas -- although it increases the maximum achievable spectral efficiency -- does not affect the maximum energy efficiency of the system significantly. The reason for this can be understood by noting that,
as Fig.~\ref{fig: energy efficiency} shows, the energy efficiency for different values of $N_t$ reaches its maximum when $P$ is small and the PAs are working in their linear region. In this region, using Proposition~\ref{prop: lower bound for EE}, it is straightforward to show that the energy efficiency is related to $P$ and $N_t$ only through their product, $N_t P$. Therefore, for each particular value of $N_t$ there is a corresponding value for $P$, where $N_t P$ leads to the same optimal energy efficiency.
This implies that in a practical system, in order to have a reasonable spectral efficiency and still perform energy efficiently, we should not increase $N_t$ unboundedly.

Fig.~\ref{fig: different beamforming approaches} shows the spectral and energy efficiency for different beamforming schemes when $N_\RF = 5$ and $N_t = 16$. In the digital beamforming scheme, the transmit beamformer is designed fully-digital and is matched to the eigen directions of the channel. As the figure illustrates, the digital scheme leads to optimal spectral and energy efficiency when $P$ is small. In the analog beamforming scheme, the baseband beamformer is not used and the RF beamformer is matched to the AoDs of the channel, i.e., $\mathbf{F}_{\BB} = \I_{N_{\RF}}$, and $\mathbf{F}_{\RF} = \mathbf{A}_t$. Observe that the analog beamforming is the optimal beamforming scheme at high $P$, both in the sense of spectral efficiency and energy efficiency. This is because in this scheme the input power is equally allocated to different  PAs and therefore the total radiated distortion power is less compared to the case where the powers are allocated unequally to different PAs (e.g. in the digital beamforming scheme).

In addition to the digital and analog beamforming, the simulation results for a hybrid and a quantized analog beamforming schemes are also plotted in Fig.~\ref{fig: different beamforming approaches}. Both schemes are implemented by revising the MATLAB code used in the simulations of~\cite{Alkhateeb2014}.
In both cases, we assume that the full channel state information
is available at the transmitter and a $4$-bit quantization level is considered for the phase shifters in the analog beamforming stage.
The simulation results show that at small $P$, where the PAs are operating linearly, the hybrid scheme outperforms the analog and quantized analog beamforming schemes.
However, at the high input powers, the analog and the quantized analog show a better performance.
Again, it is due to the equal power allocation to the different PAs in the analog and the quantized analog schemes.

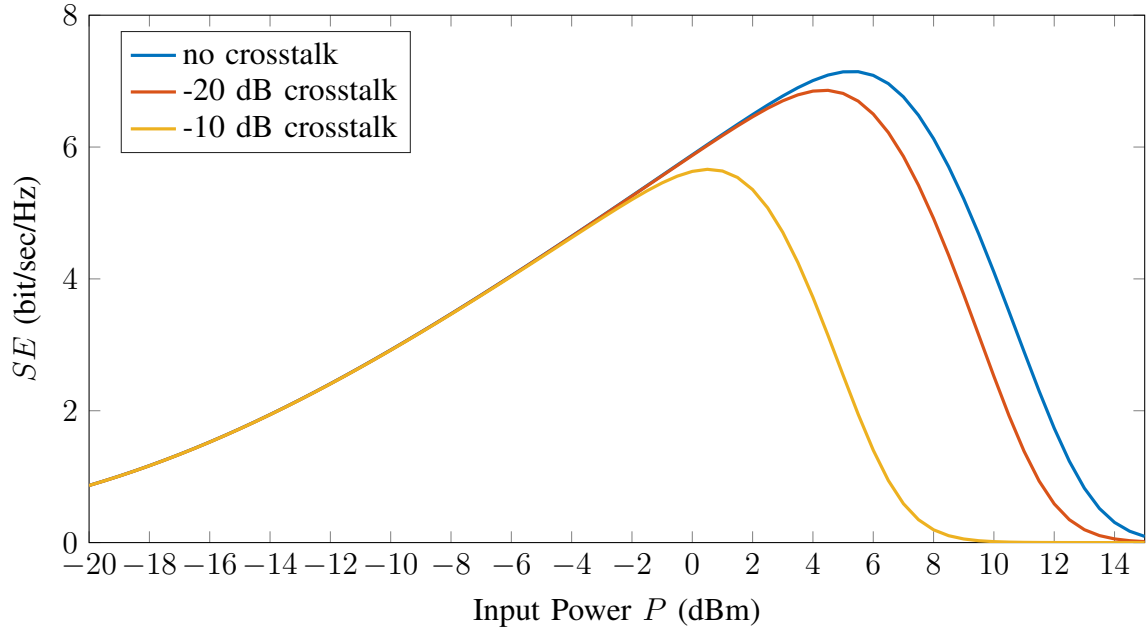
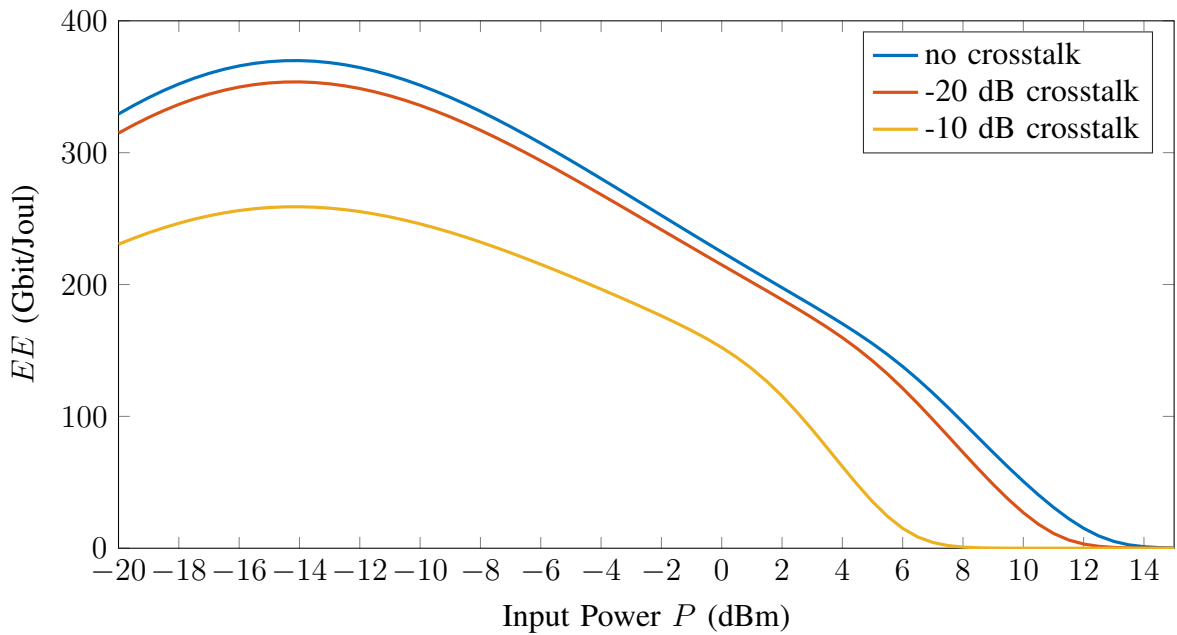
\begin{figure}[t!]
    \begin{subfigure}{1\columnwidth}
        \centering
        \input{Components/Figures/Reviewer1-Q1-SE}
        \caption{System spectral efficiency.}\label{fig: R1_Q1_S1}
    \end{subfigure}
    \begin{subfigure}{1\columnwidth}
        \centering
    \input{Components/Figures/Reviewer1-Q1-EE}
    \caption{System energy efficiency.}\label{fig: R1_Q1_S2}
    \end{subfigure}
    \caption{System performance for different levels of crosstalk power.}\label{fig: R1_Q1}
\end{figure}

The crosstalk effect in a MIMO system with nonlinear transmit PAs is studied in Fig.~\ref{fig: R1_Q1}.
In this figure, $N_t = 64$
while the rest of the simulation parameters are the same as the ones in Fig.~\ref{fig: SE vs P} and Fig.~\ref{fig: energy efficiency}.
Moreover, the entries of the crosstalk matrix, $B_{\text{TX}}$, are i.i.d. and drawn from the distribution $\mathcal{CN}(0,\sigma^2_{\text{ct}})$, where $\sigma^2_{\text{ct}}$ represents the average crosstalk power.
In this case,
it can be shown that coupling leads to an uneven allocation of the total transmit power in the antenna branches.
In other words, unlike the system with no coupling, given a fixed input power $\widetilde{P}=\sum_{n=1}^{N_t}\widetilde{P}_n$,  $\widetilde{P}_n$
is not the same for all $n=1,\dots,N_t$ when crosstalk exists.
Therefore, at high $\widetilde{P}$, more distortion power is radiated compared to the case in which
the powers are allocated equally to the antenna branches.
Fig.~\ref{fig: R1_Q1} shows the spectral efficiency and energy efficiency of the system with different level of crosstalk power.
As the figure illustrates, by increasing the crosstalk power, both spectral efficiency and energy efficiency of the system decrease.

\section{Conclusions}
\label{sec: Conclusions}
This paper investigated the spectral and energy efficiency of hybrid beamforming for mmWave systems
employing nonlinear \ac{PA}s. In order to capture the impact of nonlinearities on the spectral efficiency, a stochastic model for the transmitted distortion signal was derived. Unlike the models widely-used in the previous works, this model reflects the dependency of the spatial direction of the distortion signal to the spatial direction of the desired signal. Furthermore, a realistic power consumption model for the
transmitter's PAs was considered to find the energy efficiency of the system.

Based on the derived model, we proposed an optimization problem for maximizing the energy efficiency by designing the beamforming filters.
In the special case when the transmitter is equipped with one RF-chain, we found the closed form
solutions for the beamforming filters.

Our numerical results show that when using hybrid beamforming, increasing the transmit power level when the number of transmit antennas grows large can be counter-effective
in terms of spectral and energy efficiency.
On the other hand, with a moderate number of transmit
antennas, increasing the transmit power up to a certain threshold is beneficial for the spectral
and energy efficiency of the system.

\section*{Acknowledgments}
{We thank the Associate Editor and the anonymous Reviewers for their insightful comments, which helped
to improve the presentation and the contents of the paper.}

\section{Appendix: Proofs}\label{sec: Proofs}
\subsection*{Proof Sketch of Proposition~\ref{prop: average linear gain}}
Exploiting \eqref{eq: transmitted signal}, the transmitted signal from the $n^{th}$ antenna of the transmitter is $x_n = [\widebar{\mathbf{G}}]_{nn}\;u_n + d_n$. Therefore the average linear gain of the \ac{PA} of this antenna can be written
using the Bussgang theorem~\cite{Bussgang1952} as
\begin{equation}\label{}
\begin{split}
  [\widebar{\mathbf{G}}]_{nn} &\define \frac{\E\{x_n u_n^*\}}{\E\{|u_n|^2\}}
  \stackrel{(a)}{=} \frac{1}{P_n}\E\left\{\sum\limits_{m=0}^{M} \beta_{2m+1}\; \left|u_{n}\right|^{2m+2} \right\} \\
  & = \frac{1}{P_n}\sum\limits_{m=0}^{M} \beta_{2m+1}
  \E\left\{\left|u_{n}\right|^{2m+2}\right\},
\end{split}
\end{equation}
where (a) follows by substituting for $x_n$ from~\eqref{eq: polynomial model} and noting that by definition
$\E\{|u_n|^2\} = P_n$.
Now, by taking the expectation of the right-hand side over $u_{n}$ and considering that $u_{n}$ is a circularly symmetric complex Gaussian distributed random variable with distribution $\mathcal{CN}(0,P_n)$, the proof will be completed.

\subsection*{Proof of Proposition~\ref{prop: nonlinear distorion covariance matrix}}
We introduce a new function $\phi_m: \mathbb{C} \mapsto \mathbb{C}$, $\phi_m(a) \define \left|a\right|^{2m}a$.
In order to continue with the proof, first, we need to study the first- and second-order statistics of $\phi_{m}(a)$ using
the \emph{Isserlis' theorem} and the following lemma.
For the sake of completeness, we re-state the Isserlis' theorem below.
\begin{theorem} Isserlis' theorem~\cite{Isserlis1918}\newline
If $\left[a_1,\dots,a_K\right]^T$ is a zero-mean multivariate normal random vector, then
\begin{align}
\label{eq:Isserlis_theorem}
\E\left\{\prod\limits_{k=1}^{K}a_k\right\} \!=\!
\left\{
\begin{array}{ll}
\sum_{S}\prod_{B_{ij}\in S} \E\left\{a_i a_j\right\} & K \ \text{is even},  \\
0 & K \ \text{is odd},
\end{array}
\right.
\end{align}
where $B_{ij} = \{a_i ,a_j\}$ is an arbitrary 2-subset of $\mathcal{A} = \{a_1,\dots,a_K\}$ and $S$ runs through the list of all possible partitions of $\mathcal{A}$ into 2-subsets.
\end{theorem}
Note that although Isserlis' theorem is originally developed for real-valued random vectors, it can be extended to the case of complex Gaussian variables as well (see~\cite{Reed1962}).

\begin{lemma}
\label{lemma:1st_and_2nd-order_statistics}
Consider $a\sim\mathcal{CN}(0,\sigma_a^2)$ and $b\sim\mathcal{CN}(0,\sigma_b^2)$.
For any $m,n\in \{0,1,\dots\}$,
$\phi_{m}(a)$ and $\phi_{n}(b)$ are zero-mean random processes
and the cross-correlation between them is
\begin{equation}
\begin{split}
\label{eq:cross-correlation_terms}
&\E\left\{\phi_{m}(a) \phi_{n}^*(b)\right\} = \\
&\sum\limits_{q=1}^{\min\{m,n\}}\!\!\!\frac{(m+1)!(n+1)!}{q+1}
{{m}\choose{q}}\!{{n}\choose{q}}\sigma_{a}^{2(m-q)} \sigma_{b}^{2(n-q)}\!\left|\rho\right|^{2q}\!\rho\;,
\end{split}
\end{equation}
where $\rho = \E\left\{a b^*\right\}$.
\end{lemma}
\begin{IEEEproof}
Since the number of multiplied Gaussian terms in $\phi_{m}(a)$ and $\phi_{n}(b)$ is odd, the zero-mean property of them is proved as
an immediate implication of Isserlis' theorem.
Now, inspired by the approach in \cite{Zhou2004},
Isserlis' theorem can again be employed to find the cross-correlation between $\phi_{m}(a)$ and $\phi_{n}(b)$.
First, we define the following set
\begin{equation}
\mathcal{A} \define \{\overbrace{a,\dots,a}^{m},\overbrace{a^*,\dots,a^*}^{m+1},\overbrace{b,\dots,b}^{n+1},\overbrace{b^*,\dots,b^*}^{n}\},
\end{equation}
which contains the individual elements in the product $\phi_m(a)\phi_n(b)^*$.
Now, we form a two-way table out of the elements in $\mathcal{A}$ as,
\begin{align}
\label{tabel:partitioning}
\begin{array}{ll}
\overbrace{a \ \dots \ a}^{m+1}&\overbrace{a^* \ \dots \ a^*}^{m} \\
\underbrace{b^* \ \dots \ b^*}_{n+1}&\underbrace{b \ \dots \ b}_{n}.
\end{array}
\end{align}
Since $a$ and $b$ are circularly symmetric, the only 2-subsets that lead to non-zero expectations are $\{a,a^*\}$, $\{b,b^*\}$, $\{a, b^*\}$ and $\{a,b^*\}$. Similar to \cite{Zhou2004}, we refer to the 2-subsets that contain elements from both rows as \emph{hooking} 2-subsets.
We observe that any partition $S_{\text{nz}}$ leads to non-zero expectation if
\begin{enumerate}
\item only consists of non-zero 2-subsets,
\item has exactly $q+1$ hooking 2-subsets of the form $\{a,b^*\}$ and $q$ hooking 2-subsets of the form $\{a^*,b\}$, where $q\in\{0,\dots,\min{\{m,n\}}\}$.
\end{enumerate}
Therefore a non-zero partition, $S_{\text{nz}}$ , can be written as
\begin{equation}\label{}
  \begin{split}
  S_{\text{nz}} = \bigg\{
  &\overbrace{\{a,a^*\},\dots,\{a,a^*\}}^{m-q},
  \overbrace{\{b,b^*\},\dots,\{b,b^*\}}^{n-q}, \\
  &\overbrace{\{a,b^*\},\dots,\{a,b^*\}}^{q+1},
  \overbrace{\{a^*,b\},\dots,\{a^*,b\}}^{q}
  \bigg\},
  \end{split}
\end{equation}
and subsequently we have
\begin{equation}
\label{eq:valid_partition}
\begin{split}
&\prod_{B_{ij}\in S_{\text{nz}}} \E\left\{a_i a_j\right\}  \\
=&\E\left\{aa^*\right\}^{m-q}
\E\left\{bb^*\right\}^{n-q} \E\left\{ab^*\right\}^{q+1}
\E\left\{a^*b\right\}^{q} \\
=&\sigma_a^{2(m-q)} \sigma_b^{2(n-q)} \left|\rho\right|^{2q}\rho.
\end{split}
\end{equation}

In \cite{Zhou2004}, in a similar setup, it is shown that
the number of partitions with the similar blocks as in \eqref{eq:valid_partition} is equal to $\frac{(m+1)!(n+1)!}{q+1}{{m}\choose{q}}{{n}\choose{q}}$. Therefore, summing over all the non-zero partitions of \eqref{tabel:partitioning} leads to the result in \eqref{eq:cross-correlation_terms}.
\end{IEEEproof}


Now, we are ready to prove the proposition.
By substituting from \eqref{eq: inst. gain} and \eqref{eq: ave. gain}
into~\eqref{eq: distortion}, the distortion at $k^{th}$ antenna can be written as
\begin{equation}
\label{eq:Bussgang_distortion}
d_k = \sum_{m=0}^{M} \beta_{2m+1} \left(\phi_{m}(u_k) - P_k^{m}(m+1)! \phi_{0}(u_k)\right).
\end{equation}
Taking the exception of both sides of \eqref{eq:Bussgang_distortion} and applying Lemma~\ref{lemma:1st_and_2nd-order_statistics} proves the zero-mean property of the distortion signals.

Using \eqref{eq:Bussgang_distortion}, the cross-correlation between the distortion noise at $k$th and $j$th antennas is computed as
\begin{align}
\label{eq:cross-correlation of distortion noise}
\nonumber
&[\Cd]_{kj} = \E\left\{d_k d_j^*\right\}
=\sum_{m=1}^{M}\sum_{n=1}^{M} \beta_{2m+1}\beta_{2n+1}^* \\ \nonumber &\Bigg(\E\left\{\phi_{m}(u_k) \phi^*_{n}(u_j)\right\}
- P_{j}^{n}(n+1)! \E\left\{\phi_{m}(u_k) \phi_{0}^*(u_j)\right\} \\ \nonumber
&- P_{k}^{m}(m+1)! \E\left\{\phi_{0}(u_k) \phi_{n}^*(u_j)\right\} \\
&+ P_{k}^{m}P_{j}^{n}(m+1)!(n+1)! \E\left\{\phi_{0}(u_k) \phi_{0}^*(u_j)\right\}\Bigg).
\end{align}
Moreover, by the help of Lemma~\ref{lemma:1st_and_2nd-order_statistics}, \eqref{eq:cross-correlation of distortion noise} can be further simplified to
\begin{align}
\label{eq:cross-correlation of distortion noise_simplified}
\nonumber
\begin{split}
[\Cd]_{kj} =&
\sum\limits_{m=1}^{M}\sum\limits_{n=1}^{M}\sum\limits_{q=1}^{\min\{m,n\}}
\frac{(m+1)!(n+1)!}{q+1} {{m}\choose{q}}{{n}\choose{q}}\\
 &\times\beta_{2m+1}\beta_{2n+1}^* P_{k}^{(m-q)} P_j^{(n-q)} \left|\left[\Cu\right]_{kj}\right|^{2q} \left[\Cu\right]_{kj}
\end{split}
\end{align}
Finally by noticing that $\sum\limits_{m=1}^{M}\sum\limits_{n=1}^{M}\sum\limits_{q=1}^{\min\{m,n\}}$ is equivalent to
$\sum\limits_{q=1}^{M}\sum\limits_{m=q}^{M}\sum\limits_{n=q}^{M}$ and introducing a new function $\gamma_m(P_k)$  as in \eqref{eq: gamma},
the proof is completed.

\subsection*{Proof of Proposition~\ref{prop: maximum SE one RF chain}}
We use the following two lemmas in the proof of Proposition~\ref{prop: maximum SE one RF chain}.
\begin{lemma}\label{lemma: Nrf = 1}
Consider vector $\mathbf{z}\in \mathbb{C}^{N}$ with constant modulus entries $|z_n| = \sqrt{\alpha},~~n=1,\dots,N$ and define $\mathbf{C}_{\mathbf{z}} \define \mathbf{z}\mathbf{z}\herm$. Then,
\begin{equation}\label{eq: Cz}
  \overbrace{\left(\mathbf{C}_{\mathbf{z}} \odot \dots \odot \mathbf{C}_{\mathbf{z}} \right)} ^
{(m+1)\ \text{times}} \odot
\overbrace{\left(\mathbf{C}_{\mathbf{z}}^T \odot \dots \odot \mathbf{C}_{\mathbf{z}}^T \right)}^
{m\ \text{times}}
= \alpha^{2m}\; \mathbf{C}_{\mathbf{z}}.
\end{equation}
\end{lemma}
\begin{IEEEproof}
Note that $[\mathbf{C}_{\mathbf{z}}]_{ij} = z_i z_j$ and $[\mathbf{C}_{\mathbf{z}}\tran]_{ij} = z_i^* z_j^*$ for $i,j\in\{1,\dots,N\}$. Therefore the entry $ij$ of the left hand side of~\eqref{eq: Cz} can be written as
\begin{equation}\label{}
  \begin{split}
    (z_iz_j)^{m+1}(z_i^*z_j^*)^{m} &= |z_i|^{2m}|z_j|^{2m} z_iz_j \\
    &= \alpha^{2m}z_iz_j \\
    &= \alpha^{2m}[\mathbf{C}_{\mathbf{z}}]_{ij}.
  \end{split}
\end{equation}
This completes the proof.
\end{IEEEproof}

\begin{lemma}\label{lemma: increasing function}
Define the function $f:\mathbb{S}^{N}\mapsto \mathbb{R}^{+}$ as
\begin{equation}\label{eq: f}
  f(\mathbf{Z}) \define \log_2\det\left(\I + \left(\alpha_2\mathbf{Z}+ \I\right)^{-1}\alpha_1\mathbf{Z}\right).
\end{equation}
For any $\mathbf{Z}'\succeq \mathbf{Z}$ (that is when $\mathbf{Z}'-\mathbf{Z}$ is a \ac{PSD} matrix), we have $f(\mathbf{Z}')\ge f(\mathbf{Z})$.
\end{lemma}
\begin{IEEEproof}
Let us denote the ordered eigenvalues of $\mathbf{Z}$ and $\mathbf{Z}'$ by $\lambda_1\ge\dots\ge\lambda_N$
and $\lambda_1'\ge\dots\ge\lambda_N'$, respectively. Since $\mathbf{Z}'\succeq \mathbf{Z}$,
we know that $\lambda_n'\ge\lambda_n$ for $n=1,\dots,N$.
Moreover, note that $f(\mathbf{Z})$ can alternatively be expressed as
\begin{equation}\label{}
  f(\mathbf{Z}) = \sum_{n=1}^{N}\log_2\left(1+\frac{\alpha_1\lambda_n}{\alpha_2\lambda_n+1}\right).
\end{equation}
Now, by considering that $\log_2(1+\frac{\alpha_1\lambda_n}{\alpha_2\lambda_n+1})$ is a non-decreasing function of $\lambda_n$, we can conclude that $f(\mathbf{Z'})\ge f(\mathbf{Z})$.
\end{IEEEproof}

Observe that when $N_{\RF} = 1$, $\Cu = P\; \mathbf{F}_{\RF}\mathbf{F}_{\RF}\herm$, and $\mathbf{F}_{\RF}$ is a vector with constant modulus entries where $|[\mathbf{F}_{\RF}]_n| = \sqrt{1/N_t}$, for $n\in\{1,\dots,N_t\}$. Therefore, the input powers to all the PAs are equal, i.e., $[\Cu]_{11}=P_1=\dots=P_{N_t}=[\Cu]_{N_tN_t}=P/N_t$.
Hence, using Proposition~\ref{prop: average linear gain}, we can show that
\begin{equation}\label{eq: G_bar one RF chain}
  \widebar{\mathbf{G}} = \widebar{g}\left(\frac{P}{N_t}\right)\mathbf{I}_{N_t},
\end{equation}
and therefore
\begin{equation}\label{eq: Ctu one RF chain}
  \Ctu = \widebar{g}_{s}\left(\frac{P}{N_t}\right)
  \!P\;\mathbf{F}_{\RF}\mathbf{F}_{\RF}\herm.
\end{equation}
Moreover, using Proposition~\ref{prop: nonlinear distorion covariance matrix} and Lemma~\ref{lemma: Nrf = 1}, it is straightforward to show that in this case
\begin{equation}\label{eq: Cd one RF chain}
  \Cd = \widebar{g}_{d}\left(\frac{P}{N_t}\right)\!P\;\mathbf{F}_{\RF}\mathbf{F}_{\RF}\herm.
\end{equation}
This implies that the covariance matrices of the transmitted desired signal and distortion signal are equal up to a scaling factor and therefore the signals always have the same spatial direction.

By replacing $\alpha_1$ and $\alpha_2$ in \eqref{eq: f}
with $P/\sigma_n^2 \widebar{g}_s(P/N_t)$ and $P/\sigma_n^2 \widebar{g}_d(P/N_t)$, respectively,
we can show that $SE = f(\mathbf{H}\Cu\mathbf{H}\herm)$, for $f(.)$ defined in Lemma~\ref{lemma: increasing function}.
Hence by considering that $\mathbf{H}\Cu\mathbf{H}\herm\preceq P \widetilde{\mathbf{H}}\widetilde{\mathbf{H}}\herm$
for any $\Cu = P \mathbf{F}_{\RF}\mathbf{F}_{\RF}\herm$ which satisfies the power constraint $\tr(\Cu)\le P$
and using Lemma~\ref{lemma: increasing function}, it is straightforward to show that the maximum spectral efficiency in this case is
\begin{equation}\label{}
  \widebar{SE} =
  f(P \widetilde{\mathbf{H}}\widetilde{\mathbf{H}}\herm),
\end{equation}
which completes the proof.

\subsection*{Proof of Corollary~\ref{cor: lower-bound on SE one RF chain}}
The following lemma will be used in the proof of this corollary.
\begin{lemma}\label{lemma: h<f}
Consider function $h: \mathcal{X} \mapsto \mathbb{R}^{+}$,
where $\mathcal{X} = \{(\mathbf{Z},\mathbf{r})\in (\mathbb{S}^{N},\mathbb{C}^{N})|\mathbf{r}\herm\mathbf{r}=1\}$ and
\begin{equation}\label{eq: h}
h(\mathbf{Z},\mathbf{r}) \define \log_2 \left(1 + \frac{\alpha_1\mathbf{r}\herm\mathbf{Z}\mathbf{r}}
{\alpha_2\mathbf{r}\herm\mathbf{Z}\mathbf{r}+1} \right).
\end{equation}
then the following inequality always holds
\begin{equation}
    f(\mathbf{Z})\ge h(\mathbf{Z},\mathbf{r}),
\end{equation}
where $f(\mathbf{Z})$ is defined in Lemma~\ref{lemma: increasing function}.
Equality holds if and only if $\mathbf{Z}$ is rank one and $\mathbf{r}$ matches the  eigenvector of $\mathbf{Z}$ corresponding to its non-zero eigenvalue.
\end{lemma}
\begin{IEEEproof}
Define the ordered eigenvalues of $\mathbf{Z}$ as $\lambda_1\ge\dots\ge\lambda_N$. From \cite{Horn2012}, we know that
\begin{equation}
\mathbf{r}\herm\mathbf{Z}\mathbf{r} \le \max_{\widetilde{\mathbf{r}}}\widetilde{\mathbf{r}}\herm\mathbf{Z}
\widetilde{\mathbf{r}} = \lambda_1.
\end{equation}
Therefore, by noticing that $\log_2\left(1 + \frac{\alpha_1 \lambda_n}{1+\alpha_2 \lambda_n}\right)$ is a non-decreasing function of $\lambda_n$, we can write
\begin{equation}\label{eq: h<f}
\begin{split}
  h(\mathbf{Z},\mathbf{r}) &\le \log_2\left(1 + \frac{\alpha_1 \lambda_1}{\alpha_2\lambda_1 + 1 }\right) \\
  &\le \sum_{n=1}^{N} \log_2\left(1 + \frac{\alpha_1 \lambda_n}{\alpha_2\lambda_n + 1 }\right) \\
  &\stackrel{(a)}{=} f(\mathbf{Z}),
\end{split}
\end{equation}
where (a) is due to Lemma~\ref{lemma: increasing function}.

When $\mathbf{Z}$ is rank-one, then $\lambda_n = 0,~~\forall n>1$ and therefore the second inequality in~\eqref{eq: h<f} holds with equality.
Furthermore, when $\mathbf{r}$ is the eigenvector of $\mathbf{Z}$ corresponding to its non-zero eigenvalue then $\mathbf{r}\herm\mathbf{Z}\mathbf{r} = \lambda_1$ and the first inequality holds also with equality. This concludes the proof.
\end{IEEEproof}

\subsection*{Proof of Proposition~\ref{prop: lower bound for EE}}
First, we observe that the optimization variables $\mathbf{F}_{\text{RF}}$ and $\mathbf{F}_{\text{BB}}$ are coupled neither in the objective function nor in the constraints of the optimization problem~(P1). Therefore the optimal solution can be found by first solving~(P1) for $\mathbf{F}_{\text{RF}}$ and then solving it for $\mathbf{F}_{\text{BB}}$.

Furthermore, we notice that when $N_s = 1$ the baseband beamformer $\mathbf{F}_{\text{BB}}$ simplifies to the scalar input power $P$
as was shown in the proof of Proposition~\ref{prop: maximum SE one RF chain},
the input power, independent from the beamforming filter,
is equally divided between the PAs.
That is $P_1 = \dots = P_{N_t} = P/N_t$. Therefore, using~\eqref{eq: cons. power}, for $n = 1,\dots,N_t$, we have
\begin{equation}\label{}
  \widebar{P}_{{\rm cons},n} = \eta_{\max}\sqrt{P_{\max}P/N_t (\widebar{g}_s(P/N_t)+\widebar{g}_d(P/N_t))}.
\end{equation}

Now, we can continue with the proof. Utilizing \eqref{eq: EE definition}, (P1) is equivalent to the following problem
\begin{equation} \label{eq: P1 - Eq1}
\begin{split}
&\begin{array}{ll}
  \underset{\mathbf{F}_{\text{RF}},\mathbf{F}_{\text{BB}}}{\text{maximize}} & \frac{SE}{P_{\text{cons}}} \\
  \text{subject to}& P_{\text{cons}}\le P_0 \\
  &\left|[\mathbf{F}_{\text{RF}}]_{i,j}\right| = {\sqrt{1/N_t}},
  \quad \forall i,j
\end{array} \\
&\stackrel{(a)}{=}
\begin{array}{ll}
  \underset{\mathbf{F}_{\text{BB}}}{\text{maximize}} & \frac{\widebar{SE}}{P_{\text{cons}}}\\
  \text{subject to}& P_{\text{cons}}\le P_0,
\end{array}
\end{split}
\end{equation}
where (a) follows by noting that $\mathbf{F}_{\text{RF}}$, $\mathbf{F}_{\text{BB}}$ are uncoupled and using Proposition~\ref{prop: maximum SE one RF chain}. Since $N_s=1$, we can replace $\mathbf{F}_{\text{BB}}$ by $P$ in~\eqref{eq: P1 - Eq1} and $P_{\text{cons}}$ by
$\widebar{P}_{\text{cons}} = \sum_{n=1}^{N_t}\widebar{P}_{{\rm cons},n}$. This concludes the proof.


\bibliographystyle{Components/MetaFiles/IEEEtran}
\bibliography{Components/MetaFiles/References}
\end{document}

%% file: Components/MetaFiles/packages.tex
\usepackage{gensymb}
\usepackage{dsfont}
\usepackage{amsmath,amssymb,amsfonts,amsthm}
\usepackage{epsfig}
\usepackage{cite}
\usepackage{hhline}
\usepackage{multirow}
\usepackage{xcolor}
\usepackage{makecell}
\usepackage{subcaption}
\usepackage{capt-of}
\usepackage[labelformat=simple]{subcaption}

\usepackage{enumerate}
\usepackage{enumitem}
\usepackage{color}
\usepackage{graphicx}
\usepackage{algorithmic}
\usepackage[ruled]{algorithm}
\usepackage{booktabs}
\usepackage{bm}
\usepackage{tikz}
\usetikzlibrary{decorations.markings}
\tikzstyle arrowstyle=[scale=1]
\tikzstyle directed=[postaction={decorate,decoration={markings,
    mark=at position .65 with {\arrow[arrowstyle]{stealth}}}}]
\tikzstyle reverse directed=[postaction={decorate,decoration={markings,
    mark=at position .65 with {\arrowreversed[arrowstyle]{stealth};}}}]

\usepackage[nolist,printonlyused]{acronym} 
\usepackage{colortbl}

\usepackage[T1]{fontenc}
\usepackage[utf8]{inputenc}
\usepackage{pgfplots}
\usepackage{comment} 

\pgfplotsset{compat=newest}
\usetikzlibrary{plotmarks}
\usepgfplotslibrary{patchplots}

\usetikzlibrary{automata,arrows,positioning,calc,matrix,decorations.markings,decorations.pathreplacing}

\makeatletter
\newcommand{\vast}{\bBigg@{3}}
\newcommand{\Vast}{\bBigg@{4}}
\makeatother
\renewcommand{\IEEEQED}{\IEEEQEDopen}

\showoutput


%% file: Components/MetaFiles/acronyms.tex
\begin{acronym}[LTE-Advanced]
  \acro{2G}{Second Generation}
  \acro{3G}{3$^\text{rd}$~Generation}
  \acro{3GPP}{3$^\text{rd}$~Generation Partnership Project}
  \acro{4G}{4$^\text{th}$~Generation}
  \acro{5G}{5$^\text{th}$~Generation}
  \acro{AoA}{angle of arrival}
  \acro{AoD}{angle of departure}
  \acro{BER}{bit error rate}
  \acro{BF}{beamforming}
  \acro{BLER}{BLock Error Rate}
  \acro{BPC}{Binary Power Control}
  \acro{BPSK}{Binary Phase-Shift Keying}
  \acro{BRA}{Balanced Random Allocation}
  \acro{BS}{base station}
  \acro{CAP}{Combinatorial Allocation Problem}
  \acro{CAPEX}{Capital Expenditure}
  \acro{CBF}{Coordinated Beamforming}
  \acro{CS}{Coordinated Scheduling}
  \acro{CSI}{channel state information}
  \acro{CSIT}{channel state information at the transmitter}
  \acro{CSIR}{channel state information at the receiver}
  \acro{D2D}{device-to-device}
  \acro{DCA}{Dynamic Channel Allocation}
  \acro{DE}{Differential Evolution}
  \acro{DFT}{Discrete Fourier Transform}
  \acro{ULA}{uniform linear array}
  \acro{DIST}{Distance}
  \acro{DL}{downlink}
  \acro{DMA}{Double Moving Average}
  \acro{DMRS}{Demodulation Reference Signal}
  \acro{D2DM}{D2D Mode}
  \acro{DMS}{D2D Mode Selection}
  \acro{DPC}{Dirty Paper Coding}
  \acro{DRA}{Dynamic Resource Assignment}
  \acro{DSA}{Dynamic Spectrum Access}
  \acro{FDD}{frequency division duplexing}
  \acro{HPA}{high-power amplifier}
  \acro{LTE}{Long Term Evolution}
  \acro{LSAS}{large scale antenna system}
  \acro{LS-MIMO}{Large scale multiple-input multiple-output}
  \acro{LTE}{Long Term Evolution}
  \acro{METIS}{Mobile Enablers for the Twenty-Twenty Information Society}
  \acro{MIMO}{multiple-input multiple-output}
  \acro{MMSE}{minimum mean square error}
  \acro{MU-MIMO}{multiuser multi-input multi-output}
  \acro{OFDM}{orthogonal frequency division multiplexing}
  \acro{SU-MIMO}{single-user multiple-input multiple-output}
  \acro{MISO}{multiple-input single-output}
  \acro{mmWave}{millimeter-wave}
  \acro{MRC}{maximum ratio combining}
  \acro{MRT}{maximum ratio transmission}
  \acro{MS}{mode selection}
  \acro{MSE}{mean square error}
  \acro{MTC}{machine type communications}
  \acro{NSPS}{national security and public safety}
  \acro{NWC}{network coding}
  \acro{PC}{pilot contamination}
  \acro{PHY}{physical layer}
  \acro{QoS}{Quality of Service}
  \acro{QPSK}{Quadri-Phase Shift Keying}
  \acro{RAISES}{Reallocation-based Assignment for Improved Spectral Efficiency and Satisfaction}
  \acro{RAN}{Radio Access Network}
  \acro{RA}{Resource Allocation}
  \acro{RAT}{Radio Access Technology}
  \acro{RB}{resource block}
  \acro{RF}{radio frequency}
  \acro{SINR}{signal-to-noise-plus-interference ratio}
  \acro{SNR}{signal-to-noise ratio}
  \acro{STC}{space-time coding}
  \acro{TDD}{time division duplexing}
  \acro{UE}{user equipment}
  \acro{UL}{uplink}
  \acro{VUE}{vehicular user equipment}
  \acro{V2X}{vehicle-to-vehicle and vehicle-to-infrastructure}
  \acro{ZF}{Zero-Forcing}
  \acro{ZMCSCG}{Zero Mean Circularly Symmetric Complex Gaussian}

  \acro{RF}{radio frequency}
  \acro{FCC}{federal communications commission}
  \acro{PA}{power amplifier}
  \acro{PSD}{positive semi-definite}
  \acro{SVD}{singular value decomposition}
  \acro{EVD}{eigen value decomposition}
  \acro{NMSE}{normalized mean square error}
  \acro{LS}{least square}
  \acro{ADC}{analog-to-digital converter}
  \acro{DAC}{digital-to-analog converter}
  \acro{LNA}{low-noise amplifier}
  \acro{ML}{maximum likelihood}
  \acro{TDMA}{time division multiple access}
  \acro{FDMA}{frequency division multiple access}
  \acro{AM-AM}{amplitude-to-amplitude}
  \acro{AM-PM}{amplitude-to-phase}
  \acro{PAPR}{peak-to-average-power ratio}
  \acro{DPD}{digital predistorter}
  \acro{SISO}{single-input single-output}
  \acro{EVM}{error vector magnitude}
  \acro{CoMP}{coordinated multipoint}
  \acro{LO}{local oscillator}
  \acro{FER}{frame error rate}
  \acro{CDF}{cumulative distribution function}
  \acro{PDPR}{pilot-to-data power ratio}
  \acro{AWGN}{additive white gaussian noise}
  \acro{PPS}{pulse-per-second}
\end{acronym}

%% file: Components/MetaFiles/commands.tex

\newtheorem{theorem}{Theorem}
\newtheorem{defin}{Definition}
\newtheorem{prop}{Proposition}
\newtheorem{lemma}{Lemma}
\newtheorem{corollary}{Corollary}
\newtheorem{alg}{Algorithm}
\newtheorem{remark}{Remark}
\newtheorem{result}{Result}
\newtheorem{conjecture}{Conjecture}
\newtheorem{example}{Example}
\newtheorem{notations}{Notations}
\newtheorem{assumption}{Assumption}

\newcommand{\define}{\stackrel{\boldsymbol{\cdot}}{=}}
\newcommand{\be}{\begin{equation}}
\newcommand{\ee}{\end{equation}}
\newcommand{\ba}{\begin{array}}
\newcommand{\ea}{\end{array}}
\newcommand{\bea}{\begin{eqnarray}}
\newcommand{\eea}{\end{eqnarray}}
\newcommand{\combin}[2]{\ensuremath{ \left( \ba{c} #1 \\ #2 \ea \right) }}
\newcommand{\diag}{{\mathrm{diag}}}
\newcommand{\rank}{{\mbox{rank}}}
\newcommand{\dom}{{\mathrm{dom{\color{white!100!black}.}}}}
\newcommand{\range}{{\mbox{range{\color{white!100!black}.}}}}
\newcommand{\image}{{\mbox{image{\color{white!100!black}.}}}}
\newcommand{\herm}{^{\mbox{\scriptsize H}}}  
\newcommand{\sherm}{^{\mbox{\tiny H}}}       
\newcommand{\tran}{^{\mathrm{\scriptsize T}}}  
\newcommand{\tranIn}{^{\mbox{-\scriptsize T}}}  
\newcommand{\card}{{\mbox{\textbf{card}}}}
\newcommand{\asign}{{\mbox{$\colon\hspace{-2mm}=\hspace{1mm}$}}}
\newcommand{\ssum}[1]{\mathop{ \textstyle{\sum}}_{#1}}

\newcommand{\vbar}{\raisebox{.17ex}{\rule{.04em}{1.35ex}}}
\newcommand{\vbarind}{\raisebox{.01ex}{\rule{.04em}{1.1ex}}}
\newcommand{\D}{\ifmmode {\rm I}\hspace{-.2em}{\rm D} \else ${\rm I}\hspace{-.2em}{\rm D}$ \fi}
\newcommand{\T}{\ifmmode {\rm I}\hspace{-.2em}{\rm T} \else ${\rm I}\hspace{-.2em}{\rm T}$ \fi}
\newcommand{\B}{\ifmmode {\rm I}\hspace{-.2em}{\rm B} \else \mbox{${\rm I}\hspace{-.2em}{\rm B}$} \fi}
\newcommand{\Hil}{\ifmmode {\rm I}\hspace{-.2em}{\rm H} \else \mbox{${\rm I}\hspace{-.2em}{\rm H}$} \fi}
\newcommand{\C}{\ifmmode \hspace{.2em}\vbar\hspace{-.31em}{\rm C} \else \mbox{$\hspace{.2em}\vbar\hspace{-.31em}{\rm C}$} \fi}
\newcommand{\Cind}{\ifmmode \hspace{.2em}\vbarind\hspace{-.25em}{\rm C} \else \mbox{$\hspace{.2em}\vbarind\hspace{-.25em}{\rm C}$} \fi}
\newcommand{\Q}{\ifmmode \hspace{.2em}\vbar\hspace{-.31em}{\rm Q} \else \mbox{$\hspace{.2em}\vbar\hspace{-.31em}{\rm Q}$} \fi}
\newcommand{\Z}{\ifmmode {\rm Z}\hspace{-.28em}{\rm Z} \else ${\rm Z}\hspace{-.38em}{\rm Z}$ \fi}

\newcommand{\sgn}{\mathrm {sgn}}
\newcommand{\var}{\mathrm {var}}
\newcommand{\E}{\mathbb{E}}
\newcommand{\cov}{\mathrm {cov}}
\renewcommand{\Re}{\mathrm {Re}}
\renewcommand{\Im}{\mathrm {Im}}
\newcommand{\cum}{\mathrm {cum}}

\renewcommand{\vec}[1]{{\mathbf{#1}}}     
\newcommand{\vecsc}[1]{\mbox {\boldmath \scriptsize $#1$}}     
\newcommand{\itvec}[1]{\mbox {\boldmath $#1$}}
\newcommand{\itvecsc}[1]{\mbox {\boldmath $\scriptstyle #1$}}
\newcommand{\gvec}[1]{\mbox{\boldmath $#1$}}

\newcommand{\balpha}{\mbox {\boldmath $\alpha$}}
\newcommand{\bbeta}{\mbox {\boldmath $\beta$}}
\newcommand{\bgamma}{\mbox {\boldmath $\gamma$}}
\newcommand{\bdelta}{\mbox {\boldmath $\delta$}}
\newcommand{\bepsilon}{\mbox {\boldmath $\epsilon$}}
\newcommand{\bvarepsilon}{\mbox {\boldmath $\varepsilon$}}
\newcommand{\bzeta}{\mbox {\boldmath $\zeta$}}
\newcommand{\boldeta}{\mbox {\boldmath $\eta$}}
\newcommand{\btheta}{\mbox {\boldmath $\theta$}}
\newcommand{\bvartheta}{\mbox {\boldmath $\vartheta$}}
\newcommand{\biota}{\mbox {\boldmath $\iota$}}
\newcommand{\blambda}{\mbox {\boldmath $\lambda$}}
\newcommand{\bmu}{\mbox {\boldmath $\mu$}}
\newcommand{\bnu}{\mbox {\boldmath $\nu$}}
\newcommand{\bxi}{\mbox {\boldmath $\xi$}}
\newcommand{\bpi}{\mbox {\boldmath $\pi$}}
\newcommand{\bvarpi}{\mbox {\boldmath $\varpi$}}
\newcommand{\brho}{\mbox {\boldmath $\rho$}}
\newcommand{\bvarrho}{\mbox {\boldmath $\varrho$}}
\newcommand{\bsigma}{\mbox {\boldmath $\sigma$}}
\newcommand{\bvarsigma}{\mbox {\boldmath $\varsigma$}}
\newcommand{\btau}{\mbox {\boldmath $\tau$}}
\newcommand{\bupsilon}{\mbox {\boldmath $\upsilon$}}
\newcommand{\bphi}{\mbox {\boldmath $\phi$}}
\newcommand{\bvarphi}{\mbox {\boldmath $\varphi$}}
\newcommand{\bchi}{\mbox {\boldmath $\chi$}}
\newcommand{\bpsi}{\mbox {\boldmath $\psi$}}
\newcommand{\bomega}{\mbox {\boldmath $\omega$}}

\newcommand{\bolda}{\mbox {\boldmath $a$}}
\newcommand{\bb}{\mbox {\boldmath $b$}}
\newcommand{\bc}{\mbox {\boldmath $c$}}
\newcommand{\bd}{\mbox {\boldmath $d$}}
\newcommand{\bolde}{\mbox {\boldmath $e$}}
\newcommand{\boldf}{\mbox {\boldmath $f$}}
\newcommand{\bg}{\mbox {\boldmath $g$}}
\newcommand{\bh}{\mbox {\boldmath $h$}}
\newcommand{\bp}{\mbox {\boldmath $p$}}
\newcommand{\bq}{\mbox {\boldmath $q$}}
\newcommand{\br}{\mbox {\boldmath $r$}}
\newcommand{\bt}{\mbox {\boldmath $t$}}
\newcommand{\bu}{\mbox {\boldmath $u$}}
\newcommand{\bv}{\mbox {\boldmath $v$}}
\newcommand{\bw}{\mbox {\boldmath $w$}}
\newcommand{\bx}{\mbox {\boldmath $x$}}
\newcommand{\by}{\mbox {\boldmath $y$}}
\newcommand{\bz}{\mbox {\boldmath $z$}}

\newenvironment{Ex}
{\begin{adjustwidth}{0.04\linewidth}{0cm}
\begingroup\small
\vspace{-1.0em}
\raisebox{-.2em}{\rule{\linewidth}{0.3pt}}
\begin{example}
}
{
\end{example}
\vspace{-5mm}
\rule{\linewidth}{0.3pt}
\endgroup
\end{adjustwidth}}

\newcommand{\Hossein}[1]{{\textcolor{magenta}{\emph{**Hossein: #1**}}}}
\newcommand{\Nima}[1]{{\textcolor{magenta}{\emph{**Nima: #1**}}}}
\newcommand{\Comment}[1]{{\textcolor{blue}{\emph{**Comment: #1**}}}}
\newcommand{\Challenge}[1]{{\textcolor{red}{#1}}}
\newcommand{\NEW}[1]{{\textcolor{blue}{#1}}}


\makeatletter
\let\save@mathaccent\mathaccent
\newcommand*\if@single[3]{%
  \setbox0\hbox{${\mathaccent"0362{#1}}^H$}%
  \setbox2\hbox{${\mathaccent"0362{\kern0pt#1}}^H$}%
  \ifdim\ht0=\ht2 #3\else #2\fi
  }
\newcommand*\rel@kern[1]{\kern#1\dimexpr\macc@kerna}
\newcommand*\widebar[1]{\@ifnextchar^{{\wide@bar{#1}{0}}}{\wide@bar{#1}{1}}}
\newcommand*\wide@bar[2]{\if@single{#1}{\wide@bar@{#1}{#2}{1}}{\wide@bar@{#1}{#2}{2}}}
\newcommand*\wide@bar@[3]{%
  \begingroup
  \def\mathaccent##1##2{%
    \let\mathaccent\save@mathaccent
    \if#32 \let\macc@nucleus\first@char \fi
    \setbox\z@\hbox{$\macc@style{\macc@nucleus}_{}$}%
    \setbox\tw@\hbox{$\macc@style{\macc@nucleus}{}_{}$}%
    \dimen@\wd\tw@
    \advance\dimen@-\wd\z@
    \divide\dimen@ 3
    \@tempdima\wd\tw@
    \advance\@tempdima-\scriptspace
    \divide\@tempdima 10
    \advance\dimen@-\@tempdima
    \ifdim\dimen@>\z@ \dimen@0pt\fi
    \rel@kern{0.6}\kern-\dimen@
    \if#31
      \overline{\rel@kern{-0.6}\kern\dimen@\macc@nucleus\rel@kern{0.4}\kern\dimen@}%
      \advance\dimen@0.4\dimexpr\macc@kerna
      \let\final@kern#2%
      \ifdim\dimen@<\z@ \let\final@kern1\fi
      \if\final@kern1 \kern-\dimen@\fi
    \else
      \overline{\rel@kern{-0.6}\kern\dimen@#1}%
    \fi
  }%
  \macc@depth\@ne
  \let\math@bgroup\@empty \let\math@egroup\macc@set@skewchar
  \mathsurround\z@ \frozen@everymath{\mathgroup\macc@group\relax}%
  \macc@set@skewchar\relax
  \let\mathaccentV\macc@nested@a
  \if#31
    \macc@nested@a\relax111{#1}%
  \else
    \def\gobble@till@marker##1\endmarker{}%
    \futurelet\first@char\gobble@till@marker#1\endmarker
    \ifcat\noexpand\first@char A\else
      \def\first@char{}%
    \fi
    \macc@nested@a\relax111{\first@char}%
  \fi
  \endgroup
}
\makeatother

\newcommand{\tr}{{\rm{tr}}}  
\newcommand{\vect}{{\rm{vec}}}  
\newcommand{\cond}{\kappa}  
\newcommand{\Nb}{M}
\newcommand{\Nu}{N}
\newcommand{\Np}{L}
\newcommand{\I}{\mathbf{I}}
\newcommand{\Hh}{\widehat{\mathbf{H}}}
\newcommand{\Ht}{\widetilde{\mathbf{H}}}
\newcommand{\Gh}{\widehat{\mathbf{G}}}
\newcommand{\Gt}{\widetilde{\mathbf{G}}}
\newcommand{\Ab}{\mathbf{B}}
\newcommand{\Au}{\mathbf{U}}
\newcommand{\Rb}{\mathbf{R}^\text{B}}
\newcommand{\Ru}{\mathbf{R}^\text{U}}
\newcommand{\setU}{\mathcal{U}}
\newcommand{\setK}{\mathcal{K}}
\newcommand{\PilotEnergy}{\rho_\tau}
\newcommand{\DataEnergy}{\rho_d}
\newcommand{\NoisePower}{\sigma_z^2}
\newcommand{\CG}{\sigma^2} 
\newcommand{\Tt}{T_\tau} 
\newcommand{\Td}{T_d} 
\newcommand{\Tc}{T_c} 
\newcommand{\sx}{\mathtt{x}} 
\newcommand{\so}{\text{nPuC}} 
\newcommand{\st}{\text{PuC}} 
\newcommand{\sth}{\text{PC}} 
\newcommand{\Pt}{\breve{\mathbf{P}}}
\newcommand{\Wt}{\breve{\mathbf{W}}}
\newcommand{\Rt}{\widetilde{\mathbf{R}}}
\newcommand{\BB}{\text{BB}}
\newcommand{\RF}{\text{RF}}

\newcommand{\Po}{\mathbf{P}^{s_1}}
\newcommand{\Pth}{\mathbf{P}^{s_3}}

\newcommand{\Cu}{\mathbf{C}_{\mathbf{u}}}
\newcommand{\Ctu}{\widetilde{\mathbf{C}}_{\mathbf{u}}}
\newcommand{\Cd}{\mathbf{C}_{\mathbf{d}}}
\newcommand{\Gav}{\widebar{\mathbf{G}}}

\def\REV#1{\textcolor{blue}{#1}} 
\def\REVN#1{\textcolor{blue}{#1}} 
\newcommand{\GF}[1]{\textcolor{red}{#1}} 
\newcommand{\note}[1]{\textcolor{red}{#1}} 

\graphicspath{{Components/Figures/}}
\input{Components/MetaFiles/acronyms.tex} 
\input{Components/MetaFiles/acronyms_Nima.tex} 

%% file: Components/MetaFiles/acronyms_Nima.tex
\begin{acronym}
	\acro{PSD}{positive semi-definite}
	\acro{SVD}{singular value decomposition}
	\acro{EVD}{eigen value decomposition}
\end{acronym}

%% file: Components/Figures/transmitter_structure.tex
%
\begin{tikzpicture}
\newcommand{\drawantTX}[1]{
\draw[thick] (#1)--++(0.35,0)--++(0,0.2)--++(0.15,0.15)--++(-0.3,0)--++(0.15,-0.15);
}
\newcommand{\drawantRX}[1]{
\draw[thick] (#1)--++(-0.4,0)--++(0,0.2)--++(0.15,0.15)--++(-0.3,0)--++(0.15,-0.15);
}

\newcommand{\drawPA}[1]{
\draw[thick] (#1)--++(0.25,0)--++(0,0.2)--++(0.3,-0.2)--++(-0.3,-0.2)--++(0,0.2);
}

\newcommand{\drawdots}[1]{
\draw [fill] ($(#1)+(0,0.1)$) circle  (0.02);
\draw [fill] (#1) circle  (0.02);
\draw [fill] ($(#1)+(0,-0.1)$) circle  (0.02);
}
\tikzstyle{vecArrow} = [thick,
    decoration={markings,
    mark=at position 0 with {\arrowreversed[semithick]{open triangle 60}},
    mark=at position 1 with {\arrow[semithick]{open triangle 60}}},
    double distance=1.4pt,
    shorten >= 5.5pt, shorten <= 5.5pt,
    preaction = {decorate},
    postaction = {draw,line width=1.4pt, white,shorten >= 4.5pt}]

\tikzset{BS/.style={rectangle, draw, thick, minimum width=0.8cm, minimum height=2.5cm, rounded corners=0.5mm}}
\tikzset{PA/.style={rectangle, draw, thick, minimum width=0.5cm, minimum height=0.1cm, rounded corners=0.5mm}}

\node[BS, align=center] (PS) at (0,0){$\mathbf{F}_{\RF}$};
\node[BS, align=center] (BB) at (-1.1,0){$\mathbf{F}_{\BB}$};

\draw[thick] ($(PS)+(0.4,1)$)--++(0.3,0);
\node[PA] (PA1) at (1.0,1) {\footnotesize{PA}};
\draw[thick] ($(PS)+(0.4,-1)$)--++(0.3,0);
\node[PA] (PA2) at (1.0,-1) {\footnotesize{PA}};
\draw[thick,dashed]  (0.6,-1.4) rectangle (1.4,1.4);

\drawantTX{$(PS)+(1.3,1)$}
\drawdots{$(PS)+(1,0)$}
\drawantTX{$(PS)+(1.3,-1)$}

\draw [decorate,decoration={brace,amplitude=5pt,mirror,raise=4pt},yshift=0pt]
(1.7,-1.1) -- (1.7,1.1) node [black,midway,xshift=0.6cm] {$N_{t}$};
\draw [decorate,decoration={brace,amplitude=5pt},xshift=-4pt,yshift=0pt]
(-1.7,-0.7) -- (-1.7,0.7) node [black,midway,xshift=-0.4cm]
{$N_{s}$};

\draw[thick] (-1.5,-0.6)--++(-0.3,0);
\draw[thick] (-1.5,0.6)--++(-0.3,0);
\drawdots{-1.65,0};
\draw[thick] (-0.4,-0.8)--++(-0.3,0);
\draw[thick] (-0.4,0.8)--++(-0.3,0);
\drawdots{-0.55,0};

\node[BS, align=center] (Rx) at (4.6,0){\footnotesize{Receiver}};
\drawantRX{$(PS)+(4,1)$}
\drawdots{$(PS)+(3.65,0.2)$}
\drawantRX{$(PS)+(4,-1)$}
\draw [decorate,decoration={brace,amplitude=5pt,raise=4pt},yshift=0pt]
(3.5,-1) -- (3.5,1) node [black,midway,xshift=-0.6cm] {$N_{r}$};
\draw[thick] (5.5,-0.6)--++(-0.3,0);
\draw[thick] (5.5,0.6)--++(-0.3,0);
\drawdots{5.4,0};
\draw [decorate,decoration={brace,amplitude=5pt,mirror},xshift=-4pt,yshift=0pt]
(5.7,-0.7) -- (5.7,0.7) node [black,midway,xshift=0.5cm]
{$N_{s}$};

\node[] (s) at (-1.75,1.5) {$\mathbf{s}$};
\node[] (u) at (0.4,1.5) {$\mathbf{u}$};
\node[] (x) at (2,1.5) {$\mathbf{x}$};
\node[] (y) at (3.2,1.5) {$\mathbf{y}$};
\node[] (AS) at (1,-1.7) {\footnotesize{Amplification} };
\node[] (AS2) at (1,-2) {\footnotesize{Stage} };
\end{tikzpicture}


%% file: Components/Figures/SEvsNt.tex
%
%
\definecolor{mycolor1}{rgb}{0.00000,0.44700,0.74100}%
\definecolor{mycolor2}{rgb}{0.85000,0.32500,0.09800}%
\definecolor{mycolor3}{rgb}{0.92900,0.69400,0.12500}%
\begin{tikzpicture}

\begin{axis}[%
width=0.85\columnwidth,
height=0.425\columnwidth,
at={(0\textwidth,0\textwidth)},
scale only axis,
xmin=5,
xmax=64,
xlabel={$\text{Number of transmit antennas N}_\text{t}$},
ymin=0,
ymax=8,
ylabel={$SE$ (bit/sec/Hz)},
axis background/.style={fill=white},
legend style={at={(0.97,0.03)},anchor=south east,legend cell align=left,align=left,draw=white!15!black}
]
\addplot [color=mycolor1,solid,very thick]
  table[row sep=crcr]{%
5	0.0875419572100852\\
6	0.226110233333141\\
7	0.464120555628325\\
8	0.796871889071402\\
9	1.19352635763077\\
10	1.61681929490506\\
11	2.03853095017266\\
12	2.44150098792216\\
13	2.81626323836025\\
14	3.15929033160909\\
15	3.47189187950106\\
16	3.75733972298938\\
17	4.01761225166572\\
18	4.25312100582497\\
19	4.46501387758141\\
20	4.65647895409435\\
21	4.83129758181909\\
22	4.99159332167712\\
23	5.1377800623786\\
24	5.27063597454176\\
25	5.3926181424162\\
26	5.50671987675836\\
27	5.6143937470226\\
28	5.71524009866208\\
29	5.80875068774676\\
30	5.89575041179309\\
31	5.97780418347984\\
32	6.05556563845266\\
33	6.12833024591411\\
34	6.19535215263032\\
35	6.25717120429191\\
36	6.31532367325871\\
37	6.37092249668869\\
38	6.42393726929723\\
39	6.47397974155911\\
40	6.52143562666045\\
41	6.56745294441385\\
42	6.61284320569192\\
43	6.65732047676856\\
44	6.70001964007358\\
45	6.74059108900751\\
46	6.77950952034275\\
47	6.81730589795534\\
48	6.85379696045064\\
49	6.88833214066969\\
50	6.92071515435045\\
51	6.95159890414951\\
52	6.98186718814344\\
53	7.01180882500496\\
54	7.04110346871715\\
55	7.06954927725681\\
56	7.09753387807717\\
57	7.12564492237237\\
58	7.15394668917765\\
59	7.18186546502109\\
60	7.20883069154763\\
61	7.23485828282993\\
62	7.26036752072279\\
63	7.28552408978296\\
64	7.30999723240739\\
65	7.33344173367021\\
66	7.35604736916069\\
67	7.37842700009145\\
68	7.40099097011649\\
69	7.42360495324911\\
70	7.44592869746604\\
71	7.46794806033495\\
72	7.4900008849034\\
73	7.51228751845072\\
74	7.53452841094218\\
75	7.55622799798765\\
76	7.57720728633249\\
77	7.59772917426394\\
78	7.61808745239905\\
79	7.63822311762763\\
80	7.6578644009982\\
81	7.67697417240842\\
82	7.69589367147552\\
83	7.71499152126672\\
84	7.73427311602866\\
85	7.75343903276765\\
86	7.77228888072501\\
87	7.79093759540206\\
88	7.80958472971233\\
89	7.82816815317251\\
90	7.84637231497441\\
91	7.8639828329341\\
92	7.88112631586392\\
93	7.89809695923293\\
94	7.91501133022341\\
95	7.93173590088005\\
96	7.94815235502421\\
97	7.96438571098763\\
98	7.98069056541514\\
99	7.99714774241083\\
100	8.01356326691238\\
};
\addlegendentry{Nonlinear System};

\addplot [color=mycolor2,solid,very thick]
  table[row sep=crcr]{%
5	4.39239776468176\\
6	4.6086298300759\\
7	4.79468319545999\\
8	4.93526709666633\\
9	5.04504702595286\\
10	5.15037036700814\\
11	5.26405002367655\\
12	5.37910566186428\\
13	5.48420970227\\
14	5.57810168407107\\
15	5.66813146749288\\
16	5.75923354225618\\
17	5.84801642485534\\
18	5.92790279556079\\
19	5.99703452988852\\
20	6.05943089159207\\
21	6.11961962732727\\
22	6.17800682031037\\
23	6.23212950799475\\
24	6.28116782306101\\
25	6.327706114511\\
26	6.37500820186994\\
27	6.42359770807598\\
28	6.47130937153712\\
29	6.51629013359071\\
30	6.5589731529259\\
31	6.60089022242486\\
32	6.64225872528581\\
33	6.68152706134969\\
34	6.71728289682337\\
35	6.74995103914898\\
36	6.78125329782031\\
37	6.81230560805962\\
38	6.84276307766002\\
39	6.87188728297881\\
40	6.89995166303332\\
41	6.92816601752469\\
42	6.9573082649089\\
43	6.98684325424666\\
44	7.0155925345578\\
45	7.04303730689566\\
46	7.06964657358059\\
47	7.09595344738628\\
48	7.121669014683\\
49	7.14598639928916\\
50	7.16863727275433\\
51	7.19032169170159\\
52	7.21198900469643\\
53	7.23390787637029\\
54	7.25566548197597\\
55	7.27699052263641\\
56	7.29827260627245\\
57	7.32012072683733\\
58	7.34256472670925\\
59	7.36494005206508\\
60	7.3865984658766\\
61	7.40754133616789\\
62	7.4282102792121\\
63	7.44876951257926\\
64	7.46884662234398\\
65	7.4880574193486\\
66	7.50659641824678\\
67	7.52511056722006\\
68	7.54402462324794\\
69	7.56317794728653\\
70	7.58219240601323\\
71	7.60104171534873\\
72	7.62007551225764\\
73	7.63949602988773\\
74	7.65899667282536\\
75	7.67804560744965\\
76	7.69644985093135\\
77	7.7144842995151\\
78	7.73245488165473\\
79	7.75029441604372\\
80	7.76771226630318\\
81	7.78466571937456\\
82	7.80150932987629\\
83	7.81862461573088\\
84	7.83601105336359\\
85	7.85334978381427\\
86	7.87042760158766\\
87	7.88736187576186\\
88	7.90435783639981\\
89	7.92134739592791\\
90	7.93799934319458\\
91	7.95408932992382\\
92	7.96974834266501\\
93	7.98528085974722\\
94	8.00080583102\\
95	8.01618236552648\\
96	8.03128587051945\\
97	8.04624433393276\\
98	8.06131949850001\\
99	8.07659248334443\\
100	8.09186017431133\\
};
\addlegendentry{Linear System};

\addplot [color=mycolor3,dashed,very thick]
  table[row sep=crcr]{%
5	0.0873069613312184\\
6	0.224934453684388\\
7	0.460185064173575\\
8	0.787556391254176\\
9	1.17651715677395\\
10	1.59022189942385\\
11	2.00011696079667\\
12	2.38932029165328\\
13	2.749970314261\\
14	3.07974051274484\\
15	3.37929281536815\\
16	3.65068502772808\\
17	3.89645501640678\\
18	4.11919242111288\\
19	4.32141415604015\\
20	4.50552499853426\\
21	4.67373564329281\\
22	4.82797830481108\\
23	4.9699127565876\\
24	5.10101423498481\\
25	5.22262917917527\\
26	5.33593210361682\\
27	5.44185726497968\\
28	5.54111785418833\\
29	5.63431514333351\\
30	5.72202684337939\\
31	5.80479532961152\\
32	5.88306749080236\\
33	5.95719264650355\\
34	6.02749823139042\\
35	6.09435419177131\\
36	6.1581483526527\\
37	6.21920867919591\\
38	6.27777189267302\\
39	6.33403148349116\\
40	6.38819355516122\\
41	6.44046305194811\\
42	6.4909804950511\\
43	6.53979822120916\\
44	6.58693520751686\\
45	6.63244899374268\\
46	6.67644330825804\\
47	6.71901453555167\\
48	6.76021635604396\\
49	6.80009180697328\\
50	6.83872933607198\\
51	6.87626653060536\\
52	6.91283300504943\\
53	6.94850114072393\\
54	6.98329984154468\\
55	7.01726375988503\\
56	7.05044824818096\\
57	7.08289149448207\\
58	7.11457907676658\\
59	7.14546452580113\\
60	7.17552646233052\\
61	7.20479556936962\\
62	7.23332427434565\\
63	7.26114483223543\\
64	7.28827202458709\\
65	7.31474365385977\\
66	7.34064001394315\\
67	7.36605035026244\\
68	7.39102323655819\\
69	7.41555818295465\\
70	7.43964279826594\\
71	7.46328404885132\\
72	7.48649529868904\\
73	7.50926289812419\\
74	7.53154415045499\\
75	7.55330673667248\\
76	7.57456436354913\\
77	7.59536712562775\\
78	7.61576200859095\\
79	7.63577412975081\\
80	7.65542813601529\\
81	7.67477433446466\\
82	7.69387775289057\\
83	7.71277786031623\\
84	7.73146649870049\\
85	7.74991002102542\\
86	7.76808744215378\\
87	7.78599955241518\\
88	7.80364467544367\\
89	7.82100062369706\\
90	7.8380426210935\\
91	7.85477778269974\\
92	7.87125301278157\\
93	7.88752552013983\\
94	7.90363097747501\\
95	7.91958458341069\\
96	7.93540609382606\\
97	7.95112949340217\\
98	7.96678033185678\\
99	7.98234859297229\\
100	7.99779330766019\\
};
\addlegendentry{Lower Bound};

\end{axis}
\end{tikzpicture}%

%% file: Components/Figures/SEvsP_2.tex
%
%
\definecolor{mycolor1}{rgb}{0.00000,0.44700,0.74100}%
\definecolor{mycolor2}{rgb}{0.85000,0.32500,0.09800}%
\definecolor{mycolor3}{rgb}{0.92900,0.69400,0.12500}%
\definecolor{mycolor4}{rgb}{0.49400,0.18400,0.55600}%
\definecolor{mycolor5}{rgb}{0.46600,0.67400,0.18800}%
\definecolor{mycolor6}{rgb}{0.30100,0.74500,0.93300}%
\definecolor{mycolor7}{rgb}{0.63500,0.07800,0.18400}%
\begin{tikzpicture}

\begin{axis}[%
width=0.85\columnwidth,
height=0.425\columnwidth,
at={(0\textwidth,0\textwidth)},
scale only axis,
xmin=-20,
xmax=15,
xlabel={Input Power $P$ (dBm)},
ymin=0,
ymax=10,
ylabel={$SE$ (bit/sec/Hz)},
axis background/.style={fill=white},
legend style={at={(0.03,0.97)},anchor=north west,legend cell align=left,align=left,draw=white!15!black}
]

\addplot [color=mycolor1,solid, very thick]
  table[row sep=crcr]{%
-20	0.0213854904343761\\
-19.5	0.0239661002358897\\
-19	0.0268542718544832\\
-18.5	0.0300856945966589\\
-18	0.0336999588057383\\
-17.5	0.0377409231438836\\
-17	0.0422571018864142\\
-16.5	0.0473020694843454\\
-16	0.0529348784078153\\
-15.5	0.0592204847940037\\
-15	0.066230174673901\\
-14.5	0.074041981539403\\
-14	0.0827410837479108\\
-13.5	0.0924201677798062\\
-13	0.103179740726485\\
-12.5	0.115128372688678\\
-12	0.128382847141416\\
-11.5	0.143068194950133\\
-11	0.159317585820932\\
-10.5	0.177272049791329\\
-10	0.197080001193319\\
-9.5	0.218896538627058\\
-9	0.242882497119458\\
-8.5	0.269203232985851\\
-8	0.298027128021432\\
-7.5	0.329523807400312\\
-7	0.36386207468508\\
-6.5	0.401207576955972\\
-6	0.441720222137636\\
-5.5	0.485551377440405\\
-5	0.53284087992426\\
-4.5	0.58371388374337\\
-4	0.638277547731763\\
-3.5	0.696617522080391\\
-3	0.758794107851591\\
-2.5	0.824837810154489\\
-2	0.894743735866011\\
-1.5	0.968463812160009\\
-1	1.04589496540221\\
-0.5	1.12685991876031\\
0	1.21107463893255\\
0.5	1.29809181532863\\
1	1.38720165206999\\
1.5	1.47725760872565\\
2	1.56637350952561\\
2.5	1.65141145728061\\
3	1.72716587636977\\
3.5	1.78521634635991\\
4	1.81273835015647\\
4.5	1.79234615064756\\
5	1.70508353743793\\
5.5	1.53840224921541\\
6	1.29713220326214\\
6.5	1.00962297629282\\
7	0.720721309430428\\
7.5	0.472722569846175\\
8	0.287787848862423\\
8.5	0.165054534541603\\
9	0.0905207679946856\\
9.5	0.0480441719842539\\
10	0.0248879154176698\\
10.5	0.0126550519417969\\
11	0.00634090083715973\\
11.5	0.00313972622704394\\
12	0.00154004316580668\\
12.5	0.000750070583368203\\
13	0.000363707188071445\\
13.5	0.000176154949642998\\
14	8.55753483153676e-05\\
14.5	4.19272179962316e-05\\
15	2.08666808816482e-05\\
};
\addlegendentry{$N_t = 4$};

\addplot [color=mycolor1,dashed, very thick, forget plot]
  table[row sep=crcr]{%
-20	0.0213951061860153\\
-19.5	0.0239781777670631\\
-19	0.0268694372667105\\
-18.5	0.0301047315666648\\
-18	0.0337238475865522\\
-17.5	0.0377708889628113\\
-17	0.0422946749058738\\
-16.5	0.0473491589541408\\
-16	0.0529938641854807\\
-15.5	0.0592943300544437\\
-15	0.0663225643815737\\
-14.5	0.0741574921256516\\
-14	0.0828853904309181\\
-13.5	0.0926002970902482\\
-13	0.103404377059421\\
-12.5	0.115408229087315\\
-12	0.128731112020342\\
-11.5	0.143501068066682\\
-11	0.159854918478721\\
-10.5	0.17793810597999\\
-10	0.197904358102346\\
-9.5	0.219915146695724\\
-9	0.244138921495565\\
-8.5	0.270750100001797\\
-8	0.299927802170887\\
-7.5	0.331854326556449\\
-7	0.36671337440223\\
-6.5	0.404688039459072\\
-6	0.445958593441086\\
-5.5	0.490700109359348\\
-5	0.539079976646503\\
-4.5	0.591255372118465\\
-4	0.647370758533329\\
-3.5	0.70755548703559\\
-3	0.771921580549301\\
-2.5	0.840561771914513\\
-2	0.913547863276231\\
-1.5	0.990929462295032\\
-1	1.07273313681757\\
-0.5	1.15896201362589\\
0	1.24959582982544\\
0.5	1.34459142842033\\
1	1.44388367368185\\
1.5	1.54738674790815\\
2	1.65499577974444\\
2.5	1.76658874578307\\
3	1.88202858184196\\
3.5	2.00116543806158\\
4	2.12383901252758\\
4.5	2.24988090115826\\
5	2.37911690666437\\
5.5	2.51136925603349\\
6	2.64645868375053\\
6.5	2.78420634640008\\
7	2.92443554299209\\
7.5	3.0669732239398\\
8	3.21165127977042\\
8.5	3.35830760809128\\
9	3.50678696385244\\
9.5	3.65694160339239\\
10	3.8086317370435\\
10.5	3.96172580819507\\
11	4.11610061870948\\
11.5	4.27164132155809\\
12	4.42824130162321\\
12.5	4.58580196495911\\
13	4.74423245558454\\
13.5	4.90344931725731\\
14	5.06337611580662\\
14.5	5.22394303560367\\
15	5.38508646174123\\
};
\addlegendentry{$N_t = 8$};

\addplot [color=mycolor2,solid, very thick]
  table[row sep=crcr]{%
-20	0.041246191644209\\
-19.5	0.0461763441919849\\
-19	0.0516824323282983\\
-18.5	0.0578284631236062\\
-18	0.0646847726408148\\
-17.5	0.0723284675077002\\
-17	0.0808438535638444\\
-16.5	0.0903228389868638\\
-16	0.100865296801119\\
-15.5	0.112579369086313\\
-15	0.125581692654441\\
-14.5	0.139997523613514\\
-14	0.155960736289229\\
-13.5	0.173613670673866\\
-13	0.193106802186791\\
-12.5	0.214598208347443\\
-12	0.238252809252504\\
-11.5	0.264241362744133\\
-11	0.292739201005943\\
-10.5	0.323924703059581\\
-10	0.357977507134859\\
-9.5	0.395076477847376\\
-9	0.435397455044443\\
-8.5	0.479110823395914\\
-8	0.526378953484766\\
-7.5	0.577353575372978\\
-7	0.632173153440404\\
-6.5	0.690960335835815\\
-6	0.753819552389298\\
-5.5	0.820834830741068\\
-5	0.892067891348056\\
-4.5	0.967556567647026\\
-4	1.04731357762144\\
-3.5	1.13132564658714\\
-3	1.21955294649353\\
-2.5	1.3119287709284\\
-2	1.40835930054178\\
-1.5	1.50872321829379\\
-1	1.61287078549037\\
-0.5	1.72062174877337\\
0	1.83176104660952\\
0.5	1.94603060236444\\
1	2.06311432387153\\
1.5	2.18261142377168\\
2	2.30398973899512\\
2.5	2.42650491201336\\
3	2.54906172134088\\
3.5	2.66997898491803\\
4	2.78659925006504\\
4.5	2.89466592895579\\
5	2.98740236741442\\
5.5	3.05434660879207\\
6	3.08035769889201\\
6.5	3.04587685557122\\
7	2.93006395560959\\
7.5	2.71763270453536\\
8	2.40742392117997\\
8.5	2.01814591823501\\
9	1.58744419377436\\
9.5	1.16368551270346\\
10	0.792476972160353\\
10.5	0.502651347530804\\
11	0.299482212210546\\
11.5	0.169612742927798\\
12	0.0923944751898143\\
12.5	0.0488741994942467\\
13	0.025278904838427\\
13.5	0.0128458213175724\\
14	0.00643526108747341\\
14.5	0.00318645106076764\\
15	0.00156306952483002\\
};
\addlegendentry{$N_t = 16$};

\addplot [color=mycolor2,dashed, very thick, forget plot]
  table[row sep=crcr]{%
-20	0.0412553832348896\\
-19.5	0.0461878654415926\\
-19	0.0516968664724228\\
-18.5	0.0578465365993302\\
-18	0.0647073891743416\\
-17.5	0.0723567499180103\\
-17	0.0808791948861997\\
-16.5	0.090366964804402\\
-16	0.100920340991677\\
-15.5	0.112647965544248\\
-15	0.125667085930997\\
-14.5	0.140103701829541\\
-14	0.156092590103803\\
-13.5	0.173777182532174\\
-13	0.193309270508896\\
-12.5	0.214848511743069\\
-12	0.238561716242072\\
-11.5	0.264621892820342\\
-11	0.293207043174012\\
-10.5	0.324498698244739\\
-10	0.35868020104938\\
-9.5	0.395934751086376\\
-9	0.436443237368027\\
-8.5	0.480381899416909\\
-8	0.527919867408853\\
-7.5	0.579216643165376\\
-7	0.634419592011461\\
-6.5	0.693661520822418\\
-6	0.757058419265517\\
-5.5	0.824707438933684\\
-5	0.896685178716728\\
-4.5	0.97304633463901\\
-4	1.05382275910436\\
-3.5	1.13902295888121\\
-3	1.22863204425721\\
-2.5	1.32261212468471\\
-2	1.42090312997706\\
-1.5	1.52342402161401\\
-1	1.63007434667918\\
-0.5	1.74073607784052\\
0	1.85527567680179\\
0.5	1.97354631577447\\
1	2.09539019153399\\
1.5	2.22064086918712\\
2	2.34912559746231\\
2.5	2.4806675436783\\
3	2.61508790408256\\
3.5	2.75220785353171\\
4	2.89185030709665\\
4.5	3.03384147474022\\
5	3.17801219840721\\
5.5	3.32419906840696\\
6	3.47224532263856\\
6.5	3.62200153784858\\
7	3.77332612662822\\
7.5	3.92608565722224\\
8	4.08015501546682\\
8.5	4.23541742938206\\
9	4.39176437723877\\
9.5	4.54909539945023\\
10	4.70731783356785\\
10.5	4.86634649014891\\
11	5.02610328546622\\
11.5	5.18651684507716\\
12	5.34752209027628\\
12.5	5.50905981750909\\
13	5.67107627898955\\
13.5	5.83352277108486\\
14	5.99635523553273\\
14.5	6.15953387724998\\
15	6.32302280137619\\
};
\addlegendentry{$N_t = 32$};

\addplot [color=mycolor3,solid, very thick]
  table[row sep=crcr]{%
-20	0.0804881441809285\\
-19.5	0.0899306145949064\\
-19	0.100433726195937\\
-18.5	0.112105637797673\\
-18	0.125063078386968\\
-17.5	0.13943149328187\\
-17	0.155345068575294\\
-16.5	0.172946608395369\\
-16	0.192387239077487\\
-15.5	0.213825915085444\\
-15	0.237428703713701\\
-14.5	0.26336782948383\\
-14	0.291820464873891\\
-13.5	0.322967261633617\\
-13	0.356990626334979\\
-12.5	0.394072754704547\\
-12	0.43439345121063\\
-11.5	0.478127772685327\\
-11	0.525443546654823\\
-10.5	0.576498825648409\\
-10	0.631439347166178\\
-9.5	0.690396074396715\\
-9	0.753482894552184\\
-8.5	0.820794549445246\\
-8	0.892404866596118\\
-7.5	0.968365348985894\\
-7	1.04870416813439\\
-6.5	1.13342558930301\\
-6	1.22250984029628\\
-5.5	1.31591341760198\\
-5	1.4135698064432\\
-4.5	1.51539057550298\\
-4	1.62126679312182\\
-3.5	1.73107069977573\\
-3	1.84465756126089\\
-2.5	1.96186761729485\\
-2	2.08252802946808\\
-1.5	2.20645471780649\\
-1	2.33345395208749\\
-0.5	2.4633235251815\\
0	2.59585326918266\\
0.5	2.7308245613657\\
1	2.86800827347366\\
1.5	3.0071602899079\\
2	3.14801316635135\\
2.5	3.29026156652146\\
3	3.4335375460063\\
3.5	3.57736913415458\\
4	3.72111135462239\\
4.5	3.86383191680445\\
5	4.0041233166976\\
5.5	4.13979886916144\\
6	4.26741599682358\\
6.5	4.38157125574417\\
7	4.47396931475847\\
7.5	4.53245971678386\\
8	4.54062634662828\\
8.5	4.47895932596608\\
9	4.32852985053756\\
9.5	4.07673237994244\\
10	3.72266540567559\\
10.5	3.27931156439977\\
11	2.7718555417184\\
11.5	2.2340153665517\\
12	1.70434804481002\\
12.5	1.22202967792026\\
13	0.819843488716826\\
13.5	0.515141684476117\\
14	0.305259875653872\\
14.5	0.172390775635736\\
15	0.093781535861952\\
};
\addlegendentry{$N_t = 64$};

\addplot [color=mycolor3,dashed, very thick, forget plot]
  table[row sep=crcr]{%
-20	0.080496961815838\\
-19.5	0.0899416246057725\\
-19	0.100447461386449\\
-18.5	0.112122755908645\\
-18	0.125084389772537\\
-17.5	0.139457994167063\\
-17	0.155377980846047\\
-16.5	0.172987426983303\\
-16	0.192437788112631\\
-15.5	0.213888414097255\\
-15	0.237505845260315\\
-14.5	0.26346286967806\\
-14	0.291937328348896\\
-13.5	0.323110662549557\\
-13	0.35716620707018\\
-12.5	0.394287243902891\\
-12	0.434654842876264\\
-11.5	0.478445528034771\\
-11	0.52582882046724\\
-10.5	0.576964718912753\\
-10	0.632001187928115\\
-9.5	0.691071728880617\\
-9	0.75429311089897\\
-8.5	0.821763336793385\\
-8	0.893559912772782\\
-7.5	0.969738480806925\\
-7	1.05033185928541\\
-6.5	1.13534952205931\\
-6	1.22477752903765\\
-5.5	1.31857890434573\\
-5	1.41669444169436\\
-4.5	1.51904390198391\\
-4	1.6255275559924\\
-3.5	1.73602801574357\\
-3	1.85041229203439\\
-2.5	1.96853401260593\\
-2	2.09023573535989\\
-1.5	2.2153512935144\\
-1	2.34370811422823\\
-0.5	2.47512945853818\\
0	2.60943653797926\\
0.5	2.74645047154057\\
1	2.88599405523276\\
1.5	3.02789332512791\\
2	3.17197890295005\\
2.5	3.31808712086409\\
3	3.4660609288132\\
3.5	3.61575059342897\\
4	3.76701420209148\\
4.5	3.91971798911721\\
5	4.07373650333262\\
5.5	4.22895263753105\\
6	4.38525754063377\\
6.5	4.5425504329304\\
7	4.7007383437217\\
7.5	4.85973578919099\\
8	5.0194644065409\\
8.5	5.17985255848512\\
9	5.34083492019225\\
9.5	5.5023520588302\\
10	5.66435001402088\\
10.5	5.82677988582941\\
11	5.98959743540798\\
11.5	6.15276270210038\\
12	6.31623963969167\\
12.5	6.47999577354786\\
13	6.64400187962026\\
13.5	6.80823168567093\\
14	6.97266159459028\\
14.5	7.13727042930768\\
15	7.30203919852262\\
};
\addlegendentry{data6};

\addplot [color=mycolor4,solid, very thick]
  table[row sep=crcr]{%
-20	0.155328395910153\\
-19.5	0.172931866769025\\
-19	0.192375525749961\\
-18.5	0.213818629364941\\
-18	0.237427610694951\\
-17.5	0.263375134163384\\
-17	0.291838897708938\\
-16.5	0.323000176608827\\
-16	0.35704211258853\\
-15.5	0.39414776274548\\
-15	0.43449793474279\\
-14.5	0.478268847047889\\
-14	0.525629664906631\\
-13.5	0.57673997337481\\
-13	0.631747257185342\\
-12.5	0.690784462701769\\
-12	0.753967719062532\\
-11.5	0.821394293476526\\
-11	0.893140849421857\\
-10.5	0.969262066493545\\
-10	1.0497896674271\\
-9.5	1.13473188223932\\
-9	1.22407336249628\\
-8.5	1.31777554153119\\
-8	1.415777420042\\
-7.5	1.51799674181488\\
-7	1.62433151205236\\
-6.5	1.73466180139425\\
-6	1.84885177241219\\
-5.5	1.96675186210702\\
-5	2.0882010535202\\
-4.5	2.21302917161406\\
-4	2.34105914260272\\
-3.5	2.47210916137306\\
-3	2.60599471790876\\
-2.5	2.7425304400396\\
-2	2.88153171559997\\
-1.5	3.02281606124366\\
-1	3.16620420648651\\
-0.5	3.31152085830675\\
0	3.45859510127834\\
0.5	3.60726036682767\\
1	3.75735386658831\\
1.5	3.90871531897327\\
2	4.06118468865556\\
2.5	4.21459847879802\\
3	4.36878382136609\\
3.5	4.52354912885852\\
4	4.67866928206488\\
4.5	4.83386204031177\\
5	4.9887502696258\\
5.5	5.14280123606866\\
6	5.29522899552396\\
6.5	5.44483821193615\\
7	5.58977758220121\\
7.5	5.72716105857099\\
8	5.85251514208644\\
8.5	5.9590463084475\\
9	6.03684211472473\\
9.5	6.07237006998573\\
10	6.04896007387682\\
10.5	5.94901495466024\\
11	5.75798171663746\\
11.5	5.46870405987492\\
12	5.08391776263022\\
12.5	4.61559059629301\\
13	4.08191418914194\\
13.5	3.50413403424043\\
14	2.90508003658116\\
14.5	2.30972397366426\\
15	1.74629716538747\\
};
\addlegendentry{$N_t = 32$};

\addplot [color=mycolor4,dashed, very thick, forget plot]
  table[row sep=crcr]{%
-20	0.155336642368922\\
-19.5	0.172942094397344\\
-19	0.192388191698566\\
-18.5	0.213834289895671\\
-18	0.23744694064467\\
-17.5	0.263398949642627\\
-17	0.291868182422339\\
-16.5	0.323036112198852\\
-16	0.35708611341431\\
-15.5	0.394201515515086\\
-15	0.434563443421237\\
-14.5	0.478348483466446\\
-14	0.525726225504394\\
-13.5	0.576856742514376\\
-13	0.631888077504968\\
-12.5	0.690953813000963\\
-12	0.754170800271387\\
-11.5	0.821637123335862\\
-11	0.893430366605667\\
-10.5	0.969606245041616\\
-10	1.05019764252754\\
-9.5	1.13521408861212\\
-9	1.22464168688387\\
-8.5	1.31844349110571\\
-8	1.41656030889854\\
-7.5	1.5189118981461\\
-7	1.62539850910744\\
-6.5	1.73590271593782\\
-6	1.85029147514981\\
-5.5	1.96841834548552\\
-5	2.09012580351952\\
-4.5	2.21524759174006\\
-4	2.34361104044454\\
-3.5	2.47503931107603\\
-3	2.60935351615213\\
-2.5	2.74637467924398\\
-2	2.88592550712651\\
-1.5	3.02783195485807\\
-1	3.17192457281865\\
-0.5	3.31803963235689\\
0	3.46602003343973\\
0.5	3.61571600340311\\
1	3.76698560047288\\
1.5	3.91969503913213\\
2	4.07371885668793\\
2.5	4.22893994162078\\
3	4.38524944460945\\
3.5	4.54254659266266\\
4	4.70073842572181\\
4.5	4.85973947358777\\
5	5.01947138922399\\
5.5	5.17986255253097\\
6	5.3408476566878\\
6.5	5.50236728720317\\
7	5.66436750197546\\
7.5	5.82679941897564\\
8	5.98961881666054\\
8.5	6.15278575091159\\
9	6.3162641911718\\
9.5	6.48002167751662\\
10	6.64402899962398\\
10.5	6.80825989799262\\
11	6.97269078727318\\
11.5	7.13730050120776\\
12	7.30207005840067\\
12.5	7.46698244795121\\
13	7.63202243385204\\
13.5	7.7971763769823\\
14	7.96243207349115\\
14.5	8.1277786083657\\
15	8.2932062229995\\
};

\addplot [color=mycolor5,solid, very thick]
  table[row sep=crcr]{%
-20	0.292539569394711\\
-19.5	0.32376801928999\\
-19	0.35788202723281\\
-18.5	0.395064783151602\\
-18	0.435497209072824\\
-17.5	0.479355621317082\\
-17	0.526809270179961\\
-16.5	0.578017818607971\\
-16	0.633128829775662\\
-15.5	0.692275338871333\\
-15	0.755573586176749\\
-14.5	0.823120986307988\\
-14	0.89499440220976\\
-13.5	0.971248782438068\\
-13	1.05191620701667\\
-12.5	1.13700537156225\\
-12	1.22650152246602\\
-11.5	1.32036683878747\\
-11	1.41854124020989\\
-10.5	1.52094358584172\\
-10	1.62747321652871\\
-9.5	1.73801178413079\\
-9	1.85242530512372\\
-8.5	1.97056637289251\\
-8	2.09227646298627\\
-7.5	2.21738826806608\\
-7	2.345728003877\\
-6.5	2.47711763384825\\
-6	2.61137696740323\\
-5.5	2.74832559528399\\
-5	2.88778463373369\\
-4.5	3.02957825784064\\
-4	3.17353501237158\\
-3.5	3.3194888956991\\
-3	3.46728021868879\\
-2.5	3.61675624542765\\
-2	3.76777162625121\\
-1.5	3.92018863547724\\
-1	4.07387722637043\\
-0.5	4.22871491385743\\
0	4.38458649092943\\
0.5	4.54138357675474\\
1	4.69900398201504\\
1.5	4.85735085778943\\
2	5.01633156501976\\
2.5	5.17585615663665\\
3	5.33583529479547\\
3.5	5.496177316789\\
4	5.65678399151956\\
4.5	5.81754423593968\\
5	5.97832462655765\\
5.5	6.13895484766849\\
6	6.29920511291895\\
6.5	6.45875084853036\\
7	6.61711720407398\\
7.5	6.77359184594585\\
8	6.92708863965265\\
8.5	7.07593747466419\\
9	7.21756886746092\\
9.5	7.348063497783\\
10	7.46156434460622\\
10.5	7.54963446076906\\
11	7.60081769358912\\
11.5	7.60088862691985\\
12	7.53435744785882\\
12.5	7.38738255228509\\
13	7.15127788612151\\
13.5	6.82498358906891\\
14	6.41517001528545\\
14.5	5.93402755251329\\
15	5.39609019792408\\
};
\addlegendentry{$N_t = 64$};

\addplot [color=mycolor5,dashed, very thick, forget plot]
  table[row sep=crcr]{%
-20	0.292546923098712\\
-19.5	0.323777042639473\\
-19	0.357893075154924\\
-18.5	0.395078278882287\\
-18	0.43551365541199\\
-17.5	0.479375613311007\\
-17	0.526833509338457\\
-16.5	0.578047128766477\\
-16	0.633164174717356\\
-15.5	0.692317841832747\\
-15	0.755624551373236\\
-14.5	0.823181922633891\\
-14	0.895067049293144\\
-13.5	0.971335139262296\\
-13	1.05201856336153\\
-12.5	1.13712634256636\\
-12	1.2266440866702\\
-11.5	1.32053438008621\\
-11	1.41873759421522\\
-10.5	1.52117309125156\\
-10	1.62774077219051\\
-9.5	1.7383229126029\\
-9	1.85278622366253\\
-8.5	1.97098407293538\\
-8	2.09275879936282\\
-7.5	2.21794405935748\\
-7	2.34636714556196\\
-6.5	2.47785122613581\\
-6	2.61221745996978\\
-5.5	2.74928695152721\\
-5	2.888882517659\\
-4.5	3.03083024734977\\
-4	3.17496084359353\\
-3.5	3.3211107441838\\
-3	3.46912302491258\\
-2.5	3.6188480943442\\
-2	3.77014419387058\\
-1.5	3.92287772013529\\
-1	4.07692338916944\\
-0.5	4.23216426279552\\
0	4.38849165815311\\
0.5	4.5458049607313\\
1	4.70401136022166\\
1.5	4.86302552699338\\
2	5.02276924519322\\
2.5	5.18317101651863\\
3	5.34416564671771\\
3.5	5.50569382492091\\
4	5.66770170407199\\
4.5	5.83014048904463\\
5	5.99296603752984\\
5.5	6.15613847747061\\
6	6.31962184370273\\
6.5	6.48338373552532\\
7	6.64739499615867\\
7.5	6.81162941443184\\
8	6.97606344856043\\
8.5	7.14067597150727\\
9	7.30544803714756\\
9.5	7.47036266626822\\
10	7.63540465130548\\
10.5	7.80056037865023\\
11	7.96581766731787\\
11.5	8.13116562277798\\
12	8.29659450476127\\
12.5	8.46209560790084\\
13	8.62766115411679\\
13.5	8.79328419571234\\
14	8.95895852821415\\
14.5	9.12467861205597\\
15	9.29043950227126\\
};

\draw [color=mycolor1,solid,very thick] (-7,9) -- (-4.5,9);
\node[color=black] at (3,9) {Nonlinear System};
\draw [color=mycolor1,dotted,very thick] (-7,8) -- (-4.5,8);
\node[color=black] at (2,8) {Linear System};

\end{axis}
\end{tikzpicture}%

%% file: Components/Figures/EEvsP_Nt.tex
%
%
\definecolor{mycolor1}{rgb}{0.00000,0.44700,0.74100}%
\definecolor{mycolor2}{rgb}{0.85000,0.32500,0.09800}%
\definecolor{mycolor3}{rgb}{0.92900,0.69400,0.12500}%
\definecolor{mycolor4}{rgb}{0.49400,0.18400,0.55600}%
\definecolor{mycolor5}{rgb}{0.46600,0.67400,0.18800}%
\begin{tikzpicture}

\begin{axis}[%
width=0.85\columnwidth,
height=0.425\columnwidth,
at={(0\textwidth,0\textwidth)},
scale only axis,
xmin=-20,
xmax=15,
xlabel={Input Power $P$ (dBm)},
ymin=0,
ymax=400,
ylabel={$EE$ (Gbit/Joul)},
axis background/.style={fill=white},
legend style={legend cell align=left,align=left,draw=white!15!black}
]
\addplot [color=mycolor1,solid,very thick]
  table[row sep=crcr]{%
-20	228.00734302165\\
-19.5	237.112462818473\\
-19	246.235117338323\\
-18.5	255.335346256725\\
-18	264.371061197652\\
-17.5	273.298486782908\\
-17	282.072658731694\\
-16.5	290.647968368253\\
-16	298.978740841402\\
-15.5	307.019832808372\\
-15	314.727234380392\\
-14.5	322.058659823379\\
-14	328.97411186507\\
-13.5	335.436405449124\\
-13	341.411638328215\\
-12.5	346.869597901792\\
-12	351.784096056773\\
-11.5	356.133226322143\\
-11	359.899540254961\\
-10.5	363.070142487212\\
-10	365.636706134611\\
-9.5	367.595412158991\\
-9	368.946817647852\\
-8.5	369.695658688317\\
-8	369.850593411599\\
-7.5	369.423889669233\\
-7	368.43105938642\\
-6.5	366.890437471788\\
-6	364.822696516922\\
-5.5	362.250278197077\\
-5	359.196706335925\\
-4.5	355.685721880346\\
-4	351.740141580407\\
-3.5	347.380282367214\\
-3	342.621701186473\\
-2.5	337.471860925313\\
-2	331.925132990913\\
-1.5	325.955287144422\\
-1	319.504358320839\\
-0.5	312.466736216689\\
0	304.668057123385\\
0.5	295.841068126996\\
1	285.606362225002\\
1.5	273.47442407703\\
2	258.891447874107\\
2.5	241.343788216401\\
3	220.507502787169\\
3.5	196.394513620138\\
4	169.440149973939\\
4.5	140.515381214726\\
5	110.895505967601\\
5.5	82.2132863869097\\
6	56.3478974007207\\
6.5	35.1083390409847\\
7	19.633371203732\\
7.5	9.82736574020315\\
8	4.44496456954238\\
8.5	1.8495195006661\\
9	0.722000517876446\\
9.5	0.268873820062175\\
10	0.0966998669540041\\
10.5	0.0338722823891386\\
11	0.0116232488461104\\
11.5	0.00392375571308267\\
12	0.00130746836053558\\
12.5	0.000431370696511293\\
13	0.000141366786495996\\
13.5	4.61868292127046e-05\\
14	1.51121988614413e-05\\
14.5	4.98055358037257e-06\\
15	1.66564097102199e-06\\
};
\addlegendentry{$N_t=4$};

\addplot [color=mycolor2,solid,very thick]
  table[row sep=crcr]{%
-20	282.3028585715\\
-19.5	290.878395019255\\
-19	299.208527572947\\
-18.5	307.248042297008\\
-18	314.952851095895\\
-17.5	322.280580121385\\
-17	329.191131567993\\
-16.5	335.647204775631\\
-16	341.6147641348\\
-15.5	347.063443333693\\
-15	351.966877879458\\
-14.5	356.302960435091\\
-14	360.054016202442\\
-13.5	363.206898216866\\
-13	365.753004876953\\
-12.5	367.688224206491\\
-12	369.012811150582\\
-11.5	369.731205584921\\
-11	369.851799637374\\
-10.5	369.386663384315\\
-10	368.351238018652\\
-9.5	366.764005242548\\
-9	364.646140980479\\
-8.5	362.02116060786\\
-8	358.914561810841\\
-7.5	355.353469981226\\
-7	351.366289725187\\
-6.5	346.982364604378\\
-6	342.231645558263\\
-5.5	337.144366430299\\
-5	331.750722390599\\
-4.5	326.080543420807\\
-4	320.162949791374\\
-3.5	314.025968674835\\
-3	307.696079259745\\
-2.5	301.19763578724\\
-2	294.552090652193\\
-1.5	287.776898733693\\
-1	280.883924229808\\
-0.5	273.877088422256\\
0	266.748894700733\\
0.5	259.475374530041\\
1	252.009006155706\\
1.5	244.269484178406\\
2	236.13325973839\\
2.5	227.424985149114\\
3	217.917286234743\\
3.5	207.347826423876\\
4	195.460393860532\\
4.5	182.066328488571\\
5	167.107408678968\\
5.5	150.693607952252\\
6	133.099533288923\\
6.5	114.726456059077\\
7	96.0538378193129\\
7.5	77.6048564274634\\
8	59.9423155350501\\
8.5	43.6947414237741\\
9	29.5693629209223\\
9.5	18.2529421679008\\
10	10.1396736467924\\
10.5	5.05189272000282\\
11	2.27807110010461\\
11.5	0.946214303571145\\
12	0.369064249089146\\
12.5	0.137404576477509\\
13	0.0494200644950599\\
13.5	0.0173144100125922\\
14	0.00594286521638202\\
14.5	0.00200665770427356\\
15	0.000668793695552243\\
};
\addlegendentry{$N_t=8$};

\addplot [color=mycolor3,solid,very thick]
  table[row sep=crcr]{%
-20	329.348601649503\\
-19.5	335.795789834965\\
-19	341.753593500145\\
-18.5	347.191664276724\\
-18	352.083653361745\\
-17.5	356.407464732287\\
-17	360.145428394616\\
-16.5	363.284393592323\\
-16	365.815744354825\\
-15.5	367.735341940454\\
-15	369.043400533574\\
-14.5	369.744303935747\\
-14	369.846371919595\\
-13.5	369.361585393859\\
-13	368.305279591084\\
-12.5	366.695814191365\\
-12	364.554228710029\\
-11.5	361.903890686199\\
-11	358.770143295289\\
-10.5	355.179958046118\\
-10	351.161597273602\\
-9.5	346.744290244573\\
-9	341.957925883261\\
-8.5	336.832764403815\\
-8	331.399169505697\\
-7.5	325.687362229629\\
-7	319.727197065396\\
-6.5	313.54796042202\\
-6	307.178191084944\\
-5.5	300.645521758144\\
-5	293.976540176844\\
-4.5	287.196667519351\\
-4	280.330050859915\\
-3.5	273.399465063418\\
-3	266.426217637988\\
-2.5	259.430047344868\\
-2	252.42900337793\\
-1.5	245.439286010523\\
-1	238.475020802408\\
-0.5	231.547925417866\\
0	224.666809104943\\
0.5	217.836818142483\\
1	211.058305410286\\
1.5	204.32516238896\\
2	197.622423092228\\
2.5	190.922974635507\\
3	184.183381263841\\
3.5	177.33930359189\\
4	170.301933097497\\
4.5	162.958192765246\\
5	155.178461310271\\
5.5	146.834739125877\\
6	137.828119160844\\
6.5	128.118116632114\\
7	117.74223770598\\
7.5	106.816789381955\\
8	95.5187610523522\\
8.5	84.0569906733618\\
9	72.642942972605\\
9.5	61.4691735065794\\
10	50.7019682315337\\
10.5	40.4928672666962\\
11	31.00788997864\\
11.5	22.4627666351986\\
12	15.1365982263001\\
12.5	9.31922234028457\\
13	5.17012046810496\\
13.5	2.57498521614765\\
14	1.16140398296984\\
14.5	0.482633141081189\\
15	0.188352452412745\\
};
\addlegendentry{$N_t=16$};

\addplot [color=mycolor4,solid,very thick]
  table[row sep=crcr]{%
-20	360.221243349365\\
-19.5	363.347380900412\\
-19	365.865593640751\\
-18.5	367.771803552502\\
-18	369.066279335877\\
-17.5	369.753451817964\\
-17	369.841679207811\\
-16.5	369.34297135385\\
-16	368.272682215729\\
-15.5	366.64917945946\\
-15	364.493499496138\\
-14.5	361.828995492953\\
-14	358.680984973388\\
-13.5	355.076402666317\\
-13	351.043463322529\\
-12.5	346.611338336592\\
-12	341.809849219135\\
-11.5	336.669180270907\\
-11	331.219612214283\\
-10.5	325.491278030551\\
-10	319.513941817658\\
-9.5	313.316801107462\\
-9	306.928312748881\\
-8.5	300.376042161377\\
-8	293.686535482987\\
-7.5	286.885213872857\\
-7	279.996288977442\\
-6.5	273.042698331792\\
-6	266.046059243663\\
-5.5	259.026639499403\\
-5	252.003343035714\\
-4.5	244.993708535966\\
-4	238.013918722621\\
-3.5	231.078817908922\\
-3	224.201935110203\\
-2.5	217.395509647807\\
-2	210.670515631206\\
-1.5	204.036680866568\\
-1	197.502494453312\\
-0.5	191.075195365106\\
0	184.76073134487\\
0.5	178.563673030656\\
1	172.487061800532\\
1.5	166.532160745178\\
2	160.698066004809\\
2.5	154.981120918173\\
3	149.374061156129\\
3.5	143.864814658441\\
4	138.434907741375\\
4.5	133.057530311108\\
5	127.695550443377\\
5.5	122.300189506019\\
6	116.81160857904\\
6.5	111.162994653087\\
7	105.289279821548\\
7.5	99.1399027573616\\
8	92.6924380065607\\
8.5	85.9621305271481\\
9	79.0031953846182\\
9.5	71.9011809231401\\
10	64.7593960407037\\
10.5	57.6837879620945\\
11	50.7697715610283\\
11.5	44.0934379649331\\
12	37.708905541424\\
12.5	31.6522213209238\\
13	25.9504769812975\\
13.5	20.6344667869726\\
14	15.7539252713066\\
14.5	11.3925332924279\\
15	7.67212037193415\\
};
\addlegendentry{$N_t=32$};

\addplot [color=mycolor5,solid,very thick]
  table[row sep=crcr]{%
-20	369.837118878802\\
-19.5	369.32597087883\\
-19	368.24348702185\\
-18.5	366.608080513792\\
-18	364.44082504179\\
-17.5	361.7651026572\\
-17	358.606251372568\\
-16.5	354.991218112904\\
-16	350.948221721516\\
-15.5	346.50642984194\\
-15	341.695652707164\\
-14.5	336.546056176452\\
-14	331.087895767469\\
-13.5	325.351272927307\\
-13	319.365914356012\\
-12.5	313.16097482475\\
-12	306.764863603236\\
-11.5	300.205094315747\\
-11	293.508157773656\\
-10.5	286.699417080376\\
-10	279.803024070629\\
-9.5	272.841855931967\\
-9	265.837470666254\\
-8.5	258.810079887551\\
-8	251.778537326033\\
-7.5	244.760341319667\\
-7	237.771649529239\\
-6.5	230.827304108374\\
-6	223.940865595688\\
-5.5	217.124653865822\\
-5	210.38979457138\\
-4.5	203.7462696187\\
-4	197.202970334968\\
-3.5	190.767752089818\\
-3	184.447489218003\\
-2.5	178.248129136331\\
-2	172.174744541882\\
-1.5	166.23158249893\\
-1	160.422109042493\\
-0.5	154.749047608879\\
0	149.214409094033\\
0.5	143.81951056089\\
1	138.564978457493\\
1.5	133.450730518536\\
2	128.475928112861\\
2.5	123.638887453946\\
3	118.936933645688\\
3.5	114.366176076876\\
4	109.921178006543\\
4.5	105.594489785221\\
5	101.37601994849\\
5.5	97.2522431944002\\
6	93.2053083007011\\
6.5	89.2122346889719\\
7	85.2445787457148\\
7.5	81.2691560289098\\
8	77.2504682652146\\
8.5	73.1551789101762\\
9	68.9581864244283\\
9.5	64.6488040918928\\
10	60.2349299925778\\
10.5	55.7435132711484\\
11	51.2170355328839\\
11.5	46.7072401402089\\
12	42.2679690037609\\
12.5	37.9486374842716\\
13	33.7893778222514\\
13.5	29.818581180573\\
14	26.0530557061201\\
14.5	22.5001877001086\\
15	19.1610438292215\\
};
\addlegendentry{$N_t=64$};

\end{axis}
\end{tikzpicture}%

%% file: Components/Figures/EEvsSE_2.tex
%
%
\definecolor{mycolor1}{rgb}{0.00000,0.44700,0.74100}%
\definecolor{mycolor2}{rgb}{0.85000,0.32500,0.09800}%
\definecolor{mycolor3}{rgb}{0.92900,0.69400,0.12500}%
\definecolor{mycolor4}{rgb}{0.49400,0.18400,0.55600}%
\definecolor{mycolor5}{rgb}{0.46600,0.67400,0.18800}%
\begin{tikzpicture}

\begin{axis}[%
width=0.85\columnwidth,
height=0.425\columnwidth,
at={(0\textwidth,0\textwidth)},
scale only axis,
xmin=0,
xmax=11,
xlabel={$SE$ (bit/sec/Hz)},
ymin=0,
ymax=400,
ylabel={$EE$ (Gbit/Joul)},
axis background/.style={fill=white},
legend style={legend cell align=left,align=left,draw=white!15!black}
]
\addplot [color=mycolor1,solid, very thick, directed]
  table[row sep=crcr]{%
0.300117514661999	228.00734302165\\
0.330586311822301	237.112462818473\\
0.36363609774528	246.235117338323\\
0.399404361546099	255.335346256725\\
0.438024900441217	264.371061197652\\
0.479626380553169	273.298486782908\\
0.524330901801264	282.072658731694\\
0.572252598650004	290.647968368253\\
0.623496308352198	298.978740841402\\
0.678156336878145	307.019832808372\\
0.736315350009174	314.727234380392\\
0.798043413210377	322.058659823379\\
0.86339719906492	328.97411186507\\
0.932419375482441	335.436405449124\\
1.0051381818511	341.411638328214\\
1.0815671940644	346.869597901792\\
1.16170527318763	351.784096056773\\
1.245536686665	356.133226322143\\
1.33303138556814	359.899540254961\\
1.42414541649848	363.070142487212\\
1.51882144226621	365.636706134612\\
1.61698934102649	367.595412158991\\
1.71856684847589	368.946817647852\\
1.82346020082574	369.695658688317\\
1.93156472567405	369.850593411599\\
2.04276531055733	369.423889669233\\
2.15693665002317	368.43105938642\\
2.27394312367562	366.890437471788\\
2.39363807696635	364.822696516922\\
2.51586214231153	362.250278197077\\
2.64044001509189	359.196706335925\\
2.76717472837054	355.685721880346\\
2.89583785397303	351.740141580407\\
3.02615303546746	347.380282367214\\
3.15776857383445	342.621701186473\\
3.29021204874041	337.471860925313\\
3.4228156338812	331.925132990913\\
3.55459430526539	325.955287144422\\
3.68405054607341	319.504358320839\\
3.80887063216717	312.466736216689\\
3.92547733379164	304.668057123385\\
4.02843188334851	295.841068126996\\
4.10976759405123	285.606362225002\\
4.15851233567247	273.47442407703\\
4.16086444222412	258.891447874107\\
4.10153121872362	241.343788216401\\
3.96637718779155	220.507502787169\\
3.74580861829838	196.394513620138\\
3.43779610211906	169.440149973939\\
3.04968698881216	140.515381214726\\
2.59875174461146	110.895505967601\\
2.11175837372134	82.2132863869097\\
1.6232793780395	56.3478974007206\\
1.17147308440793	35.1083390409847\\
0.790076322752411	19.633371203732\\
0.498428379776625	9.82736574020315\\
0.296166301128718	4.44496456954239\\
0.167520361230074	1.8495195006661\\
0.0911937427963263	0.722000517876445\\
0.0482180514511669	0.268873820062175\\
0.0249310520688659	0.0966998669540041\\
0.012665424258821	0.0338722823891386\\
0.0063433338667935	0.0116232488461104\\
0.00314028564960919	0.00392375571308268\\
0.00154016974063434	0.00130746836053558\\
0.000750098865511619	0.000431370696511292\\
0.000363713451390842	0.000141366786495996\\
0.000176156329955265	4.61868292127046e-05\\
8.55756524968399e-05	1.51121988614414e-05\\
4.19272854297092e-05	4.98055358037256e-06\\
2.08666960359305e-05	1.66564097102199e-06\\
};
\addlegendentry{$N_t = 4$};

\addplot [color=mycolor2,solid, very thick, directed]
  table[row sep=crcr]{%
0.525559688589535	282.302858571501\\
0.573604070339327	290.878395019255\\
0.624982678730627	299.208527572947\\
0.67979125060907	307.248042297008\\
0.73811409588792	314.952851095895\\
0.80002316940655	322.280580121386\\
0.865577314183064	329.191131567993\\
0.934821689775048	335.647204775631\\
1.00778739359215	341.6147641348\\
1.08449127701671	347.063443333693\\
1.16493595238978	351.966877879458\\
1.24910998159975	356.302960435091\\
1.33698823242504	360.054016202442\\
1.42853238513151	363.206898216867\\
1.52369156923579	365.753004876953\\
1.62240310885721	367.688224206491\\
1.72459335463557	369.012811150582\\
1.83017858063664	369.731205584921\\
1.93906592576609	369.851799637374\\
2.05115436066311	369.386663384314\\
2.16633566250621	368.351238018652\\
2.28449538127621	366.764005242547\\
2.40551378141687	364.646140980479\\
2.52926674212837	362.021160607859\\
2.65562659727462	358.914561810841\\
2.78446289148619	355.353469981226\\
2.91564302159054	351.366289725187\\
3.04903272053772	346.982364604378\\
3.18449632208986	342.231645558264\\
3.32189671468732	337.144366430299\\
3.461094845447	331.750722390598\\
3.60194855924611	326.080543420807\\
3.74431043533028	320.162949791373\\
3.88802408543206	314.025968674835\\
4.03291805499618	307.696079259745\\
4.17879594510031	301.19763578724\\
4.32542052324032	294.552090652193\\
4.47248822478887	287.776898733693\\
4.6195882838394	280.883924229808\\
4.76613741043855	273.877088422256\\
4.91127611706455	266.748894700733\\
5.05370659421292	259.475374530041\\
5.19144609629265	252.009006155706\\
5.3214697013334	244.269484178406\\
5.43923504817581	236.13325973839\\
5.53814056259081	227.424985149114\\
5.60908457660824	217.917286234743\\
5.64043895670318	207.347826423876\\
5.61881217296907	195.460393860532\\
5.53078374633693	182.066328488571\\
5.36530965416215	167.107408678968\\
5.11601776813334	150.693607952252\\
4.78258662819709	133.099533288923\\
4.37093311270881	114.726456059077\\
3.89256081171396	96.0538378193129\\
3.36367015816845	77.6048564274634\\
2.80460845501752	59.9423155350501\\
2.24002681691808	43.6947414237741\\
1.69915981674848	29.5693629209223\\
1.2139570794154	18.2529421679008\\
0.812568729346893	10.1396736467924\\
0.509788103325548	5.05189272000282\\
0.301759463564589	2.27807110010461\\
0.170277424515593	0.946214303571145\\
0.0925754627099837	0.369064249089146\\
0.0489209528344681	0.137404576477508\\
0.0252905078566298	0.0494200644950598\\
0.0128486125589151	0.0173144100125922\\
0.00643591610019605	0.00594286521638202\\
0.00318660172010007	0.00200665770427356\\
0.00156310362280516	0.000668793695552243\\
};
\addlegendentry{$N_t = 8$};

\addplot [color=mycolor3,solid, very thick, directed]
  table[row sep=crcr]{%
0.867165971796202	329.348601649503\\
0.936523424194295	335.795789834965\\
1.00960877065221	341.753593500145\\
1.08643945816953	347.191664276724\\
1.16701881154592	352.083653361745\\
1.25133623926318	356.407464732287\\
1.33936760854923	360.145428394616\\
1.43107577235041	363.284393592323\\
1.5264112284428	365.815744354825\\
1.62531288954578	367.735341940454\\
1.72770894301401	369.043400533573\\
1.83351777934342	369.744303935747\\
1.94264897012134	369.846371919595\\
2.05500427792151	369.361585393859\\
2.17047868271481	368.305279591084\\
2.28896141137674	366.695814191365\\
2.41033695861438	364.554228710029\\
2.53448608896861	361.903890686199\\
2.66128681041511	358.770143295289\\
2.79061531050682	355.179958046118\\
2.92234684604663	351.161597273602\\
3.05635657705964	346.744290244573\\
3.19252033545506	341.957925883261\\
3.33071531830181	336.832764403814\\
3.47082069509537	331.399169505697\\
3.61271811766959	325.687362229629\\
3.7562921202722	319.727197065396\\
3.90143039534589	313.54796042202\\
4.04802392703741	307.178191084944\\
4.19596695829393	300.645521758144\\
4.34515675689783	293.976540176844\\
4.4954931283054	287.196667519351\\
4.64687759459836	280.330050859915\\
4.7992121128101	273.399465063418\\
4.95239713221298	266.426217637988\\
5.10632867275573	259.430047344868\\
5.26089392011613	252.42900337793\\
5.4159645361551	245.439286010523\\
5.57138641301609	238.475020802408\\
5.72696385608342	231.547925417866\\
5.88243501875438	224.666809104944\\
6.03743362715672	217.836818142483\\
6.19142938460859	211.058305410286\\
6.34363576206685	204.32516238896\\
6.4928693931313	197.622423092228\\
6.63734145267854	190.922974635507\\
6.77436247884622	184.183381263841\\
6.89995729836717	177.33930359189\\
7.00843023214276	170.301933097497\\
7.09200307268764	162.958192765246\\
7.14075191268266	155.17846131027\\
7.14312011921478	146.834739125877\\
7.08717024951437	137.828119160844\\
6.96241326237813	128.118116632115\\
6.76167649029022	117.74223770598\\
6.48234823128743	106.816789381955\\
6.1266218109625	95.5187610523522\\
5.70083586252002	84.0569906733618\\
5.21429665216002	72.642942972605\\
4.67802867641973	61.4691735065794\\
4.103962599786	50.7019682315337\\
3.50507895642869	40.4928672666962\\
2.89668669588898	31.00788997864\\
2.29833476056281	22.4627666351986\\
1.73501472759642	15.1365982263001\\
1.23539558124621	9.31922234028456\\
0.825100733888892	5.17012046810496\\
0.516996961890441	2.57498521614765\\
0.30585076513807	1.16140398296984\\
0.172563242905597	0.482633141081189\\
0.0938285238706141	0.188352452412745\\
};
\addlegendentry{$N_t = 16$};

\addplot [color=mycolor4,solid, very thick, directed]
  table[row sep=crcr]{%
1.34135286265565	360.221243349365\\
1.4331585881835	363.347380900412\\
1.52859382191488	365.865593640751\\
1.6275976996794	367.771803552502\\
1.73009870619368	369.066279335877\\
1.83601561147342	369.753451817964\\
1.94525845718889	369.841679207811\\
2.05772957565237	369.34297135385\\
2.17332462623251	368.272682215729\\
2.29193363604358	366.64917945946\\
2.41344203355323	364.493499496138\\
2.53773166515958	361.828995492953\\
2.66468178576031	358.680984973388\\
2.79417001490524	355.076402666317\\
2.92607325038811	351.043463322529\\
3.06026853123043	346.611338336592\\
3.19663384209612	341.809849219135\\
3.33504885139295	336.669180270906\\
3.47539557578274	331.219612214283\\
3.61755896460768	325.491278030551\\
3.76142739887044	319.513941817658\\
3.90689310085225	313.316801107462\\
4.05385245214146	306.928312748881\\
4.20220621965422	300.376042161377\\
4.35185969099806	293.686535482987\\
4.50272272206195	286.885213872857\\
4.65470970078278	279.996288977442\\
4.80773943137362	273.042698331792\\
4.96173494259717	266.046059243664\\
5.11662322155807	259.026639499403\\
5.27233487049095	252.003343035715\\
5.42880367747186	244.993708535966\\
5.58596608189304	238.013918722621\\
5.74376050039382	231.078817908922\\
5.90212645634406	224.201935110203\\
6.06100342214425	217.395509647807\\
6.22032923254761	210.670515631206\\
6.38003784946686	204.036680866568\\
6.54005613944899	197.502494453312\\
6.7002991409895	191.075195365106\\
6.86066301423759	184.76073134487\\
7.02101442578592	178.563673030656\\
7.18117444523192	172.487061800532\\
7.34089400410216	166.532160745178\\
7.49981644644795	160.698066004809\\
7.65742053952732	154.981120918173\\
7.81293448130062	149.374061156129\\
7.9652083173838	143.864814658441\\
8.11253019384767	138.434907741376\\
8.25237460394558	133.057530311108\\
8.38108510478509	127.695550443377\\
8.4935292416214	122.300189506019\\
8.58282456729793	116.81160857904\\
8.64030638486789	111.162994653087\\
8.65593909797633	105.289279821548\\
8.61929034860891	99.1399027573616\\
8.52096177542768	92.6924380065607\\
8.35409896767055	85.9621305271481\\
8.11548343898627	79.0031953846181\\
7.80586602092612	71.9011809231401\\
7.42952842334507	64.7593960407037\\
6.99330437201008	57.6837879620945\\
6.50535475398941	50.7697715610282\\
5.97401507930193	44.0934379649331\\
5.40707684419377	37.708905541424\\
4.81177366584924	31.6522213209238\\
4.19549101167533	25.9504769812975\\
3.56704800245758	20.6344667869726\\
2.93840302790762	15.7539252713066\\
2.32648618069665	11.3925332924279\\
1.75419872253181	7.67212037193415\\
};
\addlegendentry{$N_t = 32$};

\addplot [color=mycolor5,solid, very thick, directed]
  table[row sep=crcr]{%
1.94762514860852	369.837118878802\\
2.06017433783209	369.32597087883\\
2.17584701264384	368.24348702185\\
2.29453332055039	366.608080513792\\
2.41611885887564	364.44082504179\\
2.54048569425598	361.765102657201\\
2.66751335641091	358.606251372568\\
2.79707979786433	354.991218112904\\
2.92906231156623	350.948221721516\\
3.06333839847947	346.50642984194\\
3.19978657730947	341.695652707164\\
3.33828712881075	336.546056176451\\
3.47872276762314	331.087895767468\\
3.62097923544908	325.351272927307\\
3.76494581061211	319.365914356012\\
3.91051573062004	313.16097482475\\
4.05758652623473	306.764863603236\\
4.20606026762824	300.205094315747\\
4.35584372535946	293.508157773656\\
4.50684845099486	286.699417080377\\
4.65899078407456	279.803024070629\\
4.81219179365324	272.841855931967\\
4.9663771637051	265.837470666254\\
5.1214770321912	258.810079887551\\
5.2774257935021	251.778537326033\\
5.43416187331183	244.760341319667\\
5.59162748364951	237.771649529239\\
5.74976836427687	230.827304108374\\
5.90853351431734	223.940865595688\\
6.06787491555221	217.124653865822\\
6.2277472458537	210.38979457138\\
6.38810757773086	203.7462696187\\
6.54891505263915	197.202970334968\\
6.7101305160519	190.767752089818\\
6.87171609052083	184.447489218003\\
7.03363465285454	178.248129136331\\
7.19584916529502	172.174744541882\\
7.35832178643405	166.23158249893\\
7.52101265144993	160.422109042493\\
7.68387815679391	154.749047608879\\
7.84686850218643	149.214409094033\\
8.00992411822378	143.819510560891\\
8.17297041917479	138.564978457493\\
8.33591003504321	133.450730518536\\
8.49861124712342	128.475928112861\\
8.66089071061998	123.638887453946\\
8.82248761067275	118.936933645688\\
8.9830250715839	114.366176076876\\
9.1419528698098	109.921178006543\\
9.29846339972642	105.594489785222\\
9.45137095871896	101.37601994849\\
9.59894426095269	97.2522431944002\\
9.73868682362716	93.2053083007011\\
9.86707466842493	89.2122346889719\\
9.97929121298871	85.2445787457149\\
10.069045339559	81.2691560289099\\
10.1286047852548	77.2504682652146\\
10.1491856986769	73.1551789101762\\
10.1217656826768	68.9581864244283\\
10.0382238023619	64.6488040918928\\
9.8925277750456	60.2349299925778\\
9.68161113130484	55.7435132711484\\
9.40568709566339	51.2170355328839\\
9.06796695722317	46.7072401402089\\
8.67392764095793	42.2679690037609\\
8.23032634391985	37.9486374842716\\
7.74417230283783	33.7893778222514\\
7.2219001893121	29.818581180573\\
6.6689568039833	26.0530557061201\\
6.08984601237199	22.5001877001086\\
5.48850924237845	19.1610438292215\\
};
\addlegendentry{$N_t = 64$};

\end{axis}
\end{tikzpicture}%

%% file: Components/Figures/SEvsP_different_beamforming.tex
%
%
\definecolor{mycolor1}{rgb}{0.00000,0.44700,0.74100}%
\definecolor{mycolor2}{rgb}{0.85000,0.32500,0.09800}%
\definecolor{mycolor3}{rgb}{0.92900,0.69400,0.12500}%
\definecolor{mycolor4}{rgb}{0.49400,0.18400,0.55600}%
\begin{tikzpicture}

\begin{axis}[%
width=0.85\columnwidth,
height=0.425\columnwidth,
at={(0\textwidth,0\textwidth)},
scale only axis,
xmin=-20,
xmax=15,
xlabel={Input Power $P$ (dBm)},
ymin=0,
ymax=14,
ylabel={$SE$ (bit/sec/Hz)},
axis background/.style={fill=white},
legend style={at={(0.03,0.97)},anchor=north west,legend cell align=left,align=left,draw=white!15!black,legend columns=2}
]
\addplot [color=mycolor1,solid, very thick]
  table[row sep=crcr]{%
-20	0.177802673435707\\
-19.5	0.198432886265005\\
-19	0.22133189854661\\
-18.5	0.246721859877408\\
-18	0.274841078014602\\
-17.5	0.305944134917319\\
-17	0.34030179774377\\
-16.5	0.378200698492639\\
-16	0.419942759335484\\
-15.5	0.465844345501659\\
-15	0.516235133785967\\
-14.5	0.571456692182118\\
-14	0.631860774527658\\
-13.5	0.697807343013521\\
-13	0.769662340533161\\
-12.5	0.847795243657297\\
-12	0.932576435056782\\
-11.5	1.02437444102889\\
-11	1.12355308504697\\
-10.5	1.23046861167393\\
-10	1.34546683658221\\
-9.5	1.4688803777444\\
-9	1.60102602014167\\
-8.5	1.74220226172603\\
-8	1.8926870820875\\
-7.5	2.05273596760158\\
-7	2.22258021807756\\
-6.5	2.4024255504164\\
-6	2.59245100482427\\
-5.5	2.7928081489846\\
-5	3.00362056548463\\
-4.5	3.22498359785116\\
-4	3.45696432077108\\
-3.5	3.699601690233\\
-3	3.95290681884318\\
-2.5	4.2168633092737\\
-2	4.4914275625767\\
-1.5	4.77652895430689\\
-1	5.07206973393198\\
-0.5	5.37792444174514\\
0	5.69393853562416\\
0.5	6.01992574954714\\
1	6.35566342000082\\
1.5	6.70088453671687\\
2	7.05526446822875\\
2.5	7.4183989582098\\
3	7.78976771481945\\
3.5	8.1686741160436\\
4	8.55414527438534\\
4.5	8.9447665538296\\
5	9.33840892806627\\
5.5	9.73178532590472\\
6	10.1197463848082\\
6.5	10.4942123283083\\
7	10.8426787973312\\
7.5	11.1464094933095\\
8	11.3788066516903\\
8.5	11.5049291136253\\
9	11.483285315074\\
9.5	11.2704441523142\\
10	10.8280133389886\\
10.5	10.1310563284815\\
11	9.17713231072353\\
11.5	7.99481550481143\\
12	6.64912841833971\\
12.5	5.23936642224785\\
13	3.88457597963776\\
13.5	2.69675099945892\\
14	1.75087928933123\\
14.5	1.06705678179685\\
15	0.615423970302549\\
};
\addlegendentry{digital};

\addplot [color=mycolor2,dashed, very thick]
  table[row sep=crcr]{%
-20	0.159851525990391\\
-19.5	0.178451534198482\\
-19	0.199108307630641\\
-18.5	0.222025416732884\\
-18	0.247421607841121\\
-17.5	0.275530967166485\\
-17	0.306602892656591\\
-16.5	0.340901845823011\\
-16	0.378706858221627\\
-15.5	0.420310771396955\\
-15	0.46601919476756\\
-14.5	0.516149173041136\\
-14	0.571027563092504\\
-13.5	0.630989129494197\\
-13	0.696374377645947\\
-12.5	0.767527153235685\\
-12	0.844792046087881\\
-11.5	0.928511644839609\\
-11	1.01902369590663\\
-10.5	1.11665822551418\\
-10	1.22173468691608\\
-9.5	1.3345591961509\\
-9	1.45542191871573\\
-8.5	1.58459466637893\\
-8	1.72232875806754\\
-7.5	1.86885319146\\
-7	2.02437316273087\\
-6.5	2.18906896100363\\
-6	2.36309525166716\\
-5.5	2.54658074903063\\
-5	2.7396282640781\\
-4.5	2.94231509758622\\
-4	3.15469373277238\\
-3.5	3.37679276497317\\
-3	3.60861798829964\\
-2.5	3.85015353984311\\
-2	4.10136297881528\\
-1.5	4.36219014722775\\
-1	4.63255961366044\\
-0.5	4.91237643082969\\
0	5.20152482153812\\
0.5	5.49986521325747\\
1	5.80722871356813\\
1.5	6.12340756331195\\
2	6.4481391637327\\
2.5	6.78107968320538\\
3	7.12176056959898\\
3.5	7.46951681824457\\
4	7.82336849264451\\
4.5	8.18182530184151\\
5	8.54256658301568\\
5.5	8.90192622396157\\
6	9.25409118226812\\
6.5	9.58992811056807\\
7	9.89544041024506\\
7.5	10.150104854057\\
8	10.3257622410418\\
8.5	10.3871504989332\\
9	10.2951259849185\\
9.5	10.0127487804679\\
10	9.51300935820639\\
10.5	8.7860979563493\\
11	7.8447484796375\\
11.5	6.72803485214882\\
12	5.50391663459705\\
12.5	4.26594929709814\\
13	3.11674506978002\\
13.5	2.14065928617699\\
14	1.38186819796198\\
14.5	0.840587946454896\\
15	0.484696213928602\\
};
\addlegendentry{hybrid};

\addplot [color=mycolor4,dashdotted, very thick]
  table[row sep=crcr]{%
-20	0.121865449329183\\
-19.5	0.13634436398655\\
-19	0.152494012669202\\
-18.5	0.170495417230291\\
-18	0.190546517222549\\
-17.5	0.212863158302057\\
-17	0.237680009624979\\
-16.5	0.265251377890632\\
-16	0.29585188207076\\
-15.5	0.329776949977992\\
-15	0.367343096024749\\
-14.5	0.408887939180762\\
-14	0.454769921607175\\
-13.5	0.505367692021201\\
-13	0.561079123712453\\
-12.5	0.622319945321745\\
-12	0.689521972846139\\
-11.5	0.763130943481658\\
-11	0.843603965284359\\
-10.5	0.931406610478654\\
-10	1.02700969371687\\
-9.5	1.13088578881743\\
-9	1.24350554766461\\
-8.5	1.36533389237163\\
-8	1.49682615603029\\
-7.5	1.63842424817745\\
-7	1.79055291852581\\
-6.5	1.95361618677153\\
-6	2.12799399780797\\
-5.5	2.31403915095744\\
-5	2.51207453944996\\
-4.5	2.72239072290158\\
-4	2.94524384149548\\
-3.5	3.18085386635107\\
-3	3.42940316637518\\
-2.5	3.69103535757667\\
-2	3.96585438569892\\
-1.5	4.25392377553976\\
-1	4.55526595762446\\
-0.5	4.86986155002787\\
0	5.19764842182148\\
0.5	5.53852028106008\\
1	5.89232439143313\\
1.5	6.25885778907891\\
2	6.63786097928702\\
2.5	7.02900743075089\\
3	7.43188606288204\\
3.5	7.84597201541259\\
4	8.27057774885108\\
4.5	8.70477102826186\\
5	9.14723709557098\\
5.5	9.59604705641279\\
6	10.0482701836073\\
6.5	10.499331962058\\
7	10.9419752263757\\
7.5	11.3646512569262\\
8	11.7492231500647\\
8.5	12.0681587603973\\
9	12.2821033778869\\
9.5	12.3397271339443\\
10	12.1821417377486\\
10.5	11.7527980015458\\
11	11.0111053716353\\
11.5	9.94658593324448\\
12	8.59127011187945\\
12.5	7.02849158138917\\
13	5.39278839698338\\
13.5	3.8498500767038\\
14	2.54971302123578\\
14.5	1.57329779027555\\
15	0.913874496398642\\
};
\addlegendentry{analog};

\addplot [color=mycolor3,dotted, very thick]
  table[row sep=crcr]{%
-20	0.0794631540893841\\
-19.5	0.0890089471555667\\
-19	0.0996818171364081\\
-18.5	0.111610002941164\\
-18	0.124935169620059\\
-17.5	0.139813540286295\\
-17	0.156417060000251\\
-16.5	0.17493457655909\\
-16	0.195573019053319\\
-15.5	0.218558550602449\\
-15	0.244137666988845\\
-14.5	0.272578208155668\\
-14	0.304170244970844\\
-13.5	0.339226799600037\\
-13	0.378084354653601\\
-12.5	0.421103104395134\\
-12	0.468666901158744\\
-11.5	0.521182852131939\\
-11	0.579080526163272\\
-10.5	0.642810737464255\\
-10	0.71284388302912\\
-9.5	0.789667823104199\\
-9	0.873785308659909\\
-8.5	0.965710975859841\\
-8	1.06596794407069\\
-7.5	1.17508406994344\\
-7	1.2935879243822\\
-6.5	1.42200457069946\\
-6	1.56085122997416\\
-5.5	1.71063292285474\\
-5	1.8718381753563\\
-4.5	2.04493486949696\\
-4	2.23036630813424\\
-3.5	2.42854754760482\\
-3	2.63986203242085\\
-2.5	2.86465854404026\\
-2	3.10324845111908\\
-1.5	3.35590322168093\\
-1	3.62285212736751\\
-0.5	3.90428003381421\\
0	4.20032512403887\\
0.5	4.51107633402021\\
1	4.83657017472016\\
1.5	5.17678644402807\\
2	5.53164204645832\\
2.5	5.90098165387727\\
3	6.28456311249265\\
3.5	6.68203408043807\\
4	7.09289393663659\\
4.5	7.51643079858023\\
5	7.95161628174379\\
5.5	8.39692840145824\\
6	8.85005269107584\\
6.5	9.30737928709093\\
7	9.76316688674348\\
7.5	10.2081903170626\\
8	10.6276667989774\\
8.5	10.9983836260477\\
9	11.285452203533\\
9.5	11.4401783303829\\
10	11.4017747537254\\
10.5	11.1056296643432\\
11	10.4983724584441\\
11.5	9.55642019371961\\
12	8.30306832959865\\
12.5	6.81955460246173\\
13	5.24373477724851\\
13.5	3.74608362886019\\
14	2.47976860478042\\
14.5	1.52763128112685\\
15	0.884969151053108\\
};
\addlegendentry{quantized analog};

\end{axis}

\begin{axis}[%
width=0.283\columnwidth,
height=0.142\columnwidth,
at={(0.1\columnwidth,0.05\columnwidth)},
scale only axis,
xmin=-14.5,
xmax=-10.5,
xtick={-14,-13,-12,-11},
xticklabels={{},{},{},{}},
ymin=0.272578208155668,
ymax=1.23046861167393,
ytick={0.4,0.6,0.8,1,1.2},
yticklabels={{},{},{},{},{}},
axis background/.style={fill=white}
]
\addplot [color=mycolor1,solid, very thick,forget plot]
  table[row sep=crcr]{%
-14.5	0.571456692182118\\
-14	0.631860774527658\\
-13.5	0.697807343013521\\
-13	0.769662340533161\\
-12.5	0.847795243657297\\
-12	0.932576435056782\\
-11.5	1.02437444102889\\
-11	1.12355308504697\\
-10.5	1.23046861167393\\
};
\addplot [color=mycolor2,dashed, very thick,forget plot]
  table[row sep=crcr]{%
-14.5	0.516149173041136\\
-14	0.571027563092504\\
-13.5	0.630989129494197\\
-13	0.696374377645947\\
-12.5	0.767527153235685\\
-12	0.844792046087881\\
-11.5	0.928511644839609\\
-11	1.01902369590663\\
-10.5	1.11665822551418\\
};
\addplot [color=mycolor3,dotted, very thick,forget plot]
  table[row sep=crcr]{%
-14.5	0.272578208155668\\
-14	0.304170244970844\\
-13.5	0.339226799600037\\
-13	0.378084354653601\\
-12.5	0.421103104395134\\
-12	0.468666901158744\\
-11.5	0.521182852131939\\
-11	0.579080526163272\\
-10.5	0.642810737464255\\
};
\addplot [color=mycolor4,dashdotted, very thick,forget plot]
  table[row sep=crcr]{%
-14.5	0.408887939180762\\
-14	0.454769921607175\\
-13.5	0.505367692021201\\
-13	0.561079123712453\\
-12.5	0.622319945321745\\
-12	0.689521972846139\\
-11.5	0.763130943481658\\
-11	0.843603965284359\\
-10.5	0.931406610478654\\
};
\end{axis}
\end{tikzpicture}%

%% file: Components/Figures/EEvsP_different_beamforming.tex
%
%
\definecolor{mycolor1}{rgb}{0.00000,0.44700,0.74100}%
\definecolor{mycolor2}{rgb}{0.85000,0.32500,0.09800}%
\definecolor{mycolor3}{rgb}{0.92900,0.69400,0.12500}%
\definecolor{mycolor4}{rgb}{0.49400,0.18400,0.55600}%
\begin{tikzpicture}

\begin{axis}[%
width=0.85\columnwidth,
height=0.425\columnwidth,
at={(0\textwidth,0\textwidth)},
scale only axis,
xmin=-20,
xmax=15,
xlabel={Input Power $P$ (dBm)},
ymin=0,
ymax=250,
ylabel={$EE$ (Gbit/Joul)},
axis background/.style={fill=white},
legend style={at={(0.5,0.03)},anchor=south,legend cell align=left,align=left,draw=white!15!black, legend columns=2}
]
\addplot [color=mycolor1,solid,very thick]
  table[row sep=crcr]{%
-20	67.643775630035\\
-19.5	71.2699970328232\\
-19	75.0483338368897\\
-18.5	78.9785387338712\\
-18	83.0592572246245\\
-17.5	87.2878997769694\\
-17	91.6605186478272\\
-16.5	96.1716935620019\\
-16	100.814430807401\\
-15.5	105.580080502182\\
-15	110.458276770751\\
-14.5	115.436905301444\\
-14	120.502102233687\\
-13.5	125.638287540903\\
-13	130.828235062552\\
-12.5	136.053180140155\\
-12	141.292964489926\\
-11.5	146.526216572392\\
-11	151.73056437546\\
-10.5	156.882876288302\\
-10	161.959524677719\\
-9.5	166.936665941667\\
-9	171.79053024545\\
-8.5	176.497713865238\\
-8	181.035467073195\\
-7.5	185.381970783244\\
-7	189.516595706226\\
-6.5	193.420138496344\\
-6	197.075030258192\\
-5.5	200.465513772716\\
-5	203.577786838192\\
-4.5	206.400110157437\\
-4	208.922879186537\\
-3.5	211.138660246199\\
-3	213.042191936794\\
-2.5	214.630353439831\\
-2	215.902101569014\\
-1.5	216.858378370871\\
-1	217.501990552107\\
-0.5	217.837460855401\\
0	217.870849449268\\
0.5	217.609540014162\\
1	217.061979809271\\
1.5	216.237354480362\\
2	215.145164924296\\
2.5	213.794652291866\\
3	212.193983688856\\
3.5	210.34905852432\\
4	208.26171420041\\
4.5	205.926988622945\\
5	203.328928747989\\
5.5	200.434236088597\\
6	197.18290476625\\
6.5	193.475210348723\\
7	189.155539676527\\
7.5	183.996436420356\\
8	177.690984709505\\
8.5	169.865604187935\\
9	160.122749286898\\
9.5	148.111825594639\\
10	133.61682277546\\
10.5	116.653445558255\\
11	97.5802222724064\\
11.5	77.2199585930992\\
12	56.9309913070534\\
12.5	38.4703493637005\\
13	23.4962478881911\\
13.5	12.8746308238507\\
14	6.34022353033083\\
14.5	2.83603973852834\\
15	1.17115134867428\\
};
\addlegendentry{digital};

\addplot [color=mycolor2,dashed,very thick]
  table[row sep=crcr]{%
-20	61.1607187795194\\
-19.5	64.4579605735001\\
-19	67.8963262618961\\
-18.5	71.4760235541352\\
-18	75.196274015871\\
-17.5	79.0551891173793\\
-17	83.0496484128874\\
-16.5	87.175183660771\\
-16	91.4258731586666\\
-15.5	95.7942508955345\\
-15	100.271235266317\\
-14.5	104.846082012737\\
-14	109.506365717141\\
-13.5	114.237993573973\\
-13	119.025254305232\\
-12.5	123.850904004221\\
-12	128.696289438268\\
-11.5	133.541507984471\\
-11	138.365601990887\\
-10.5	143.146784029177\\
-10	147.862688308453\\
-9.5	152.490642516906\\
-9	157.00795359494\\
-8.5	161.392200449154\\
-8	165.621526400831\\
-7.5	169.674924218713\\
-7	173.532506893494\\
-6.5	177.175757840487\\
-6	180.58775493075\\
-5.5	183.753363609815\\
-5	186.659395325805\\
-4.5	189.294728512674\\
-4	191.650390415516\\
-3.5	193.719599055002\\
-3	195.49776555169\\
-2.5	196.982457801096\\
-2	198.173327021888\\
-1.5	199.071998879531\\
-1	199.681930559523\\
-0.5	200.008234101154\\
0	200.057464162006\\
0.5	199.837364633195\\
1	199.356562321545\\
1.5	198.624185906335\\
2	197.649372401135\\
2.5	196.440597976248\\
3	195.004729923169\\
3.5	193.345633985215\\
4	191.462076014295\\
4.5	189.344519065707\\
5	186.970238210786\\
5.5	184.296000608231\\
6	181.247549274119\\
6.5	177.705682269906\\
7	173.490523997053\\
7.5	168.349239573348\\
8	161.957166634832\\
8.5	153.944524497043\\
9	143.955413125992\\
9.5	131.732190513862\\
10	117.206097527605\\
10.5	100.577906476756\\
11	82.3905904261185\\
11.5	63.5994897638611\\
12	45.5779012417927\\
12.5	29.8889163539908\\
13	17.7505658595962\\
13.5	9.51249743465491\\
14	4.61642022479818\\
14.5	2.04797185485791\\
15	0.841743518478194\\
};
\addlegendentry{hybrid};

\addplot [color=mycolor4,dashdotted,very thick]
  table[row sep=crcr]{%
-20	46.2843522824341\\
-19.5	48.8870456537678\\
-19	51.6193780520446\\
-18.5	54.4849390498687\\
-18	57.4869173104431\\
-17.5	60.6280040527041\\
-17	63.910287467638\\
-16.5	67.3351389411488\\
-16	70.9030924575296\\
-15.5	74.6137191455656\\
-15	78.4654995667316\\
-14.5	82.4556970013412\\
-14	86.580235622325\\
-13.5	90.8335880062412\\
-13	95.2086768588585\\
-12.5	99.6967960678841\\
-12	104.287556182205\\
-11.5	108.968859111565\\
-11	113.72690621885\\
-10.5	118.546243041146\\
-10	123.409842657444\\
-9.5	128.299228281937\\
-9	133.194634089341\\
-8.5	138.075201676568\\
-8	142.919208043161\\
-7.5	147.704319634599\\
-7	152.40786592392\\
-6.5	157.007125268821\\
-6	161.479615404514\\
-5.5	165.80338091872\\
-5	169.957270379602\\
-4.5	173.921196404466\\
-4	177.676372805908\\
-3.5	181.205523961741\\
-3	184.49306264839\\
-2.5	187.525233672616\\
-2	190.290221647707\\
-1.5	192.778222095391\\
-1	194.98147561029\\
-0.5	196.894264974869\\
0	198.512874695087\\
0.5	199.835511204108\\
1	200.862179592183\\
1.5	201.594508589233\\
2	202.035508717266\\
2.5	202.18923751052\\
3	202.060327964967\\
3.5	201.653307846411\\
4	200.971591613012\\
4.5	200.015953273304\\
5	198.782172394177\\
5.5	197.257366900181\\
6	195.41426711479\\
6.5	193.202354727041\\
7	190.534499828101\\
7.5	187.267872134872\\
8	183.1794540694\\
8.5	177.94111825541\\
9	171.108050573587\\
9.5	162.143689294522\\
10	150.502970814495\\
10.5	135.775683060157\\
11	117.869545364172\\
11.5	97.2129223597995\\
12	74.9518732427661\\
12.5	53.0195159818449\\
13	33.7914687579919\\
13.5	19.1747877891746\\
14	9.68199918334843\\
14.5	4.40027459841016\\
15	1.83452212070935\\
};
\addlegendentry{analog};

\addplot [color=mycolor3,dotted,very thick]
  table[row sep=crcr]{%
-20	30.1800111318808\\
-19.5	31.9146632538273\\
-19	33.7423962660158\\
-18.5	35.6670244068273\\
-18	37.6923066860681\\
-17.5	39.8219022714675\\
-17	42.0593186832665\\
-16.5	44.4078523244246\\
-16	46.8705210021923\\
-15.5	49.4499882802645\\
-15	52.1484797472427\\
-14.5	54.9676916024567\\
-14	57.9086923466851\\
-13.5	60.9718188202916\\
-13	64.1565683453459\\
-12.5	67.4614892838269\\
-12	70.8840728941333\\
-11.5	74.4206499165421\\
-11	78.0662958002269\\
-10.5	81.8147488493077\\
-10	85.6583457606554\\
-9.5	89.587979002981\\
-9	93.5930802063369\\
-8.5	97.6616331712862\\
-8	101.780219267424\\
-7.5	105.934096896772\\
-7	110.107315399739\\
-6.5	114.282862353636\\
-6	118.442841746509\\
-5.5	122.568679100713\\
-5	126.641348367612\\
-4.5	130.641614401718\\
-4	134.550284113854\\
-3.5	138.348459036428\\
-3	142.017782016955\\
-2.5	145.540671064202\\
-2	148.900533948193\\
-1.5	152.081957913728\\
-1	155.070869688665\\
-0.5	157.854661701729\\
0	160.42228087731\\
0.5	162.764276294595\\
1	164.872801038802\\
1.5	166.741561228032\\
2	168.365700695658\\
2.5	169.741601913992\\
3	170.866570461705\\
3.5	171.738348392762\\
4	172.354365907044\\
4.5	172.710582106034\\
5	172.799670765642\\
5.5	172.608155919171\\
6	172.111868904085\\
6.5	171.268762727184\\
7	170.007707111223\\
7.5	168.211591874326\\
8	165.693525214682\\
8.5	162.167628076224\\
9	157.223210632196\\
9.5	150.323641725667\\
10	140.862010140305\\
10.5	128.299189119987\\
11	112.38094151094\\
11.5	93.3996388876169\\
12	72.4375461208681\\
12.5	51.4433972136621\\
13	32.8574916827271\\
13.5	18.6579625680893\\
14	9.41639996596186\\
14.5	4.27255231884738\\
15	1.77649719972502\\
};
\addlegendentry{quantized analog};

\end{axis}
\end{tikzpicture}%

%% file: Components/Figures/Reviewer1-Q1-SE.tex
%
%
\definecolor{mycolor1}{rgb}{0.00000,0.44700,0.74100}%
\definecolor{mycolor2}{rgb}{0.85000,0.32500,0.09800}%
\definecolor{mycolor3}{rgb}{0.92900,0.69400,0.12500}%
\definecolor{mycolor4}{rgb}{0.49400,0.18400,0.55600}%
\begin{tikzpicture}

\begin{axis}[%
width=0.85\columnwidth,
height=0.425\columnwidth,
at={(0\textwidth,0\textwidth)},
scale only axis,
xmin=-20,
xmax=15,
xlabel={Input Power $P$ (dBm)},
ymin=0,
ymax=8,
ylabel={$SE$ (bit/sec/Hz)},
axis background/.style={fill=white},
legend style={at={(0.03,0.97)},anchor=north west,legend cell align=left,align=left,draw=white!15!black}
]
\addplot [color=mycolor1,solid,very thick]
  table[row sep=crcr]{%
-20	0.867165971796202\\
-19.5	0.936523424194295\\
-19	1.00960877065221\\
-18.5	1.08643945816953\\
-18	1.16701881154592\\
-17.5	1.25133623926318\\
-17	1.33936760854923\\
-16.5	1.43107577235041\\
-16	1.5264112284428\\
-15.5	1.62531288954578\\
-15	1.72770894301401\\
-14.5	1.83351777934342\\
-14	1.94264897012134\\
-13.5	2.05500427792151\\
-13	2.17047868271481\\
-12.5	2.28896141137674\\
-12	2.41033695861438\\
-11.5	2.53448608896861\\
-11	2.66128681041511\\
-10.5	2.79061531050682\\
-10	2.92234684604663\\
-9.5	3.05635657705964\\
-9	3.19252033545506\\
-8.5	3.33071531830181\\
-8	3.47082069509537\\
-7.5	3.61271811766959\\
-7	3.7562921202722\\
-6.5	3.90143039534589\\
-6	4.04802392703741\\
-5.5	4.19596695829393\\
-5	4.34515675689783\\
-4.5	4.4954931283054\\
-4	4.64687759459836\\
-3.5	4.7992121128101\\
-3	4.95239713221298\\
-2.5	5.10632867275573\\
-2	5.26089392011613\\
-1.5	5.4159645361551\\
-1	5.57138641301609\\
-0.5	5.72696385608342\\
0	5.88243501875438\\
0.5	6.03743362715672\\
1	6.19142938460859\\
1.5	6.34363576206685\\
2	6.4928693931313\\
2.5	6.63734145267854\\
3	6.77436247884622\\
3.5	6.89995729836717\\
4	7.00843023214276\\
4.5	7.09200307268764\\
5	7.14075191268266\\
5.5	7.14312011921478\\
6	7.08717024951437\\
6.5	6.96241326237813\\
7	6.76167649029022\\
7.5	6.48234823128743\\
8	6.1266218109625\\
8.5	5.70083586252002\\
9	5.21429665216002\\
9.5	4.67802867641973\\
10	4.103962599786\\
10.5	3.50507895642869\\
11	2.89668669588898\\
11.5	2.29833476056281\\
12	1.73501472759642\\
12.5	1.23539558124621\\
13	0.825100733888892\\
13.5	0.516996961890441\\
14	0.30585076513807\\
14.5	0.172563242905597\\
15	0.0938285238706141\\
};
\addlegendentry{no crosstalk};

\addplot [color=mycolor2,solid,very thick]
  table[row sep=crcr]{%
-20	0.866024593261672\\
-19.5	0.935317375278153\\
-19	1.00833720462354\\
-18.5	1.08510170126459\\
-18	1.16561435286825\\
-17.5	1.2498647167691\\
-17	1.33782879126326\\
-16.5	1.42946953898991\\
-16	1.52473754264432\\
-15.5	1.62357177187583\\
-15	1.72590043991534\\
-14.5	1.831641929115\\
-14	1.94070576595738\\
-13.5	2.05299362794385\\
-13	2.16840036682271\\
-12.5	2.28681503460582\\
-12	2.40812190053421\\
-11.5	2.53220144844279\\
-11	2.65893134477553\\
-10.5	2.78818736781881\\
-10	2.91984428860943\\
-9.5	3.05377669351256\\
-9	3.1898597377156\\
-8.5	3.32796981785101\\
-8	3.46798515054124\\
-7.5	3.6097862415838\\
-7	3.75325622725563\\
-6.5	3.89828106394328\\
-6	4.04474953359148\\
-5.5	4.19255301806622\\
-5	4.34158497189695\\
-4.5	4.49173998435374\\
-4	4.6429122594245\\
-3.5	4.79499324154065\\
-3	4.9478679526277\\
-2.5	5.10140934486031\\
-2	5.25546955355619\\
-1.5	5.40986626135839\\
-1	5.56436131235502\\
-0.5	5.71862702878783\\
0	5.87219309850757\\
0.5	6.02436312254119\\
1	6.17408487623197\\
1.5	6.31975290408725\\
2	6.45891957074919\\
2.5	6.58789953345528\\
3	6.7012882768988\\
3.5	6.79149538141413\\
4	6.84851415916929\\
4.5	6.8602478052625\\
5	6.81365585067389\\
5.5	6.69667004293073\\
6	6.50036532794569\\
6.5	6.22061247941127\\
7	5.85866581500377\\
7.5	5.42069253016255\\
8	4.91665899012866\\
8.5	4.35910274397633\\
9	3.7624117468682\\
9.5	3.14325307769242\\
10	2.52225862144958\\
10.5	1.9258728912402\\
11	1.38582942784683\\
11.5	0.933207915668604\\
12	0.587422690696579\\
12.5	0.347756662204203\\
13	0.195768069134168\\
13.5	0.106057117899512\\
14	0.0558507330124665\\
14.5	0.0287997834094411\\
15	0.0146150988508208\\
};
\addlegendentry{-20 dB crosstalk};

\addplot [color=mycolor3,solid,very thick]
  table[row sep=crcr]{%
-20	0.865941770112182\\
-19.5	0.935212822802891\\
-19	1.00820688168401\\
-18.5	1.08494096154007\\
-18	1.1654178682717\\
-17.5	1.2496263918564\\
-17	1.3375416674492\\
-16.5	1.42912568718431\\
-16	1.52432794267258\\
-15.5	1.62308617672916\\
-15	1.72532722245702\\
-14.5	1.83096790830523\\
-14	1.93991600888281\\
-13.5	2.05207122284762\\
-13	2.16732616077091\\
-12.5	2.2855673271597\\
-12	2.40667608145476\\
-11.5	2.53052956245869\\
-11	2.65700155888758\\
-10.5	2.78596330505125\\
-10	2.91728417424423\\
-9.5	3.0508322319516\\
-9	3.18647459421403\\
-8.5	3.32407750970994\\
-8	3.4635060410093\\
-7.5	3.60462315052897\\
-7	3.74728788244787\\
-6.5	3.8913521440263\\
-6	4.03665527983871\\
-5.5	4.18301511987772\\
-5	4.33021333519422\\
-4.5	4.47797154009302\\
-4	4.62591230897902\\
-3.5	4.77349565928667\\
-3	4.91991603572405\\
-2.5	5.06393711338515\\
-2	5.20363286185105\\
-1.5	5.3359983760554\\
-1	5.45640704784372\\
-0.5	5.55794894135013\\
0	5.63081926113379\\
0.5	5.66212310361707\\
1	5.63659404332799\\
1.5	5.5385503617485\\
2	5.35481970827302\\
2.5	5.07768555628481\\
3	4.70679356166327\\
3.5	4.24961801665725\\
4	3.72089958038368\\
4.5	3.14178443281058\\
5	2.53937185146682\\
5.5	1.94682223152419\\
6	1.4021860279789\\
6.5	0.942160810707361\\
7	0.590199230355275\\
7.5	0.347160294633713\\
8	0.194126911194269\\
8.5	0.104553894227168\\
9	0.0548206447532277\\
9.5	0.0281967803082951\\
10	0.0142988663776138\\
10.5	0.00717307464959098\\
11	0.00356802655322442\\
11.5	0.0017630323629473\\
12	0.000866781431808481\\
12.5	0.000424726099913427\\
13	0.00020783282280227\\
13.5	0.000101814776505491\\
14	5.01012102695608e-05\\
14.5	2.48779101775209e-05\\
15	1.25444372289429e-05\\
};
\addlegendentry{-10 dB crosstalk};

\end{axis}
\end{tikzpicture}%

%% file: Components/Figures/Reviewer1-Q1-EE.tex
%
%
\definecolor{mycolor1}{rgb}{0.00000,0.44700,0.74100}%
\definecolor{mycolor2}{rgb}{0.85000,0.32500,0.09800}%
\definecolor{mycolor3}{rgb}{0.92900,0.69400,0.12500}%
\definecolor{mycolor4}{rgb}{0.49400,0.18400,0.55600}%
\begin{tikzpicture}

\begin{axis}[%
width=0.85\columnwidth,
height=0.425\columnwidth,
at={(0\textwidth,0\textwidth)},
scale only axis,
xmin=-20,
xmax=15,
xlabel={Input Power $P$ (dBm)},
ymin=0,
ymax=400,
ylabel={$EE$ (Gbit/Joul)},
axis background/.style={fill=white},
legend style={legend cell align=left,align=left,draw=white!15!black}
]
\addplot [color=mycolor1,solid,very thick]
  table[row sep=crcr]{%
-20	329.348601649503\\
-19.5	335.795789834965\\
-19	341.753593500145\\
-18.5	347.191664276724\\
-18	352.083653361745\\
-17.5	356.407464732287\\
-17	360.145428394616\\
-16.5	363.284393592323\\
-16	365.815744354825\\
-15.5	367.735341940454\\
-15	369.043400533574\\
-14.5	369.744303935747\\
-14	369.846371919595\\
-13.5	369.361585393859\\
-13	368.305279591084\\
-12.5	366.695814191365\\
-12	364.554228710029\\
-11.5	361.903890686199\\
-11	358.770143295289\\
-10.5	355.179958046118\\
-10	351.161597273602\\
-9.5	346.744290244573\\
-9	341.957925883261\\
-8.5	336.832764403815\\
-8	331.399169505697\\
-7.5	325.687362229629\\
-7	319.727197065396\\
-6.5	313.54796042202\\
-6	307.178191084944\\
-5.5	300.645521758144\\
-5	293.976540176844\\
-4.5	287.196667519351\\
-4	280.330050859915\\
-3.5	273.399465063418\\
-3	266.426217637988\\
-2.5	259.430047344868\\
-2	252.42900337793\\
-1.5	245.439286010523\\
-1	238.475020802408\\
-0.5	231.547925417866\\
0	224.666809104943\\
0.5	217.836818142483\\
1	211.058305410286\\
1.5	204.32516238896\\
2	197.622423092228\\
2.5	190.922974635507\\
3	184.183381263841\\
3.5	177.33930359189\\
4	170.301933097497\\
4.5	162.958192765246\\
5	155.178461310271\\
5.5	146.834739125877\\
6	137.828119160844\\
6.5	128.118116632114\\
7	117.74223770598\\
7.5	106.816789381955\\
8	95.5187610523522\\
8.5	84.0569906733618\\
9	72.642942972605\\
9.5	61.4691735065794\\
10	50.7019682315337\\
10.5	40.4928672666962\\
11	31.00788997864\\
11.5	22.4627666351986\\
12	15.1365982263001\\
12.5	9.31922234028457\\
13	5.17012046810496\\
13.5	2.57498521614765\\
14	1.16140398296984\\
14.5	0.482633141081189\\
15	0.188352452412745\\
};
\addlegendentry{no crosstalk};

\addplot [color=mycolor2,solid,very thick]
  table[row sep=crcr]{%
-20	314.840038245118\\
-19.5	321.012842738249\\
-19	326.718193593326\\
-18.5	331.927012560047\\
-18	336.614041997702\\
-17.5	340.75808831945\\
-17	344.342188989854\\
-16.5	347.35370296253\\
-16	349.784326797817\\
-15.5	351.630040783357\\
-15	352.890991112938\\
-14.5	353.571315505876\\
-14	353.678920544111\\
-13.5	353.22521946941\\
-13	352.224839248719\\
-12.5	350.695305434802\\
-12	348.656712791754\\
-11.5	346.131388898964\\
-11	343.143557071308\\
-10.5	339.719004008623\\
-10	335.884756671867\\
-9.5	331.668772017866\\
-9	327.099642432783\\
-8.5	322.206318993464\\
-8	317.017854048263\\
-7.5	311.563164025999\\
-7	305.870812824879\\
-6.5	299.968815565035\\
-6	293.884461860718\\
-5.5	287.644157017621\\
-5	281.2732785994\\
-4.5	274.796044508106\\
-4	268.23538689586\\
-3.5	261.612823580802\\
-3	254.948314736395\\
-2.5	248.260086784743\\
-2	241.56439663631\\
-1.5	234.875196197762\\
-1	228.203637371096\\
-0.5	221.557329088323\\
0	214.939218078404\\
0.5	208.345914741123\\
1	201.765235063397\\
1.5	195.172714131002\\
2	188.526953771732\\
2.5	181.764066573316\\
3	174.792405334404\\
3.5	167.490318019668\\
4	159.711306024146\\
4.5	151.301035030298\\
5	142.126996195527\\
5.5	132.11413205252\\
6	121.272910990691\\
6.5	109.706965841732\\
7	97.597240115897\\
7.5	85.1715040626791\\
8	72.6733003049712\\
8.5	60.3431657773325\\
9	48.4228505128465\\
9.5	37.1859423542093\\
10	26.9758610020763\\
10.5	18.2026024837739\\
11	11.2477476293694\\
11.5	6.28995708412149\\
12	3.17205789535224\\
12.5	1.45208873643046\\
13	0.611998428127353\\
13.5	0.241508509669439\\
14	0.0906375075464891\\
14.5	0.0327596011741938\\
15	0.0115114708348091\\
};
\addlegendentry{-20 dB crosstalk};

\addplot [color=mycolor3,solid,very thick]
  table[row sep=crcr]{%
-20	230.544395572682\\
-19.5	235.063671318854\\
-19	239.24067617485\\
-18.5	243.054142849081\\
-18	246.485605476732\\
-17.5	249.519578051735\\
-17	252.143676865947\\
-16.5	254.348686870275\\
-16	256.128573593402\\
-15.5	257.480443775609\\
-15	258.404459138803\\
-14.5	258.903708678297\\
-14	258.984045506808\\
-13.5	258.653894607019\\
-13	257.924037875781\\
-12.5	256.807382606332\\
-12	255.318719101027\\
-11.5	253.474472485837\\
-11	251.292453055341\\
-10.5	248.791608647238\\
-10	245.991781643629\\
-9.5	242.913472209761\\
-9	239.57760826029\\
-8.5	236.005321288889\\
-8	232.217725438898\\
-7.5	228.235694756036\\
-7	224.079630014708\\
-6.5	219.769201164479\\
-6	215.323043233884\\
-5.5	210.758370792676\\
-5	206.090456274814\\
-4.5	201.331886882027\\
-4	196.49146839952\\
-3.5	191.572576541941\\
-3	186.570665116015\\
-2.5	181.469537697491\\
-2	176.235931195437\\
-1.5	170.812101234972\\
-1	165.106770331422\\
-0.5	158.986481559041\\
0	152.27229715044\\
0.5	144.749718117306\\
1	136.199171518268\\
1.5	126.446431257242\\
2	115.418503684523\\
2.5	103.181935753156\\
3	89.9489933825172\\
3.5	76.0588514917914\\
4	61.9553467707361\\
4.5	48.1754332284502\\
5	35.3397085966438\\
5.5	24.1108604030981\\
6	15.0740668255388\\
6.5	8.53640108719843\\
7	4.36244627048356\\
7.5	2.02461105008518\\
8	0.86524036279426\\
8.5	0.346145264350956\\
9	0.131606877817192\\
9.5	0.0481413590136515\\
10	0.0171013245164357\\
10.5	0.0059406263416181\\
11	0.00202857593571926\\
11.5	0.000683692971618948\\
12	0.000228179791358131\\
12.5	7.56331022384653e-05\\
13	2.49698330440719e-05\\
13.5	8.23683838005933e-06\\
14	2.72523348223151e-06\\
14.5	9.08832279350113e-07\\
15	3.07506546198634e-07\\
};
\addlegendentry{-10 dB crosstalk};

\end{axis}
\end{tikzpicture}%

%% file: Manuscript-r02-Arxiv.bbl
\begin{thebibliography}{10}
\providecommand{\url}[1]{#1}
\csname url@samestyle\endcsname
\providecommand{\newblock}{\relax}
\providecommand{\bibinfo}[2]{#2}
\providecommand{\BIBentrySTDinterwordspacing}{\spaceskip=0pt\relax}
\providecommand{\BIBentryALTinterwordstretchfactor}{4}
\providecommand{\BIBentryALTinterwordspacing}{\spaceskip=\fontdimen2\font plus
\BIBentryALTinterwordstretchfactor\fontdimen3\font minus
  \fontdimen4\font\relax}
\providecommand{\BIBforeignlanguage}[2]{{%
\expandafter\ifx\csname l@#1\endcsname\relax
\typeout{** WARNING: IEEEtran.bst: No hyphenation pattern has been}%
\typeout{** loaded for the language `#1'. Using the pattern for}%
\typeout{** the default language instead.}%
\else
\language=\csname l@#1\endcsname
\fi
#2}}
\providecommand{\BIBdecl}{\relax}
\BIBdecl

\bibitem{Larsson2014}
E.~G. Larsson, O.~Edfors, F.~Tufvesson, and T.~L. Marzetta, ``Massive {MIMO}
  for next generation wireless systems,'' \emph{IEEE Commun. Mag.}, vol.~52,
  no.~2, pp. 186--195, Feb. 2014.

\bibitem{Han2015}
S.~Han, C.~l.~I, Z.~Xu, and C.~Rowell, ``Large-scale antenna systems with
  hybrid analog and digital beamforming for millimeter wave {5G},'' \emph{IEEE
  Commun. Mag.}, vol.~53, no.~1, pp. 186--194, Jan. 2015.

\bibitem{Rangan2014}
S.~Rangan, T.~S. Rappaport, and E.~Erkip, ``Millimeter-wave cellular wireless
  networks: Potentials and challenges,'' \emph{Proc. IEEE}, vol. 102, no.~3,
  pp. 366--385, Mar. 2014.

\bibitem{Pi:11}
Z.~Pi and F.~Khan, ``An introduction to millimeter-wave mobile broadband
  systems,'' \emph{IEEE Communications Magazine}, vol.~49, no.~6, pp. 101--107,
  2011.

\bibitem{Hur2013}
S.~Hur, T.~Kim, D.~J. Love, J.~V. Krogmeier, T.~A. Thomas, and A.~Ghosh,
  ``Millimeter wave beamforming for wireless backhaul and access in small cell
  networks,'' \emph{IEEE Trans. Commun.}, vol.~61, no.~10, pp. 4391--4403, Oct.
  2013.

\bibitem{Ayach2014}
O.~E. Ayach, S.~Rajagopal, S.~Abu-Surra, Z.~Pi, and R.~W. Heath, ``Spatially
  sparse precoding in millimeter wave {MIMO} systems,'' \emph{IEEE Trans.
  Wireless Commun.}, vol.~13, no.~3, pp. 1499--1513, Mar. 2014.

\bibitem{Alkhateeb2014}
A.~Alkhateeb, O.~E. Ayach, G.~Leus, and R.~W. Heath, ``Channel estimation and
  hybrid precoding for millimeter wave cellular systems,'' \emph{IEEE J. Sel.
  Top. Signal Process.}, vol.~8, no.~5, pp. 831--846, Oct. 2014.

\bibitem{Kutty2016}
S.~Kutty and D.~Sen, ``Beamforming for millimeter wave communications: An
  inclusive survey,'' \emph{IEEE Communications Surveys Tutorials}, vol.~18,
  no.~2, pp. 949--973, Sep. 2016.

\bibitem{Noh2016}
S.~Noh, M.~D. Zoltowski, and D.~J. Love, ``Training sequence design for
  feedback assisted hybrid beamforming in massive {MIMO} systems,'' \emph{IEEE
  Trans. Commun.}, vol.~64, no.~1, pp. 187--200, Jan. 2016.

\bibitem{Song2017}
J.~Song, J.~Choi, and D.~J. Love, ``Common codebook millimeter wave beam
  design: Designing beams for both sounding and communication with uniform
  planar arrays,'' \emph{IEEE Trans. Commun.}, vol.~65, no.~4, pp. 1859--1872,
  Apr. 2017.

\bibitem{He2017}
S.~He, J.~Wang, Y.~Huang, B.~Ottersten, and W.~Hong, ``Codebook-based hybrid
  precoding for millimeter wave multiuser systems,'' \emph{IEEE Trans. Signal
  Process.}, vol.~65, no.~20, pp. 5289--5304, Oct. 2017.

\bibitem{Shokri-Ghadikolaei2016}
H.~Shokri-Ghadikolaei, F.~Boccardi, C.~Fischione, G.~Fodor, and M.~Zorzi,
  ``Spectrum sharing in {mmWave} cellular networks via cell association,
  coordination, and beamforming,'' \emph{IEEE J. Sel. Areas Commun.}, vol.~34,
  no.~11, pp. 2902--2917, Nov. 2016.

\bibitem{Studer2010}
C.~Studer, M.~Wenk, and A.~Burg, ``{MIMO} transmission with residual
  transmit-{RF} impairments,'' in \emph{Proc. Int. ITG Workshop Smart Antennas
  (WSA)}, Feb. 2010, pp. 189--196.

\bibitem{Qi2010}
J.~Qi and S.~Aissa, ``Analysis and compensation of power amplifier nonlinearity
  in {MIMO} transmit diversity systems,'' \emph{IEEE Trans. Veh. Technol.},
  vol.~59, no.~6, pp. 2921--2931, July 2010.

\bibitem{Studer2011}
C.~Studer, M.~Wenk, and A.~Burg, ``System-level implications of residual
  transmit-{RF} impairments in {MIMO} systems,'' in \emph{Proc. 5th European
  Conf. Antennas and Propagation (EUCAP)}, Apr. 2011, pp. 2686--2689.

\bibitem{Bjoernson2012}
E.~Bj{\"o}rnson, P.~Zetterberg, and M.~Bengtsson, ``Optimal coordinated
  beamforming in the multicell downlink with transceiver impairments,'' in
  \emph{Proc. IEEE Global Communications Conf. (GLOBECOM)}, Dec. 2012, pp.
  4775--4780.

\bibitem{Ghannouchi2009}
F.~M. Ghannouchi and O.~Hammi, ``Behavioral modeling and predistortion,''
  \emph{IEEE Microwave Mag.}, vol.~10, no.~7, pp. 52--64, Dec. 2009.

\bibitem{Qi2012}
J.~Qi and S.~Aissa, ``On the power amplifier nonlinearity in {MIMO} transmit
  beamforming systems,'' \emph{IEEE Trans. Commun.}, vol.~60, no.~3, pp.
  876--887, March 2012.

\bibitem{Zavjalov:16}
S.~V. Zavjalov, D.~K. Fadeev, and S.~V. Volvenko, ``Influence of input power
  backoff of nonlinear power amplifier on {BER} performance of optimal {SEFDM}
  signals,'' in \emph{$8^{th}$ International Congress on Ultra Modern
  Telecommunications and Control Systems and Workshops (ICUMT)}, 2016.

\bibitem{Schenk2008}
T.~Schenk, \emph{{RF} imperfections in high-rate wireless systems: impact and
  digital compensation}.\hskip 1em plus 0.5em minus 0.4em\relax Springer
  Science \& Business Media, 2008.

\bibitem{Lucciardi:16}
J.-A. Lucciardi, N.~Thomas, M.-L. Boucheret, C.~Poulliat, and G.~Mesnager,
  ``Trade-off between spectral efficiency increase and papr reduction when
  using ftn signaling: Impact of non linearities,'' in \emph{IEEE International
  Conference on Communications (ICC)}, 2016.

\bibitem{Ochiai2013}
H.~Ochiai, ``An analysis of band-limited communication systems from amplifier
  efficiency and distortion perspective,'' \emph{IEEE Trans. Commun.}, vol.~61,
  no.~4, pp. 1460--1472, Apr. 2013.

\bibitem{Bjornson2013}
E.~Bj\"ornson, P.~Zetterberg, M.~Bengtsson, and B.~Ottersten, ``Capacity limits
  and multiplexing gains of {MIMO} channels with transceiver impairments,''
  \emph{IEEE Commun. Lett.}, vol.~17, no.~1, pp. 91--94, Jan. 2013.

\bibitem{Bjoernson2014}
E.~Bj\"ornson, J.~Hoydis, M.~Kountouris, and M.~Debbah, ``Massive {MIMO}
  systems with non-ideal hardware: Energy efficiency, estimation, and capacity
  limits,'' \emph{IEEE Trans. Inf. Theory}, vol.~60, no.~11, pp. 7112--7139,
  Nov. 2014.

\bibitem{Bjoernson2014a}
E.~Bj\"{o}rnson, M.~Bengtsson, and B.~Ottersten, ``Optimal multiuser transmit
  beamforming: A difficult problem with a simple solution structure [lecture
  notes],'' \emph{IEEE Signal. Proc. Mag.}, vol.~31, no.~4, pp. 142--148, July
  2014.

\bibitem{Brandt2014}
R.~Brandt, E.~Bj{\"o}rnson, and M.~Bengtsson, ``Weighted sum rate optimization
  for multicell {MIMO} systems with hardware-impaired transceivers,'' in
  \emph{Proc. Acoustics, Speech and Signal Processing (ICASSP) 2014 IEEE
  International Conference on}, May 2014, pp. 479--483.

\bibitem{Moghadam2012}
N.~N. Moghadam, P.~Zetterberg, P.~H\"andel, and H.~Hjalmarsson, ``Correlation
  of distortion noise between the branches of {MIMO} transmit antennas,'' in
  \emph{Proc. Indoor and Mobile Radio Communications - (PIMRC) 2012 IEEE 23rd
  Int. Symp. Personal}, Sep. 2012, pp. 2079--2084.

\bibitem{Bjoernson2013}
E.~Bj{\"o}rnson and E.~Jorswieck, ``Optimal resource allocation in coordinated
  multi-cell systems,'' \emph{Foundations and Trends{\textregistered} in
  Communications and Information Theory}, vol.~9, no. 2-3, 2013.

\bibitem{Bjoernson2015}
E.~Bj{\"o}rnson, L.~Sanguinetti, J.~Hoydis, and M.~Debbah, ``Optimal design of
  energy-efficient multi-user {MIMO} systems: Is massive {MIMO} the answer?''
  \emph{IEEE Trans. Wireless Commun.}, vol.~14, no.~6, pp. 3059--3075, June
  2015.

\bibitem{Sabbaghian2013}
M.~Sabbaghian, A.~I. Sulyman, and V.~Tarokh, ``Analysis of the impact of
  nonlinearity on the capacity of communication channels,'' \emph{IEEE Trans.
  Inf. Theory}, vol.~59, no.~11, pp. 7671--7683, Nov 2013.

\bibitem{Fozooni2015}
M.~Fozooni, M.~Matthaiou, E.~Bj\"{o}rnson, and T.~Q. Duong, ``Performance
  limits of {MIMO} systems with nonlinear power amplifiers,'' in \emph{Proc.
  2015 IEEE Global Communications Conference (GLOBECOM)}, Dec 2015, pp. 1--7.

\bibitem{Wu2016}
M.~Wu, D.~Wuebben, A.~Dekorsy, P.~Baracca, V.~Braun, and H.~Halbauer,
  ``Hardware impairments in millimeter wave communications using {OFDM} and
  {SC-FDE},'' in \emph{Proc. Smart Antennas (WSA), 2016 International ITG
  Workshop on}, March 2016, pp. 1--8.

\bibitem{Khansefid2016}
A.~Khansefid, H.~Minn, Q.~Zhan, N.~Al-Dhahir, H.~Huang, and X.~Du, ``Waveform
  parameter design and comparisons for millimeter-wave massive {MIMO} systems
  with {RF} distortions,'' in \emph{Proc. IEEE Globecom Workshops (GC Wkshps)},
  Dec 2016, pp. 1--6.

\bibitem{Yan2017}
H.~Yan and D.~Cabric, ``Digital predistortion for hybrid precoding architecture
  in millimeter-wave massive {MIMO} systems,'' in \emph{Proc. 42nd IEEE Int.
  Conf. on Acoustics, Speech and Signal Process.}, Mar. 2017.

\bibitem{Persson2013}
D.~Persson, T.~Eriksson, and E.~G. Larsson, ``Amplifier-aware multiple-input
  multiple-output power allocation,'' \emph{IEEE Communications Letters},
  vol.~17, no.~6, pp. 1112--1115, June 2013.

\bibitem{Persson2014}
------, ``Amplifier-aware multiple-input single-output capacity,'' \emph{IEEE
  Trans. Commun.}, vol.~62, no.~3, pp. 913--919, March 2014.

\bibitem{Ying2015}
D.~Ying, D.~J. Love, and B.~M. Hochwald, ``Closed-loop precoding and capacity
  analysis for multiple-antenna wireless systems with user radiation exposure
  constraints,'' \emph{IEEE Trans. Wireless Commun.}, vol.~14, no.~10, pp.
  5859--5870, Oct. 2015.

\bibitem{Ying2017}
------, ``Sum-rate analysis for multi-user {MIMO} systems with user exposure
  constraints,'' \emph{IEEE Trans. Wireless Commun.}, vol.~16, no.~11, pp.
  7376--7388, Nov. 2017.

\bibitem{Castellanos2016}
M.~R. Castellanos, D.~J. Love, and B.~M. Hochwald, ``Hybrid precoding for
  millimeter wave systems with a constraint on user electromagnetic radiation
  exposure,'' in \emph{2016 50th Asilomar Conference on Signals, Systems and
  Computers}, Nov. 2016, pp. 296--300.

\bibitem{Mollen2016}
C.~Moll\'{e}n, E.~G. Larsson, and T.~Eriksson, ``Waveforms for the massive
  {MIMO} downlink: Amplifier efficiency, distortion and performance,''
  \emph{IEEE Trans. Commun.}, vol.~64, no.~12, pp. 5050 -- 5063, Dec. 2016.

\bibitem{Mendez-Rial2016}
R.~Mendez-Rial, C.~Rusu, N.~Gonzalez-Prelcic, A.~Alkhateeb, and R.~W. Heath,
  ``Hybrid {MIMO} architectures for millimeter wave communications: Phase
  shifters or switches?'' \emph{IEEE Access}, vol.~4, pp. 247--267, 2016.

\bibitem{Kenington2000}
P.~B. Kenington, \emph{High-Linearity {RF} Amplifier Design}.\hskip 1em plus
  0.5em minus 0.4em\relax Artech House Publishers, 2000.

\bibitem{Cripps2002}
S.~C. Cripps, \emph{Advanced Techniques in {RF} Power Amplifier Design}.\hskip
  1em plus 0.5em minus 0.4em\relax Artech House Publishers, 2002.

\bibitem{Mollen2017}
\BIBentryALTinterwordspacing
C.~Moll{\'{e}}n, U.~Gustavsson, T.~Eriksson, and E.~G. Larsson. (2017, Nov.)
  Spatial characteristics of distortion radiated from antenna arrays with
  transceiver nonlinearities. [Online]. Available:
  \url{http://arxiv.org/abs/1711.02439}
\BIBentrySTDinterwordspacing

\bibitem{Akdeniz2014}
M.~R. Akdeniz, Y.~Liu, M.~K. Samimi, S.~Sun, S.~Rangan, T.~S. Rappaport, and
  E.~Erkip, ``Millimeter wave channel modeling and cellular capacity
  evaluation,'' \emph{IEEE J. Sel. Areas Commun.}, vol.~32, no.~6, pp.
  1164--1179, Jun. 2014.

\bibitem{Dardari2000}
D.~Dardari, V.~Tralli, and A.~Vaccari, ``A theoretical characterization of
  nonlinear distortion effects in {OFDM} systems,'' \emph{IEEE Trans. Commun.},
  vol.~48, no.~10, pp. 1755--1764, Oct. 2000.

\bibitem{Raeesi2017}
O.~Raeesi, A.~Gokceoglu, P.~C. Sofotasios, M.~Renfors, and M.~Valkama,
  ``Modeling and estimation of massive mimo channel non-reciprocity:
  Sparsity-aided approach,'' in \emph{2017 25th European Signal Processing
  Conference (EUSIPCO)}, Aug 2017, pp. 2596--2600.

\bibitem{Goldsmith1997}
A.~J. Goldsmith and P.~P. Varaiya, ``Capacity of fading channels with channel
  side information,'' \emph{IEEE Trans. Inf. Theory}, vol.~43, no.~6, pp.
  1986--1992, Nov. 1997.

\bibitem{Faulkner1992}
M.~Faulkner and T.~Mattsson, ``Spectral sensitivity of power amplifiers to
  quadrature modulator misalignment,'' \emph{IEEE Trans. Veh. Technol.},
  vol.~41, no.~4, pp. 516--525, Nov 1992.

\bibitem{Bussgang1952}
J.~J. Bussgang, ``Crosscorrelation function of amplitude-distorted {Gaussian}
  signals,'' Research Lab. Electron, M.IT., Cambridge, MA, Tech. Rep. 216,
  March 1952.

\bibitem{Isserlis1918}
L.~Isserlis, ``On a formula for the product-moment coefficient of any order of
  a normal frequency distribution in any number of variables,''
  \emph{Biometrika}, vol.~12, no. 1/2, pp. 134--139, 1918.

\bibitem{Reed1962}
I.~Reed, ``On a moment theorem for complex gaussian processes,'' \emph{IRE
  Transactions on Information Theory}, vol.~8, no.~3, pp. 194--195, April 1962.

\bibitem{Zhou2004}
\BIBentryALTinterwordspacing
G.~T. Zhou and R.~Raich, ``Spectral analysis of polynomial nonlinearity with
  applications to {RF} power amplifiers,'' \emph{EURASIP Journal on Advances in
  Signal Processing}, vol. 2004, no.~12, p.~1, Sep. 2004. [Online]. Available:
  \url{http://dx.doi.org/10.1155/S1110865704312114}
\BIBentrySTDinterwordspacing

\bibitem{Horn2012}
R.~A. Horn and C.~R. Johnson, \emph{Matrix analysis}.\hskip 1em plus 0.5em
  minus 0.4em\relax Cambridge university press, 2012.

\end{thebibliography}
